%% file: datapaper.tex
\documentclass[fleqn,usenatbib,useAMS]{mnras}

\newcommand{\kms}{\ensuremath{\rm ~km~s^{-1}}\xspace} 
\newcommand{\woi}{\ensuremath{W_{80}}\xspace}
\newcommand{\Hb}{H$\beta$\xspace}
\newcommand{\Ha}{H$\alpha$\xspace}
\newcommand{\oiii}{[\ion{O}{iii}]\xspace}
\newcommand{\nii}{[\ion{N}{ii}]\xspace}
\newcommand{\sii}{[\ion{S}{ii}]\xspace}

\newcommand{\jha}{\ensuremath{j_{\rm H\alpha}}\xspace}
\newcommand{\lha}{\ensuremath{L_{\rm H\alpha}}\xspace}
\newcommand{\wha}{\ensuremath{W_{\lambda} (H\alpha)}\xspace}
\newcommand{\lagn}{\ensuremath{L_{\rm AGN}}\xspace}
\newcommand{\mdot}{\ensuremath{\dot{M}}\xspace}

\usepackage[T1]{fontenc}
\usepackage{ae,aecompl}

\usepackage{graphicx}
\usepackage{natbib}
\usepackage{amsmath}	
\usepackage{amssymb}	
\usepackage{multicol}        
\usepackage{bm}		
\usepackage{hyperref}
\usepackage{xcolor}

\usepackage{xspace}

\title[AGNIFS survey of local AGN]{AGNIFS survey of local AGN: GMOS-IFU data and outflows in 30 sources}

\author[D. Ruschel-Dutra]{
    D. Ruschel-Dutra$^{1}$\thanks{Contact e-mail:
        \href{mailto:daniel.ruschel@ufsc.br}
        {daniel.ruschel@ufsc.br}},
    T. Storchi-Bergmann$^{2}$,
    A. Schnorr-M\"uller$^{2}$,
    R. A. Riffel$^{3}$,
    \and
    B. Dall'Agnol de Oliveira$^{2}$,
    D. Lena$^{4,5,6}$,
    A. Robinson$^7$,
    N. Nagar$^8$
    and 
    M. Elvis$^9$
    \\
    $^{1}$Departamento de F\'isica, Universidade Federal de Santa Catarina, P.O. Box 476, 88040-900, Florian\'opolis, SC, Brazil \\
    $^{2}$Departamento de Astronomia, Universidade Federal do Rio Grande do Sul, IF, CP 15051, 91501-970, Porto Alegre, RS, Brazil\\
    $^{3}$Departamento de F\'isica, Centro de Ci\^encias Naturais e Exatas, Universidade Federal de Santa Maria, 97105-900,\\
    Santa Maria, RS, Brazil\\
    $^{4}$SRON Netherlands Institute for Space Research, Sorbonnelaan 2, NL-3584 CA Utrecht, the Netherlands\\
    $^5$Department of Astrophysics/IMAPP, Radboud University Nijmegen, PO Box 9010, NL-6500 GL Nijmegen, the Netherlands\\
    $^{6}$ ASML Netherlands B.V., De Run 6501, 5504 DR Veldhoven, The Netherlands \\
    $^{7}$School of Physics and Astronomy, Rochester Institute of Technology, 85 Lomb Memorial Dr., Rochester, NY 14623, USA\\
    $^{8}$Universidad de Concepci\'on, Departamento de Astronom\'ia, Casilla 160-C, Concepci\'on, Chile\\
    $^{9}$ Harvard-Smithsonian Center for Astrophysics, 60 Garden  Street, Cambridge, MA 02138, USA
}

\date{Last updated \today}

\pubyear{2020}

\begin{document}
\label{firstpage}
\pagerange{\pageref{firstpage}--\pageref{lastpage}}
\maketitle

\begin{abstract}

    We analyse optical datacubes of the inner kiloparsec  of 30 local ($z\le0.02$) active galactic nuclei (AGN) hosts that our research group, AGNIFS, has collected over the past decade via observations with the integral field units of the Gemini Multi-Object Spectrographs.
    Spatial resolutions range between 50\,pc and 300\,pc and spectral coverage is from 4800\AA\ or 5600\AA\ to 7000\AA, at velocity resolutions of $\approx$50\,\kms.
    We derive maps of the gas excitation and kinematics,
        determine the AGN ionisation axis -- which has random orientation relative to the galaxy, and the kinematic major axes of the emitting gas.
        We find
    that rotation dominates the gas kinematics in most cases, but is disturbed by the presence of inflows and outflows.
    Outflows have been found in 21 nuclei, usually along the ionisation axis. 
    The gas velocity dispersion is traced by \woi (velocity width encompassing 80 per cent of the line flux), 
    adopted as a tracer of outflows.
        In 7 sources \woi is enhanced perpendicularly to the ionisation axis, indicating lateral expansion of the outflow.
We have estimated mass-outflow rates $\dot{M}$ and powers $\dot{E}$, finding median values of $\log\,[\dot{M}/({\rm\,M_\odot\,yr^{-1}})]=-2.1_{-1.0}^{+1.6}$ and $\log\,[\dot{E}/({\rm\,erg\,s^{-1}})]=38.5_{-0.9}^{+1.8}$, respectively. 
    Both quantities show a mild correlation with the AGN luminosity (\lagn).
    $\dot{E}$ is of the order of 0.01 \lagn\ for 4 sources, but much lower for the majority (9) of the sources, with a median value of
$\log\,[\dot{E}/\lagn]=-5.34_{-0.9}^{+3.2}$
    indicating that typical outflows in the local Universe are unlikely to significantly impact their host galaxy evolution.
\end{abstract}




\section{Introduction}
\label{sec:intro}

The discovery of correlations between the mass of the central supermassive black hole (SMBH) and various properties of the host galaxy, such as the host spheroid mass and stellar velocity dispersion \citep{Ferrarese2000, Gebhardt2000, Gultekin2009, Kormendy2013, Bosch2016}, or the similar evolution of the cosmic star formation rate (SFR) density and the black hole accretion rate density \citep{madau2014} points to the growth of SMBHs being closely linked to the stellar mass assembly of their host galaxies.
It is believed that this link emerges due to both the mass transfer to the inner region of the galaxy that also feeds the SMBH \citep{Storchi-Bergmann2019} and  the regulating effect of feedback from the triggered active galactic nuclei (AGN) on the star formation in the host galaxy \citep{Harrison2017}.
The role of AGN feedback is supported by cosmological simulations and models of galaxy evolution \citep{Springel2005b,Vogelsberger2014,Schaye2015}: feedback is required to reproduce observables such as the shape of the galaxy luminosity function, the colour bimodality of the galaxy population in the local Universe, and the low star formation efficiency in the most massive
galaxies  \citep{Alexander2012,Fabian2012,Harrison2017}.

AGN feedback consists of the injection of mechanical energy (through radio jets) and/or radiative energy (through accretion radiation coupling to the gas on small scales and launching outflows) on the host interstellar and circumgalactic medium by the AGN.
These two modes of energy injection are referred to as mechanical and radiative AGN feedback, respectively.
Examples of mechanical AGN feedback in action have been found in dense environments in the local Universe, i.e. in the surroundings of elliptical galaxies in the centres of galaxy groups and clusters, where radio jets driven by low-luminosity AGN activity heat the circumgalactic medium through the injection of mechanical energy, reducing the cooling rate of the hot gas and maintaining star formation in the central galaxy at low levels \citep{Mcnamara2012}. 
Radiative AGN feedback, on the other hand, has been mostly associated with luminous AGN (i.e. Quasars, L$_{\textrm{X}}$\,$\gtrsim$\,10$^{45}$\,erg\,s$^{-1}$) at intermediate to high redshifts ($z$\,$\gtrsim$\,0.5), where galaxy-wide AGN-driven winds (with typical extensions of 1--10\,kpc) have been observed both in warm ionised and cold molecular gas \citep{Cicone2014,Harrison2014,Leung2017,Vayner2017, RebeccaDavies2020, herrera-camus2019}. 
These winds can sweep away or heat up large amounts of gas, shutting off star formation in the host galaxy \citet{dallagnol2021}.

A number of studies focusing on the star formation properties of Quasars, however, paint a different picture: Quasars are found to have similar star formation rates to main sequence galaxies at the same redshift \citep{Harrison2012,Stanley2015,Schulze2019,Ramasawmy2019}, implying no evidence of enhancement or suppression of star formation.
These apparently conflicting results can be reconciled if the time scale for suppression of star-formation is longer than the time-scale of AGN activity \citep{Hickox2014,Harrison2017,Rodighiero2019}.
In this case, theoretical predictions will need to be combined with observations to identify the effects of AGN feedback (see \citealt{Scholtz2018} for a discussion).
Thus, it is critical to constrain how the energy radiated by the AGN couples to the host interstellar medium and determine the efficiency of this coupling, so that AGN feedback is properly implemented in numerical simulations and semi-analytical models. 

While the most energetic AGN-driven winds are observed in distant Quasars, they are not the most adequate targets to quantify the impact of AGN radiative feedback on the evolution of the general galaxy population.
This is because Quasars are not representative of the bulk of the AGN population, which is comprised mainly of moderate luminosity AGNs (L$_{\textrm{X}}$\,$\approx$\,10$^{42}$--10$^{44}$\,erg\,s$^{-1}$, i.e. Seyfert galaxies), as evidenced by high-redshift X-Ray surveys \citep{Brandt2015}.
Ideally one should study AGNs with redshifts in the range $z$\,$\sim$\,1--3, as this is the epoch where both the star formation rate density and black hole accretion density peak.
However, this is not feasible, as in moderate luminosity AGNs winds typically do not extend beyond the inner 1\,kpc, so they would be unresolved with current observational facilities.
It is thus necessary to find local Universe analogues that still serve as subjects for the study of feedback mechanisms.

In order to be able to resolve the inner kpc of moderately active galaxies, our group AGNIFS (AGN Integral Field Spectroscopy) has observed over the years several AGN hosts using Gemini IFS in the optical \citep[e.g.][]{fathi2006, Storchi-Bergmann2007, Barbosa2009, SchnorrMuller2014, Lena2015, SchnorrMuller2016, SchnorrMuller2017,SchnorrMuller2017a, brum2017, Slater2018, Freitas2018, Humire2018, munoz-vergara2019, soto-pinto2019} and in the near-IR \citep{Storchi-Bergmann2009, Storchi-Bergmann2010, Riffel2017, Riffel2018a, schoenell2019}.
While we have so far focused on individual galaxies, in this paper we present data for 30 objects observed in the optical using the Gemini Multi-Object Spectrograph Integral Field Unit (GMOS-IFU).
For these we homogeneously derive maps of the gas emission-line fluxes, flux ratios and kinematic properties.
Our main goal is to probe the gas excitation and kinematics within the inner kiloparsec of the host galaxies at spatial resolution down to tens to hundreds of parsecs in order to resolve the relevant processes of feeding and feedback of the AGN at the nucleus.
In this first paper, we also report the values we have obtained for the galaxy photometric major axis, kinematic major axes of the gas and stellar kinematics, as well as mass-outflow rates and powers obtained from the measurements of emission-line profile widths at 80\% intensity (\woi) when such values exceed 600\,\kms.
We also compare our results with those of previous studies relating these quantities to the AGN luminosity \citep[e.g.][]{Fiore2017}.

This paper is structured as follows:
in section \ref{sec:sample} we discuss the sample,
section \ref{sec:data} deals with the observations and data reduction procedures,
emission line analysis and fitting is discussed in section \ref{sec:measures},
general results are shown in section \ref{sec:results},
outflow estimates and comparison to AGN properties are shown in section \ref{sec:discussion}
and finally we present our conclusions in section \ref{sec:conclusions}.

\section{The Sample}
\label{sec:sample}

\begin{figure}
    \centering
    \includegraphics[width=\columnwidth]{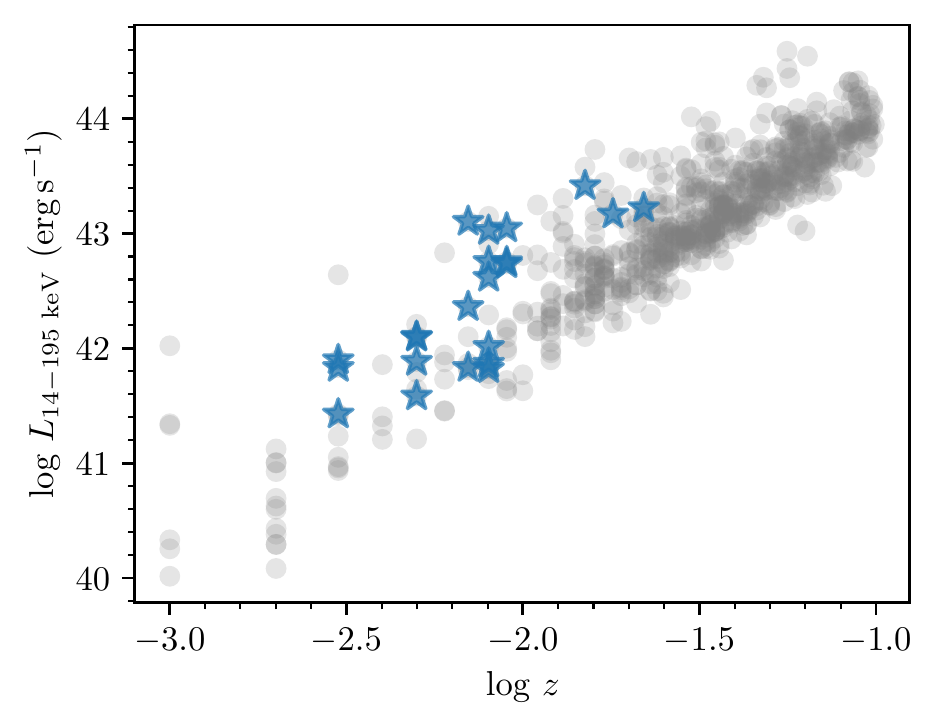}
    \caption{
        The location of the 22 galaxies of our sample which are listed in the Swift/BAT 105 month sample are shown as blue stars in this luminosity vs. redshift plane.
        Grey circles are Swift/BAT sources, blue stars are the 22 targets that overlap that sample in this paper.
    }
    \label{fig:sample_bat}
\end{figure}

The sample comprises 30 AGN, primarily in late-type galaxies, observed with the Gemini instruments GMOS-IFUs (North and South), between 2010 and 2017, and limited to a redshift of $z \le 0.01$, except for three sources (Mrk\,1058, Mrk\,6 and Mrk\,79) that are at $z\approx 0.02$.
Out of the 30 galaxies in the sample, 22 have a counterpart in the 105 month Swift/BAT (hereafter SB105) catalogue \citep{Oh2018}.
A target in our sample is considered to have a counterpart in SB105 if it lies within 15 arcmin of an X-Ray source.
Incidentally, all the 22 counterparts in SB105 are classified as AGN.
The remaining eight galaxies without a bright X-ray counterpart are either LINERS or Seyfert 2's (Sy2).
Figure \ref{fig:sample_bat} shows the location of our sample galaxies (\emph{blue stars}) in the redshift-luminosity plane, along with other sources from the SB105 catalogue (\emph{grey circles}).
The 19 shared targets below $z = 0.01$ in our sample correspond to nearly a third of all Seyfert and LINER galaxies detected by Swift/BAT in that volume.

\begin{figure}
  \centering
  \includegraphics[width=\columnwidth]{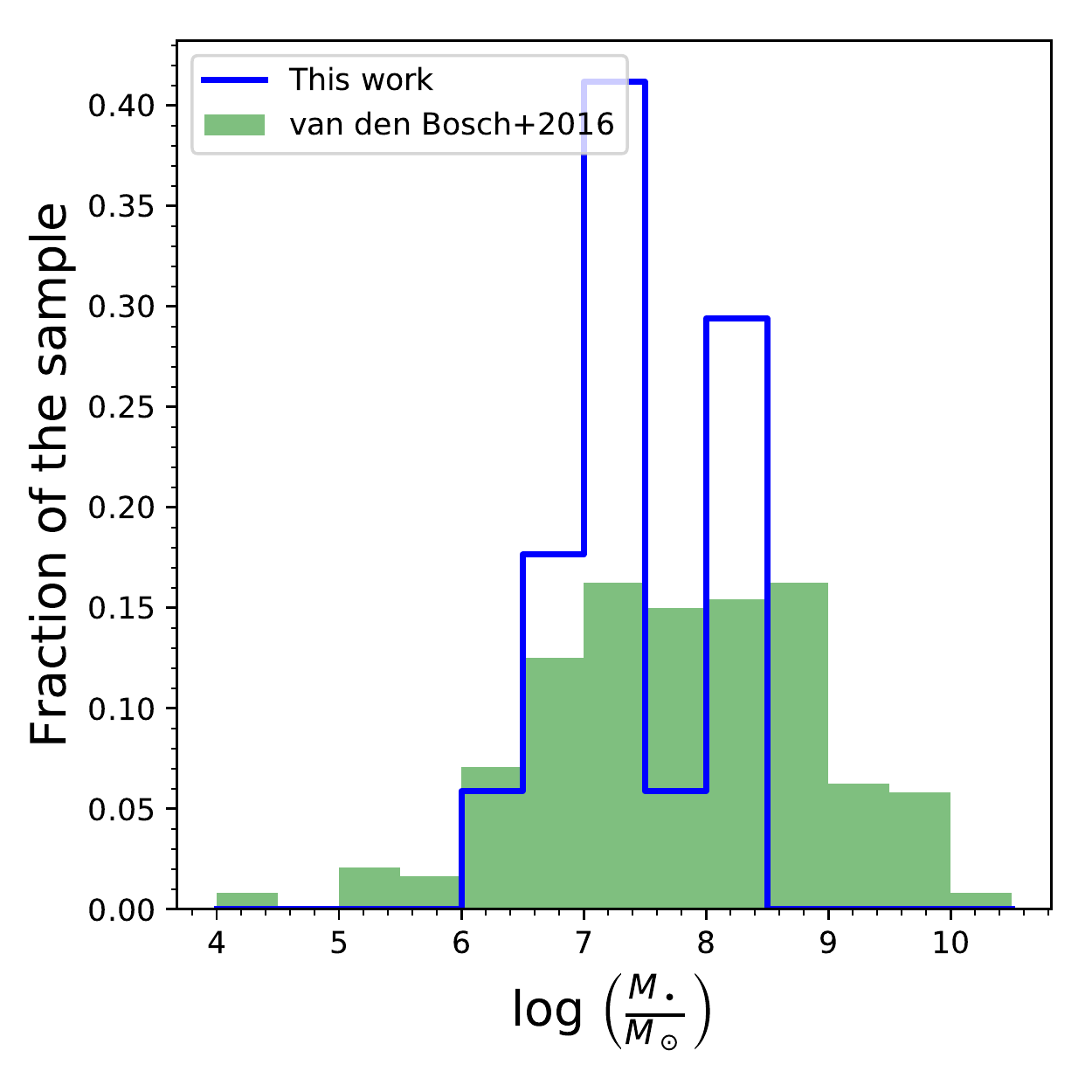}
  \caption{
    Histogram of the black hole masses of our sample galaxies compared to that compiled by \citet{Bosch2016}. Our sample shows a narrower distribution -- covering masses in the range $10^6-10^{8.5}$\,M$_\odot$ and missing the high mass end of the distribution ($10^{8.5}-10^{10}$\,M$_\odot$).
  }
  \label{fig:masshist}
\end{figure}

With the advent of recent surveys on black hole masses ($M_\bullet$) we can assess how well are we sampling the population of SMBHs in terms of their masses. 
In \autoref{fig:masshist} we compare the $M_\bullet$ distribution of our sample and the complete sample from \citet{Bosch2016}.
We can see that, in comparison to that sample, our targets are lacking in the high mass end.

Basic properties of the galaxies studied in this paper are listed in Table \ref{tab:sample}.
The nuclear activity type is based on \citet{VeronCetty2010}, and our reevaluation based on the nuclear spectra, where S1, S2, L1 and L2 represent Seyfert 1, Seyfert 2, Liner 1 and Liner 2, respectively.
Most of the distances were taken from \citet{Tully2013}, which is a compilation of measurements based on six different, redshift independent, methods.
When more then one method was available, the average was adopted.
No redshift independent distance method could be found for the galaxies MCG~-05-23-015 and MCG~-06-30-16; for these we based our distance calculation on the redshift with respect to the cosmic microwave background. 

\input{sample_table.tex}

Projected scales were evaluated from the distances assuming the cosmological parameters from \citet{Hinshaw2013} ($H_0 = 69.32\,\,{\rm km\,s^{-1}\,Mpc^{-1}}$, $\Omega_\Lambda = 0.7135$ and $\Omega_m = 1 - \Omega_\Lambda$) and a flat $\Lambda$CDM model.
They range from 32 pc per arcsec for NGC~1566 up to 449 pc per arcsec for Mrk~1058.
Regarding the environment, only two the galaxies in our sample are known to be interacting with another galaxy: NGC~3227 \citep{Mundell2004} and NGC~3786 \citep{Noordermeer2005}.
Additionally, the galaxy NGC~2110 shows a prominent dust lane, which may be an indication of a recent merger \citep{Drake2003}.

Finally, we point out that many galaxies in this sample have been individually presented in previous papers of the AGNIFS group, as pointed out in the Introduction.
The main difference between those studies and the present paper is that here we performed an homogeneous analysis of the whole sample, with a uniform methodology. Our aim is to use as few target specific assumptions as possible, in order to form an internally consistent picture across this sample of galaxies.

\section{Observations and data reduction}
\label{sec:data}

\subsection{Observations}

The data used in this study comes from many observing runs, although with similar setups, obtained with the Gemini Multi-Object Spectrographs (GMOS) integral field units (IFUs) \citep{AllingtonSmith2002} both at the northern
and southern Gemini telescopes. These IFUs consist of up to 1500 lenslets that feed the light, via fiber optic cables, to the diffraction grating. Each lenslet, which is hexagonal in shape, has a projected diameter in the plane of the sky
of 0$\farcs$18.

GMOS has two modes of IFS observation: a ``single slit'' mode which provides a field of view (hereafter FoV) of $3.5 \times 5$~arcsec, and spectral coverage of $\sim 2000$\AA, and a ``two slit'' mode which trades roughly half the spectral coverage for a doubled FoV.
The slits mentioned here are not actual slits, but rather the result of arranging the fibers in a straight line.
In the single slit mode each exposure produces 500 on-source spectra, and 250 sky spectra.
The latter are used to remove atmospheric emission from the science spectra.
These numbers are doubled in the two slit mode.

The gratings used in these observations, namely B600 and R400, have a resolving power of $R \sim 1800$, which translates to a velocity resolution of $\sim 50$\,\kms.
\oiii and \Hb lines are available only for the 13 galaxies of the sample observed in single slit mode, covering the wavelength range $\approx$ 4800--7000\AA, and identified in Table\,\ref{tab:obslog} with the symbol $\dagger$ adjacent to its name while the remaining targets were observed in two-slit mode and have spectra in the range $\approx$ 5600--7000\AA. Angular resolutions vary between 0$\farcs$6 and 1$\farcs$0, depending on the seeing.

\input{obslog_table.tex}

\begin{figure*}
    \centering
    \begin{minipage}[t]{0.45\textwidth}
    \includegraphics[width=\columnwidth]{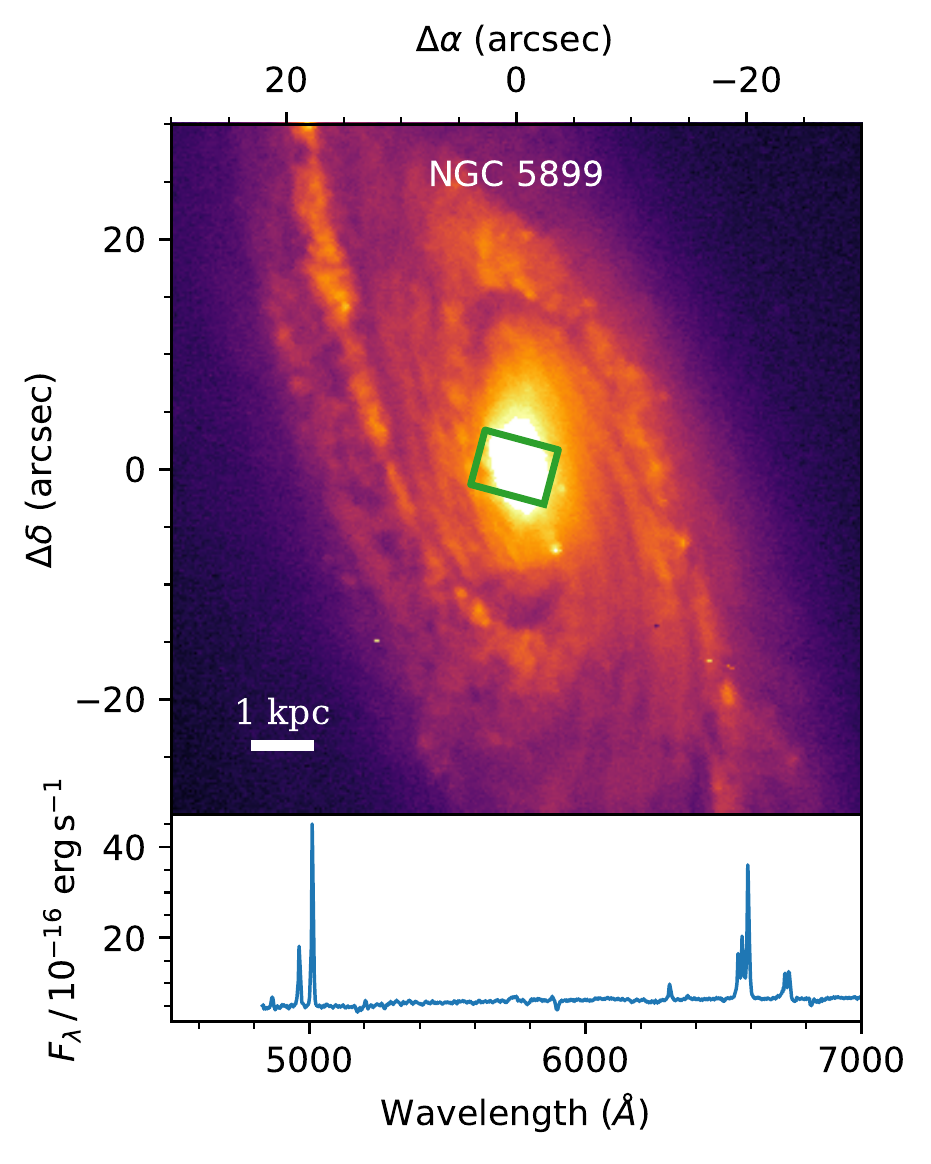}
    \end{minipage}
    \begin{minipage}[t]{0.45\textwidth}
    \includegraphics[width=\columnwidth]{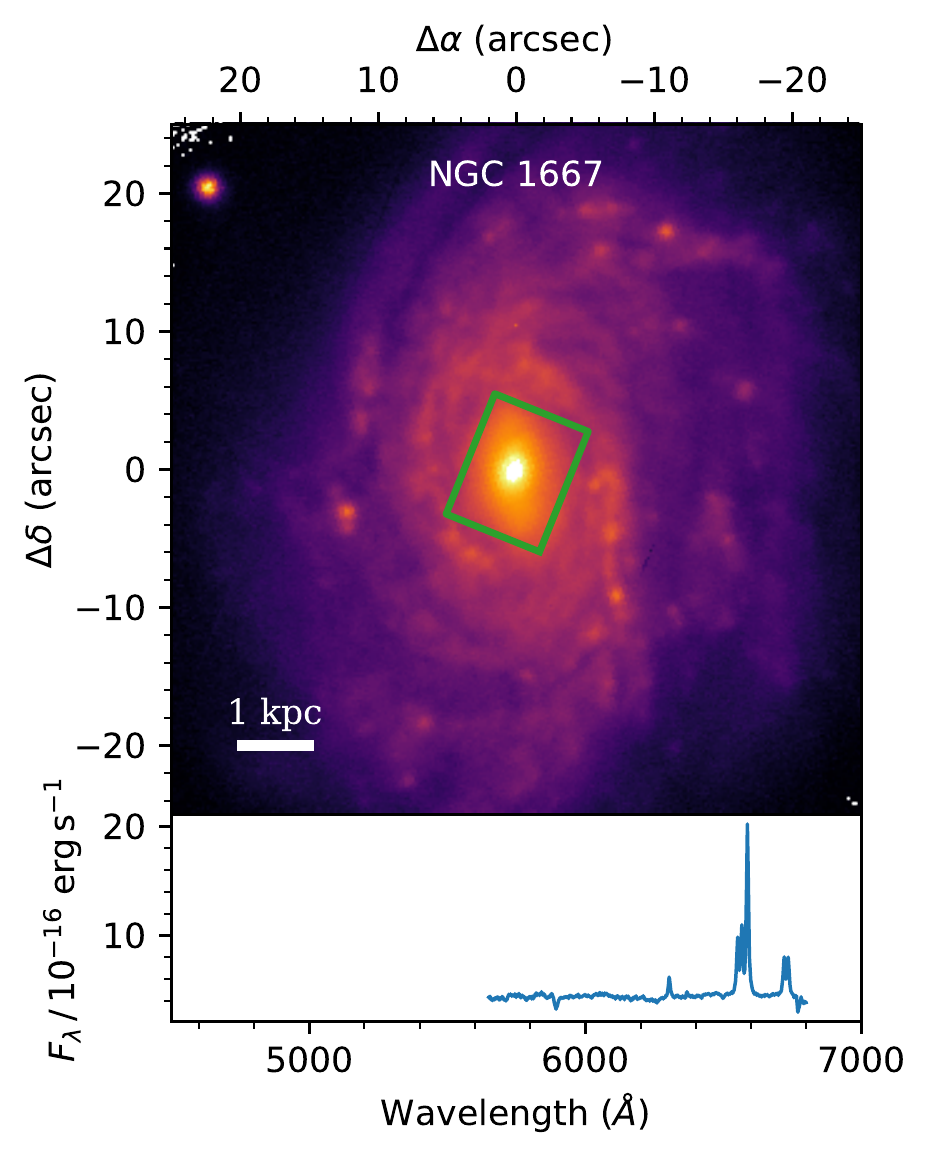}
    \end{minipage}
    \caption{
        \emph{Upper panels}: acquisition images of NGC\,5899 and NGC\,1667, where the rectangle shows the GMOS-IFU FoV and the white bar shows the scale at the distance of the galaxy. \emph{Lower panels}: the summed spectrum of the central 1 arcsec radius in arbitrary flux units and wavelengths in \AA.
    }
    \label{fig.acqspec_n1068}
\end{figure*}

Figure \ref{fig.acqspec_n1068} shows examples of the IFU's FoV superimposed on the $r$ band acquisition images. The bottom panel displays the nuclear spectrum of each galaxy, summed
over a circular aperture with a 1\,arcsec\ radius.
The acquisition images with the superimposed IFU FoVs of all other galaxies in the sample are shown in \autoref{fig:acqspec1} of the Appendix \ref{sec:acquisition_plots} (available as online supplementary material).

\subsection{Data reduction}

Data reduction was based on the package provided by the Gemini observatory for IRAF. Additionally, we have developed a publicly available automated pipeline named {\sc gireds}\footnote{\url{https://github.com/danielrd6/gireds}} to process the raw data through the many tasks of the Gemini package, and also perform quality control checks along the reduction.
The reduction follows standard procedures of bias subtraction, flat-fielding and wavelength calibration based on arc lamp spectra.
Spectra from each fiber were extracted using apertures identified in the flat-field images, which were taken within two hours from the science images. Relative flux calibration was achieved using spectrophotometric stars observed in the same semester of the science observations, and the same instrumental setup.

In order to facilitate posterior analyses of the data cubes, GMOS' original IFU matrix, which is composed of hexagonal lenslets with a diameter of 0$\farcs$18, was interpolated into an image with square spaxels, each having a side of 0$\farcs$1.
This process causes a minor oversampling of the data but allows the direct application of standard image analysis tools over the data cube.
The task GFCUBE was used in re-sampling the data cube, which includes a correction for differential atmospheric refraction for each wavelength plane.

Gemini's world coordinate system (WCS) uncertainty and repeatability is comparable to the IFU field of view ($\sim 5$ arcsec), thus one cannot rely on the WCS data for combining data cubes with spatial dithering. Since these
observations were not performed using adaptive optics, the spatial resolution element is seeing limited to a FWHM of 0$\farcs6$ at best. Therefore, registering of different exposures of the same galaxy based on the peak of continuum
emission would most probably degrade the spatial resolution. The best results were achieved by combining data cubes
from different observations based on instrumental offset coordinates.

\subsection{Flux calibration}

The majority of the data cubes in our sample do not have an associated observation
of a standard spectro-photometric star. We therefore resorted to a method of
absolute flux calibration based on the acquisition images. The method consists
in matching stars of known magnitude, taken from an astrometric catalog, to
point sources that appear within the field of view of the acquisition camera.

Acquisition images were first reduced using the Gemini IRAF package for GMOS
image reduction. Using the Astrometry.net software \citep{Lang2010}, we have
re-generated the astrometric calibration of each acquisition image, in order to
ensure that we are matching the correct sources to the correct stars in the
catalogue. Aperture photometry was performed using the {\sc photutils} package
\citep{Bradley2018}, with stellar sources being identified by a Python
implementation of the {\sc daofind} algorithm \citep{Stetson1987}.  Stellar
magnitudes and positions were taken from the USNO-B catalogue
\citep{Monet2003}, which has typical R band magnitude uncertainties of
0.25 mag. In principle, performing flux calibration with the acquisition images
should account for all low frequency atmospheric effects, since the acquisition
is taken within two hours or less of all the science exposures. 

Zero point magnitudes for the acquisition images were evaluated as the median of equation \ref{magzp} for all the $i$ stellar sources identified in the acquisition image. Sources with FWHM which
differed by more than 20\% of the median FWHM were rejected. An iterative
sigma-clipping algorithm was also used to reject outliers above the $3\sigma$
level.

\begin{equation}
    \mu_i = 2.512 \log I_i + m_i
    \label{magzp}
\end{equation}

\noindent where $\mu_i$ is the magnitude zero point, $I_i$ is the background subtracted instrumental flux from the aperture photometry, and $m_i$ is the magnitude in the USNO-B catalogue.
Since all acquisition images were taken with the filters G0326 and G0303 (for Gemini's South and North respectively), we chose to use the relatively similar R band magnitudes of USNO-B.
The flux ratio between the Johnson R band and GMOS' filters are reasonably stable for a wide variety of stellar spectra, which we tested with the MILES library \citep{falcon-barroso2011}, with a standard deviation of 12 and 5 per cent for G0303 and G0326 respectively.

Once we have the estimate for the conversion of instrumental units to R band magnitudes, we proceeded to perform a spectrophotometric analysis of the data cube.
A nuclear spectrum was extracted from the IFS data, using a circular aperture with 1 arcsec of radius, centred on the peak continuum emission.
The equivalent R band magnitude of this spectrum was estimated by multiplying the spectrum by the filter transmission curve and integrating over wavelength.
Comparing the latter to the magnitude obtained within an identical aperture in the acquisition image yielded a conversion factor from the relative flux to the absolute flux.

\section{Emission line fitting}

\label{sec:measures}

\begin{figure}
    \centering
    \includegraphics[width=\columnwidth]{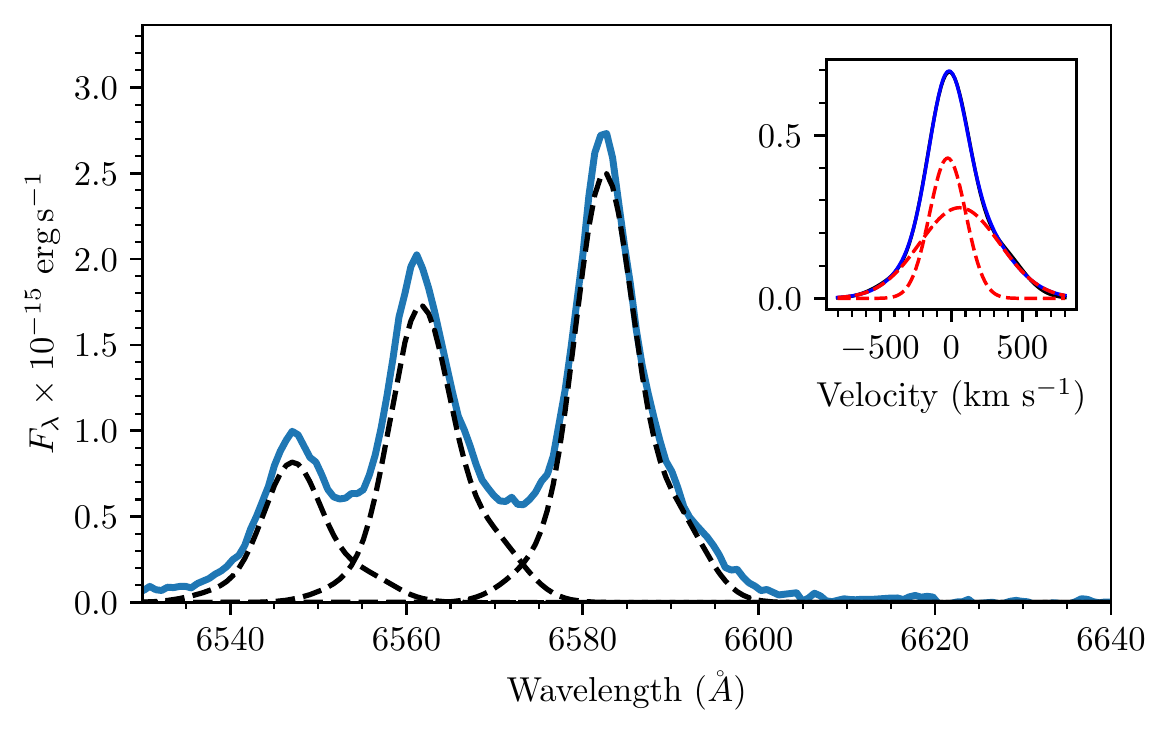}
    \caption{
        Emission line modelling example (nuclear region of NGC\,2110) showing only the narrow component of \Ha and the \nii lines.
        In this example $h_3 = 0.06$ and $h_4 = 0.09$, indicating the presence of a red wing in the profile.
        This line profile can also be represented by a two-Gaussian fit, which is shown in the inset plot.
    }
    \label{fig:h3mexample}
\end{figure}

The emission line measurements are based on the fit of a combination of Gauss-Hermite polynomials \citep{vanderMarel1993,Riffel2010} up to the fourth order with a number of constraints among related spectral lines. Gauss-Hermite
polynomials have the advantage of reduced dimensionality in comparison to multi-component Gaussian fits.
However the function itself is not physically motivated, and therefore the quantitative interpretation of the results is comparatively more complex. 
In addition to the amplitude, mean and standard deviation parameters of a single Gaussian curve, we fitted the coefficients $h_3$ and $h_4$ for the third and fourth order elements, respectively.
The effect of the $h_3$ coefficient is to produce an asymmetric profile, with positive values having a blue wing, and negative values having a red wing.
The $h_4$ profile, on the other hand, produces a symmetric effect of broadening the base of the profile for positive value, or the top for negative values.
An example of such a fit and its interpretation in light of a double-gaussian profile is shown in \autoref{fig:h3mexample}.

Profile fitting of emission lines was done by a in-house developed algorithm.
This code is part of a Python based package of spectral analysis routines, named {\sc ifscube} \citep{ifscube}, which is publicly available on the internet.
{\sc ifscube} allows the fitting of Gaussian or Gauss-Hermite profiles, with or without constraints or bounds.
The fitting algorithm includes integrated support for pixel-by-pixel uncertainties, weights and flags, subtraction of stellar population spectra, pseudo-continuum fitting, signal-to-noise ratio evaluation and equivalent width measurements.
Model fitting relies on {\sc scipy's} \citep{2020SciPy-NMeth} routines for non-linear numerical minimization.
{\sc ifscube} also supports user interaction via a human-readable configuration file, allowing even those that are unfamiliar with the Python language to use it.
The stellar population contribution was fit to the spectrum with spectral synthesis code pPXF \citep{Cappellari2004, cappellari2017} and the MILES simple stellar population models \citep{vazdekis2010}.
For the few galaxies in which the signal-to-noise ratio in the stellar continuum was not high enough to reliably constrain the stellar population the continuum was represented by a smooth polynomial function.

In type 1 AGNs the broad component of the Hydrogen lines was fitted by a combination of three Gaussian curves, in tandem with two Gaussian curves for each of the narrow lines. The central wavelength, relative flux and width of each of the three Gaussian components was fit only once, using the summed spectrum of all the spaxels within 1 arcsec from the continuum centre. Since the broad H lines originate in the same unresolved source, we can apply the same model for all the spectra with the broad line contribution, with only a scaling factor for the flux. Having constrained the broad components in this way, the narrow components were fitted again over the whole datacube.

Figure \ref{fig:h3mexample} shows an example of the emission line modelling for the nuclear spectrum of NGC\,2110.
In this example each individual component is shown as a dashed black line and the observed spectrum, minus the stellar population, as a solid blue line.
The broad \Ha component was omitted here to emphasise the profiles of the narrow lines.
A representation of the Gauss-Hermite polynomial fit in terms of a two component Gaussian model is shown in the inset at the upper right of Fig. \ref{fig:h3mexample}.

Assuming the narrow line region to be in ionisation equilibrium, and to be well represented by the case B  recombination scenario \citep{osterbrock}, we imposed the corresponding constraint on the flux ratio of the \nii lines F[N\,{\sc ii}]$\lambda6548$/F[N\,{\sc ii}]$\lambda6583$=1/3. Kinematic parameters for the \nii and \sii lines are kept the same, namely the velocities corresponding to the line centres and the velocity dispersion with respect to the rest frame, as well as the $h_3$ and $h_4$ parameters.
These last two parameters were also limited to values between $-0.2$ and $+0.2$.
When the blue portion of the spectrum was available, the following contraints were also used: fixed kinematics and between \oiii lines and between \Ha and \Hb, and also a fixed ratio of $f_{5007} / f_{4959} = 3$ for the \oiii lines.
The flux ratio between the \sii lines at $\lambda\lambda 6716, 6731${\AA} was constrained within the lower and upper electron density limits: $0.41 < f_{6716}/f_{6731} < 1.45$.

\subsection{Velocity dispersion via \woi}

Traditional quantities to represent the velocity dispersion of a quasi-Gaussian profile, such as the Gaussian $\sigma$ or the FWHM, fail to capture the complex kinematic picture that is commonly encountered in AGN, in particular at high velocity dispersion. 
Therefore we employed the \woi index \citep{zakamska2014}, which is the width, in velocity scale, that encompasses 80\% of the flux of a given emission line. 
This index has the advantage of being independent from the assumed line profile, since it is measured directly on the observed profile. 
Assumptions about the line profile are only important when measuring the \woi index of a blended line, as neighbouring lines have to be subtracted prior to the integration.

The \woi evaluation begins with the subtraction of the continuum -- the fitted stellar population templates or a local continuum in the cases we could not fit the stellar population -- and also of
neighbouring emission lines, when applicable. Then a cumulative integral is calculated and normalised, so that the \woi will be the difference between the velocity at 90\% and 10\% of the total flux. The integration limits for the cumulative integral are set at $\pm 5\sigma$ from the line centre, which was evaluated during the profile fitting process.
For a Gaussian profile \woi is about 10\% larger than the FWHM, or $\woi = 2.56\times \sigma$.
Normal rotation velocities and velocity dispersions for a galactic potential of even the most massive galaxies limits the \woi to 600 km s$^{-1}$. Emission lines with $\woi > 600$ \kms are therefore a signature of gas in unbound orbits, most probably (and assumed to be the case here) outflowing \citep[e.g.][]{Sun2017, Harrison2014}.

Some emission line profiles, most notably the Hydrogen lines of the Balmer series, are affected by the underlying stellar absorption, which influences both the flux and the centre of the ionised gas features.
This has been taken care of with the fit and subtraction of the stellar population contribution, except for the Markarian galaxies for which the stellar continuum is too weak and this fit and subtraction was not possible.
For these galaxies, with no stellar population subtraction, we impose a lower limit on the equivalent width of H$\alpha$ emission of $\wha > 6$\AA\ when estimating the \Ha luminosities.
The reason is that this value corresponds to approximately twice the maximum absorption \wha of any simple stellar population \citep[e.g. ][]{Bruzual2003}, thus corresponding to a possible maximum error of about $30$\% in the emission-line flux.

\section{Results}
\label{sec:results}

Results in the form of maps obtained from the emission line fits, as well as of some derived properties, are shown in Fig. \ref{fig.example_maps}, Figs.\,\ref{fig:data_mrk348} to \ref{fig:data_ngc7213} of the Appendix and in Table \ref{tab:vsys}.
Fig. \ref{fig.example_maps} is an example, showing the galaxy Mrk\,348, with a spectral coverage of $\approx$4800--7000\AA, however almost half of our sample is limited to the range $\approx$5600--7000\AA.
The former allows the mapping of the [O\,{\sc iii}] and H$\beta$ emission-line properties besides those of H$\alpha$+[N\,{\sc ii}] and [S\,{\sc ii}] of the latter. 

The maps in the above figures show: the flux distribution of a selected ionised gas emission line (\nii and \oiii when available); the radial velocity of the ionised gas, given by the central wavelength of the emission line fit; the \woi map; the equivalent width of the narrow component of \Ha; the flux ratios between \nii (\oiii when available) and \Ha (\Hb when available); and the electron density of the ionised gas. The latter is based on the flux ratio between the \sii $\lambda\lambda 6716,6731${\AA} lines (see section \ref{sec:density}). All figures were rotated so that North is up and East is to the left.
In order to remove high frequency noise from the images, we have convolved them with a 2D Gaussian kernel, with $\sigma = 0.2$\,arcsec, corresponding to approximately the size of the lenslet in the IFU array, and two spaxels of the datacube.
The cross in the centre of the images marks the peak of stellar continuum emission, which we adopted as corresponding to the galaxy nucleus. 
In order to better visualize the flux distributions of the emission lines, we have defined the ``strong emission region" (SER) as that enclosed by an isophote with a flux level 1/10$^{th}$ of the peak flux;  shown in the  in the flux maps (upper left panel of the figures) as the dashed light green contour.
Only continuously connected emitting regions with origin at the galactic nucleus are considered, thus excluding ionised gas clouds that are not directly connected to the nucleus.

We have also included in the figures a dashed red line showing the orientation of the photometric major axis (hereafter PMA) determined over a 2MASS K-band image except for the galaxies NGC\,1068, NGC\,1566, NGC\,4593 which were individually evaluated based on DSS images, and NGC\,5728 which follows \citet{Erwin2004}.
The orientation of the PMA is also listed in Table \ref{tab:vsys}.

The magenta dashed line in the figures shows the approximate direction of elongation of the area dominated by outflows when present, as visually inferred from the gas kinematic and flux distribution maps (these latter showing the orientation of the ionisation axis).
The adopted outflow PA is listed in the 8$^{th}$ column of Table \ref{tab:outflow}.
In some cases for which there is no clear indication of outflow along the ionisation axis, but there is enhancement in \woi perpendicular to the ionisation axis, we have adopted this PA as corresponding to the outflow.
We have also included in the \woi maps a dashed circle showing the distance at which we have calculated the mass-outflow rates and powers, which are also listed in the 2$^{th}$ and 4$^{th}$ columns of Table \ref{tab:outflow}.

    Table \ref{tab:vsys} also includes the kinematic major axis derived from the \nii and \oiii velocity fields, which is represented by the blue continuous line in Fig. \ref{fig.example_maps}.
    Position angles for the kinematic axes were inferred based on the assumption of a symmetric velocity field, and using the {\sc fit\_kinematic\_pa} code \citep{krajnovic2006}.
    Although not as justifiable as in the \nii case, assuming a symmetric \oiii velocity field is still informative for our purposes, if only in a comparative sense.
    Furthermore, a symmetric velocity field does not necessarily imply rotation, since it could also be the result of a biconical outflow.
Fits of the \nii velocity field were used to derive the gas systemic velocities of each galaxy, which were subtracted from the velocity maps and are listed in the second column of table \ref{tab:vsys}.
These same velocity field fits also give the kinematic major axis for each emission line (third and sixth columns).

\input{vsys_table.tex}

\begin{figure*}
    \centering
    \includegraphics[width=.8\textwidth, page=1]{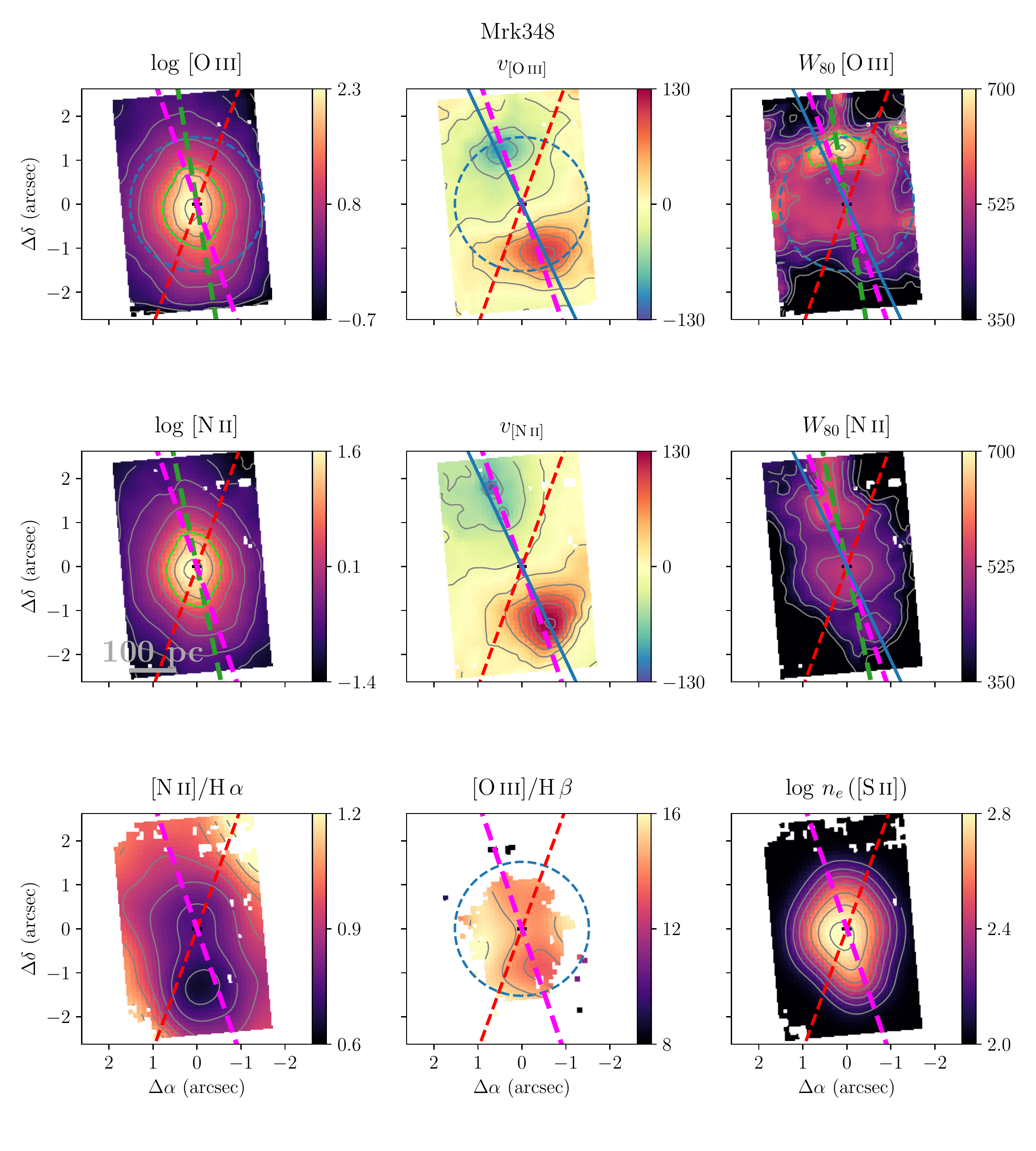}
    \caption{
        Summary of results for Mrk\,348.
        The bar in the lower left of the \nii flux panels represents the projected scale.
        The colour coding is in units of $\log_{10}(F \times 10^{-15}\,\,{\rm erg\,s^{-1}\,cm^{-2}})$ for the line fluxes, ${\rm km\,s^{-1}}$ for the velocities and \woi, {\AA} for the equivalent width and $\log_{10}\,(n_e\times{\rm\,cm^3})$.
        The systemic velocity of the galaxy, inferred from the ionised gas velocity field, has been subtracted from the radial velocity of the emission lines.
        Markings in the maps are as follows:
        \emph{red dashed line:} orientation of the photometric major axis;
        \emph{blue dashed line} orientation of kinematic major axis;
        \emph{green contours:} in the \nii and \oiii flux panels enclose the region with 10 per cent of the total flux and in the \woi map the region with values higher than 600\,\kms
        \emph{blue dashed circle:} the distance used to estimate the mass-outflow rate and kinetic power;
        \emph{green dashed line:} ionisation axis;
        \emph{magenta dashed line:} direction of elongation of the outflow dominated area.
        This figure is also displayed in Appendix \ref{sec:data_plots} (available as online supplementary material).
    }
    \label{fig.example_maps}
\end{figure*}

We now discuss the global properties of the sample.

\subsection{Flux maps and excitation}

Inspection of the flux maps in  Fig. \ref{fig.example_maps} and Figs. \ref{fig:data_mrk348} to \ref{fig:data_ngc7213} of the Appendix reveals extended emission over most of the FoV and some degree of collimation along a direction that we identify as the ionisation axis. 
The orientation of the ionisation axis varies; for the following 10 galaxies this orientation is similar to the PMA: Mrk\,607, NGC\,1365, NGC\,1566, NGC\,1667, NGC\,2110, MCG-05-23-016, NGC\,3516, NGC\,3786, NGC\,4593, MCG-06-30-015. 
Colimated gas emission along a direction distinct from that of the PMA is observed in 15 galaxies: Mrk\,348, Mrk\,1058, NGC\,1068, NGC\,1358, NGC\,1386, Mrk\,6, Mrk\,79, NGC\,2787, NGC\,3081, NGC\,3227, NGC4501, NGC\,5728, NGC\,5899, NGC\,6300, NGC\,6814.
In the case of the following 5 galaxies: NGC\,3783, NGC\,3982, NGC\,4180, NGC\,4450 and NGC\,7213, the orientation of the ionisation axis is not clear.

The line ratios \oiii/\Hb and \nii/\Ha are all consistent with AGN excitation over most of the FoV.
Only occasionally their values indicate that ionisation from young stars becomes dominant in the vicinity of the AGN  (i.e. in the galaxies NGC\,1566 and NGC\,2110).

\subsection{Velocity fields and \woi maps}

Most velocity fields are dominated by a rotation component, consistent with the ``S" (spiral) morphology type of the galaxies as listed in the first column of Table\,\ref{tab:sample}.
    This rotation pattern can be seen in the individual velocity maps in Appendix \ref{sec:data_plots} (available as online supplementary material), and it orientation was determined based on the assumption of a symmetric velocity field (see section \ref{sec:results}).
But in most cases the rotation pattern is disturbed due to the presence of non-circular motions, as described below.
The nature of this non-circular component has been investigated in previous studies by our group for a number of individual cases, as discussed in the appendix \ref{sec:individual}, being associated to inflows along nuclear spirals and/or to outflows.
In this section we point out signatures of outflows in the gas velocity fields and \woi maps, using \woi as an indicator of the mechanical feedback of the outflows in the host galaxy.
A further analysis of the gas kinematics is deferred to a forthcoming paper where we will present modelling of the gas velocity fields and their comparison with the stellar velocity field. 

Inspection of the velocity fields and \woi maps show that in 5 cases -- Mrk\,348, NGC\,1068, NGC\,3227, NGC\,3516 and NGC\,5728 -- increased \woi values -- reaching \woi$\ge$600\kms surround regions of blue and redshifts to both side of the nucleus where steep velocity gradients are observed along the ionisation axis and can be interpreted as due to outflows.
We interpret the increase in \woi as compression of the surrounding gas by the passing outflow. There are some cases in which an increase in \woi is also observed in association to blue and redshifts along the ionisation axis but which do not produce \woi$\ge$\,600\kms: Mrk\,1058, NGC\,4501 and NGC\,6814.

Increase in the values of \woi, reaching \woi$\ge$600\kms, not along but approximately perpendicularly to the ionisation axis have been found in another 5 cases: NGC\,1386, NGC\,2110, Mrk\,6, Mrk\,79 and NGC\,5899.
Our interpretation in these cases is a lateral expansion of the surrounding gas by an outflow or radio jet.
The cases in which the increase of \woi is observed only perpendicularly to the ionisation axis can be interpreted as due to the fact that the outflow or jet is launched at an angle to the galaxy plane, and does not have much gas to compress along its path, only at its base in the galaxy plane.
In the cases of Mrk\,607 and NGC\,3081, we also observe an increase in \woi perpendicularly to the ionisation axis, but it does not reach the 600\kms threshold for its feedback to be considered as significant.

In three cases -- NGC\,1667, NGC\,2787 and NGC\,4180 -- \woi$\ge$600\kms is observed in a small patch close to the nucleus, bringing the total number of galaxies for which we have evaluated the feedback power of the outflows to 13 of the 30 galaxies of our sample.

In all the cases for which \woi$\ge$\,600\kms, we suggest that the increase in \woi traces mechanical feedback from the AGN, justifying its use in the quantification of the power of the outflow.
In support to this interpretation we find also a notable correspondence between regions of high \woi and high density, which will be further discussed in section \ref{sec:discussion}.

We find additional signatures of outflows with \woi$\le$\,600\kms in NGC\,1365 and NGC\,4593, bringing the total number of galaxies with signatures of outflows but with \woi$\le$\,600\kms, to 7.
Thus, considering all signatures of outflows, we find them in 21 of the 30 galaxies of our sample.

\subsection{Determining the ionisation axis}
\label{sec:determining_the_ionisation_axis}

\begin{figure}
    \centering
    \includegraphics[width=\columnwidth]{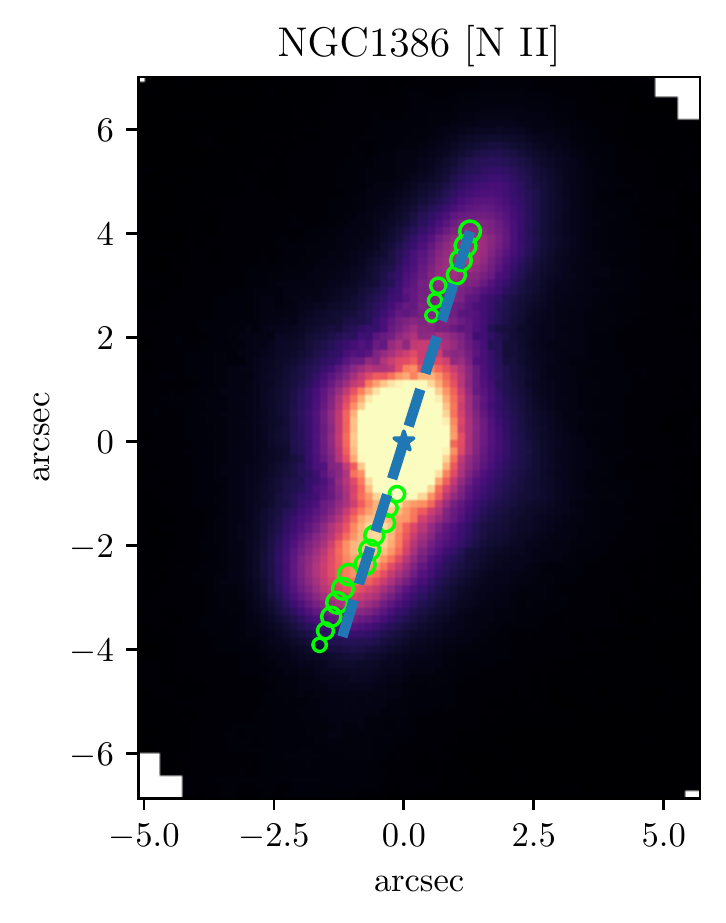}
    \caption{
        Example of the method for determining the ionisation axis.
        The image is a \nii flux map for the galaxy NGC~1386.
        Green circles mark peak positions, with their sizes representing the prominence of that peak.
        A blue star marks the centre of the galaxy, and the ionisation axis is shown as the blue dashed line.
    }
    \label{fig:polar_image}
\end{figure}

The ionisation axes were determined from the flux maps of the \nii and \oiii lines.
In order to quantify the orientation in an objective manner, we developed a method that searches for peaks in the flux map in polar coordinates.
An example of this method, applied on the galaxy NGC~1386, is shown in figure \ref{fig:polar_image}.

\begin{figure}
    \centering
    \includegraphics[width=\columnwidth]{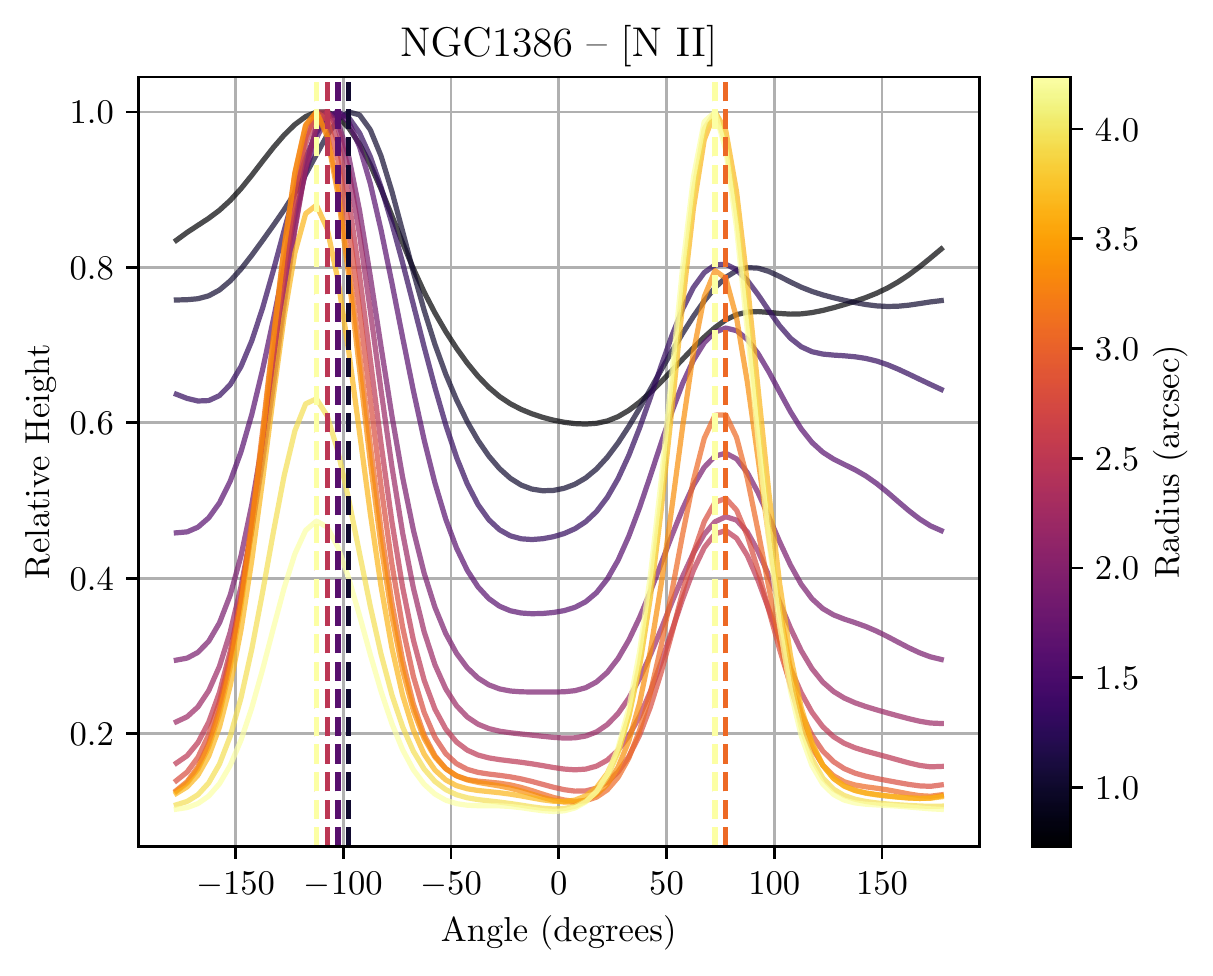}
    \caption{
        Relative flux map in polar coordinates for various radii, based on the image shown in figure \ref{fig:polar_image}.
        Each line represents a radius, identified by the colour shown in the vertical bar to the right.
        Vertical dashed lines mark the peak positions.
        The angle in the horizontal axis is measured counter-clockwise from a line extending from the centre to the right in the frame of the image.
    }
    \label{fig:polar_peaks}
\end{figure}

First the image is transformed to polar coordinates based on a given centre position, and bins of angle and radius.
We then have, for each radius, a separate flux as a function of the angle, which are represented by the lines in figure \ref{fig:polar_image}.
Best results were achieved by using 72 bins in angle, and a step in radii of 3 pixels ($0.27$ arcsec), with further smoothing by convolution with a Gaussian kernel.
After that each curve was normalised with respect to its maximum value.
A final step, essential to avoid artificial border effects, was repeating the curves once in each angular direction, effectively wrapping the plot, and guaranteeing that pixel at zero degrees matches the pixel at 360 degrees.
Figure \ref{fig:polar_peaks} shows the resulting curves for the \nii flux map of NGC~1386 shown in figure \ref{fig:polar_image}.

The peaks in flux for each radius are identified\footnote{We have used the {\sc find\_peaks} function of the {\sc scipy} package to implement the peak finding method.} by selecting those points which have lower values on either side, that are at a minimum distance of 120 degrees from another peak, and that have a prominence of at least $0.3$.
This prominence is the difference between the height of the peak in question and the lowest point between this peak and its closest neighbour.
After the direction of the peak in emission is identified for each radius, the general direction of the ionisation axis is determined by a weighted mean, where the weight is the product between the distance from the centre of the galaxy and the prominence (see above) of the peak.

This method for determining the ionisation axis has two main advantages: it is objective, although a little complicated at first, and it is not limited to a particular configuration of the emission profile.
Fitting ellipses to isophotes, for instance, is challenging if the emission is one sided, or if it is dominated by a spiral structure.

\section{Discussion}
\label{sec:discussion}

    \subsection{Ionisation and kinematic axes}%
    \label{sub:ionisation_axis}

\begin{figure}
    \centering
    \includegraphics[width=\linewidth]{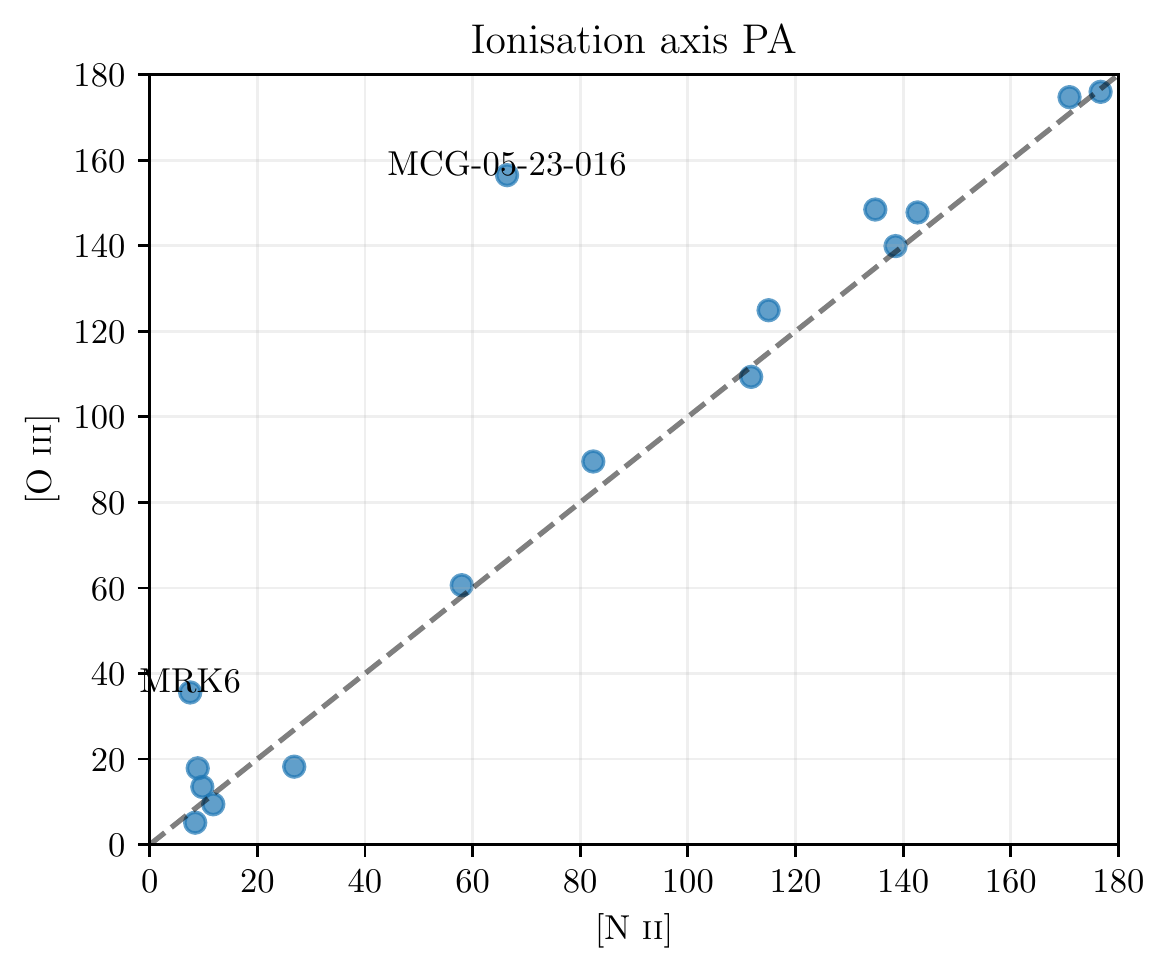}
    \caption{
        Comparison between the ionisation axis for the \nii and \oiii lines.
        There is a general agreement between the two, with only two galaxies differing by more than 15 degrees.
        Only the 16 galaxies with spectral coverage reaching down to the \oiii~5007{\AA} line are shown here, although we have measured the \nii ionisation axis for the whole sample.
    }
    \label{fig:ionisation_pa}
\end{figure}

    Fig \ref{fig:ionisation_pa} shows the correlation between the ionisation axis PA based on the emission of \nii vs \oiii (see \autoref{sec:determining_the_ionisation_axis}).
    It is clear that there is a very good agreement between the two, which is to be expected, since the flux intensity of both lines is following the same ionised structure.
    There are only two galaxies which differ by a reasonable amount from this direct correspondence: MCG~-05-23-016 and Mrk~6, and they both have almost circular emission profiles, which reduces their significance.

\begin{figure}
    \centering
    \includegraphics[width=\linewidth]{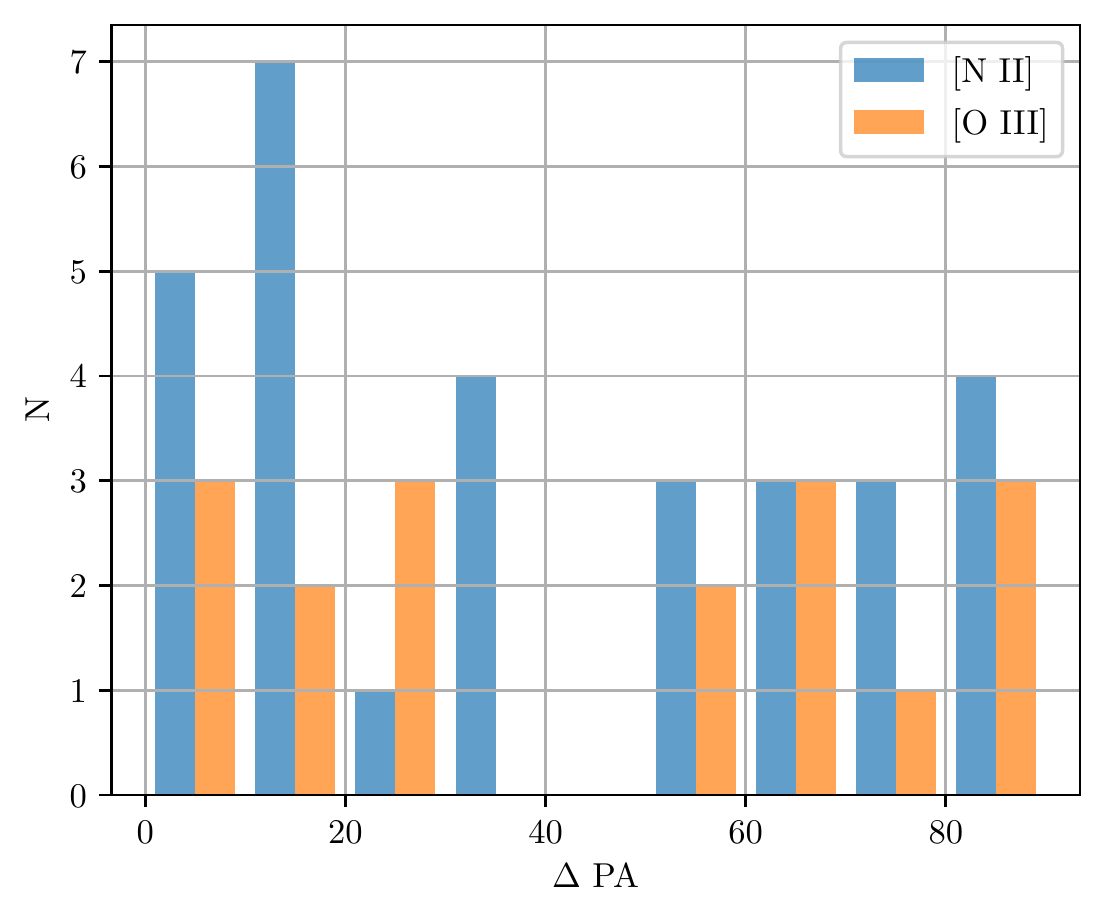}
    \caption{
        Histogram of difference between the photometric major axis position angle, and the orientation of the ionisation axis.
        If there was an alignment between the plane of the AGN and that of the host galaxy's disk there should be a concentration at high values of $\Delta~{\rm PA}$.
        Both distributions, either considering the \nii or \oiii emission, are indistinguishable from an uniform distribution (p-value $> 0.3$).
    }
    \label{fig:photometric_ionisation_pa}
\end{figure}

    The direction of the ionisation axis is
    a tracer of the orientation of the AGN's central engine, being perpendicular to the plane of the accretion disk and the dusty torus.
    Some previous studies have shown that there is no relation between the the orientation of the ionisation axis and that of the plane of the galaxy \citep{schmitt_results_2003}, although others argue for a preferable orientation of the AGN axis perpendicular to the galaxy plane, which would mean that the AGN plane would be preferably aligned with the galaxy plane \citep{he_morphology_2018}.
    Here we use the orientation of the photometric major axis PMA (Table \ref{tab:vsys})  as an indicator of the orientation of the galaxy plane.
    If the AGN's plane is aligned with the disk of the host galaxy, then the ionisation axis should preferentially be found in a direction perpendicular to the PMA, otherwise there should be not preferred relative orientation between the two.

    We investigate this possibility in Fig. \ref{fig:photometric_ionisation_pa}, which is an histogram of the modulus of the difference between the PMA and the ionisation axis, for both \nii and \oiii emission lines.
    The results are compatible with there being no preferential orientation of the AGN with respect to the disk of the host galaxy.
    A KS test comparing the measured distribution of $\Delta~{\rm PA}$ against a uniform distribution with the same dispersion returns p-values $> 0.3$, meaning that the current sample is statistically indistinguishable from random orientations for the ionisation axis relative to the galaxy plane.

\begin{figure}
    \centering
    \includegraphics[width=\linewidth]{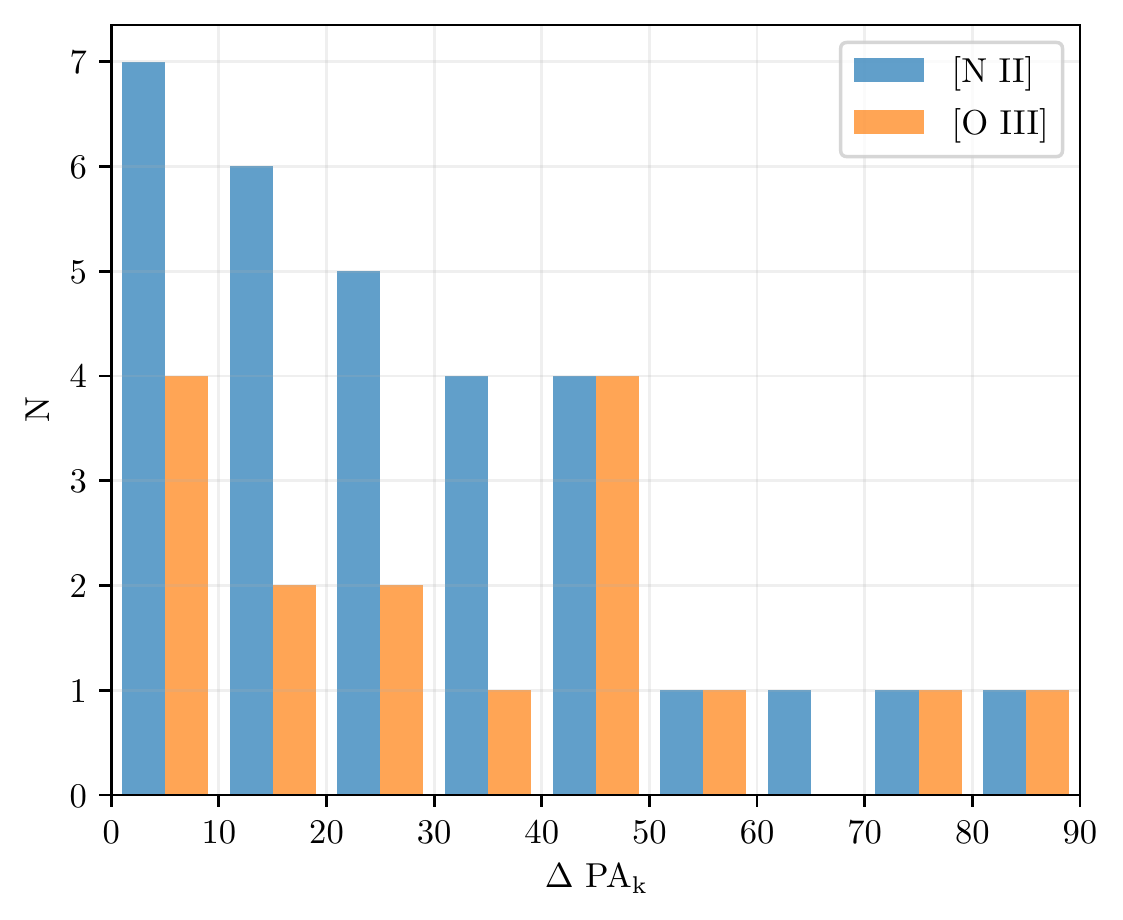}
    \caption{
        Difference between the orientation of the photometric major axis PMA and the gas kinematic major axes \oiii and \nii KMA at the central kiloparsec (Table \ref{tab:vsys})
        Since the PMA admits two rotation directions, the differences shown here are restricted to 90 degrees.
        Blue bars represent $\Delta~{\rm PA_k}$ for the \nii velocity field, while orange bars refer to the \oiii velocity field.
    }
    \label{fig:delta_pa_kinematic_hist}
\end{figure}

    The PMA is also a good proxy for the orientation of the large scale kinematics of the host galaxy (dominated by rotation in the galaxy plane), since all of our targets are disk galaxies. 
    By comparing the orientation of the velocity fields probed by our measurements
    with that of the PMA we can asses the misalignment between the ionised gas kinematics within the FoV of our measurements with that of the large scale kinematics of the galaxy.
    In order to investigate this, we present in Figure \ref{fig:delta_pa_kinematic_hist} two histograms, showing the difference between the orientations of the kinematic major axes KMA of \nii and \oiii velocity fields (Table \ref{tab:vsys}) and that of the PMA.
    We note that the PMA is the same for two rotation directions, causing this analysis to be restricted to a misalignment of 90 degrees or less; if a source has $\Delta~{\rm PA}_{\rm k}=180$~degrees it would show as  $\Delta~{\rm PA_k}=0$, but we have only one such case in our sample, Mrk~607 \citep{Freitas2018}.

    The analysis of Fig. \ref{fig:delta_pa_kinematic_hist} reveals that, for the \nii velocity field, there is a concentration towards low values of $\Delta~{\rm PA_{k}}$, meaning that for most sources the \nii velocity field is dominated by co-planar rotation with the galactic disk.
    But, when considering the \oiii velocity field, the distribution of $\Delta~{\rm PA_{k}}$ is skewed towards higher values, which we interpret as a consequence of its closer connection with the AGN outflow that is oriented at random directions relative to the galaxy plane, as discussed above.

\subsection{AGN Bolometric Luminosities}

In order to relate the AGN properties with its total luminosity, we have 
estimated the bolometric luminosities from X-Ray fluxes in the 14-195\,keV band when available, and from the 2-10\,keV band otherwise. Conversion between X-Ray luminosity and bolometric luminosity follows band specific correction formulae, both based on \citet{Marconi2004}. For the 2-10 keV band specifically, we used equation 21 from \citet{Marconi2004}.

\begin{equation}
    \log \left[ \frac{L_{12}}{L_{2-10~{\rm keV}}} \right] =
 1.54 + 0.24 L_{12} + 0.012L_{12}^2-0.0015L_{12}^3
\end{equation}

\noindent where $L_{12} = \log(L_{\rm AGN}) - 12$ and $L_{\rm AGN}$ is the bolometric luminosity in units of $L_\odot$. However, for the majority of targets the 14-195 keV flux from the Swift-BAT survey was available, and the bolometric correction followed equation 5 from \citet{Ichikawa2017}:

\begin{equation}
    L_{\rm AGN} = 0.0378 (\log L_{14-195\,\,{\rm keV}})^2
    - 2.03 \log L_{14-195\,\,{\rm keV}} 
    + 61.6
    \label{bol_hard}
\end{equation}
The bolometric luminosities obtained using the above two equations are listed in the last column of Table \ref{tab:outflow}.

\subsection{Gas kinematics}
\label{sec:gas_kinematics}

For the remainder of this section 
we adopt the hypothesis that the signature of mechanical feedback of outflows onto the surrounding medium is an increase in \woi, and from there we calculate the associated mass-outflow rates and kinetic powers. 
    This analysis based on general criteria differs from the one presented in section \ref{sec:results}, where we discussed outflows considering more aspects of the velocity field.

In order to quantify the outflows, we need to identify the spaxels in which the ionised gas kinematics is not compatible with disk rotation. A spaxel is defined as being part of an outflow, or having its nebular emission dominated by outflowing gas according to the following criteria:

\begin{itemize}
    \item It has $\wha > 6${\AA} to ensure an accurate value for \lha. 
    Keeping in mind that, in a fraction of galaxies, the stellar spectrum has not been subtracted, this limit in \wha means that the \Ha emission typically has $\wha \geq 9${\AA}, since the \wha associated with the absorption is expected to be close to $3${\AA}. 
    \item The velocity dispersion measured by \woi of the \oiii or \nii emission lines must be above a limit which excludes reasonable expectations for bound orbits in the galaxy's potential. For this limit we chose $\woi > 600$\kms. 
        Although the \oiii $5007${\AA} line is a better proxy for the gas ionised by radiation from the AGN, we decided to include also measurements using the \nii line to avoid limiting the size of our sample to those for which such measurements could be made.
    \item It has at least two more neighbouring spaxels also classified as having outflows. This last criterion ensures that isolated spurious spaxels are not included in the outflow mask. Since the FWHM of the point spread function typically spans 3 to 6 spaxels, isolated detections must be false positives.
\end{itemize}

Using these simple and very general criteria, we reach the conclusion that 13 out of the 30 AGN's in our sample display signatures of outflows in ionised gas via \woi.
These galaxies can be identified in Table \ref{tab:outflow} as those for which we show the mass outflow rates $\dot{M}$ and outflow powers $\dot{E}$.

\subsection{Gas densities in the outflows}
\label{sec:density}

We have used the \sii lines $\lambda\lambda 6716,6731${\AA} to estimate the electron density at each spaxel of our sample galaxies.
The actual computation followed the equations of \citet{Proxauf2014}.
The only galaxies for which we do not show electron density estimates are NGC\,3227  and NGC\,3786 due to an observational problem that precluded the use of the \sii lines.
For all spectra we adopted a fixed standard electron temperature of $10^4$ K.
If temperatures in the outflows were higher, this would increase the estimated densities.

\begin{figure}
    \centering
    \includegraphics[width=\columnwidth]{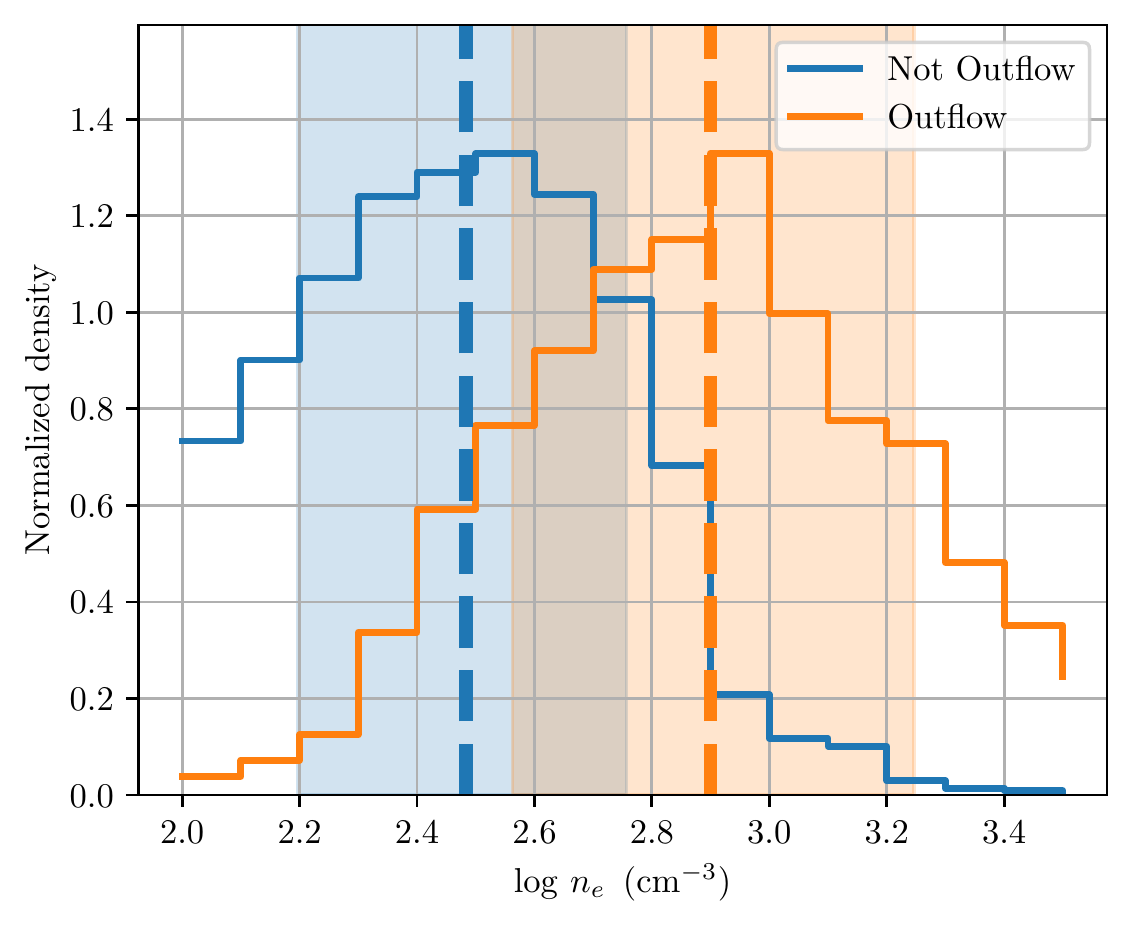}
    \caption{
        Histograms of electron density for spectra from spaxels covering
        (\emph{orange}) and not covering (\emph{blue}) outflows, in units of
        probability density.
        Median values are shown as vertical dashed lines, and the shaded regions encompass the 16 and 84 percentiles.
    }
    \label{fig:density_histogram}
\end{figure}

Figure \ref{fig:density_histogram} shows the histogram of electron densities for spectra identified as having outflows (corresponding to spaxels with \woi$\ge$600\,\kms) as compared to those without outflows. 
The histograms are given in units of probability density, which means that the integral of the histogram equals 1.
Median values are $305_{-147}^{+266}\,\,{\rm cm^{-3}}$ and $794_{-429}^{+969}\,\,{\rm cm^{-3}}$ for the non-outflow and outflow samples respectively, where the given intervals are the distances from the median to the 16 and 84 percentiles.
Spectra for which only upper or lower limits could be given, due to the saturation of the line ratio, are not included in these statistics.
The number of sample points are 14604 and 2110 for non-outflows and outflows, respectively.

We performed a Kolmogorov-Smirnov (KS) test in order to ascertain the statistical significance of the difference in distributions of electron densities, arriving at a value {\bf $D = 0.51$} (a value of $0$ indicates that both distributions are drawn from the same sample).
Given the large size of this sample of spectra, we can confirm the rejection of the null hypothesis with a confidence, given by the p-value, of $p <  0.1\%$.
Therefore we conclude that the gas in outflow  is on average denser than the rest of the narrow line region, by a factor of 1.7.
This result supports that the criteria listed in section \ref{sec:discussion} are indeed selecting a physically distinct portion of the emitting gas.

\begin{figure}
        \centering
        \includegraphics[width=\linewidth]{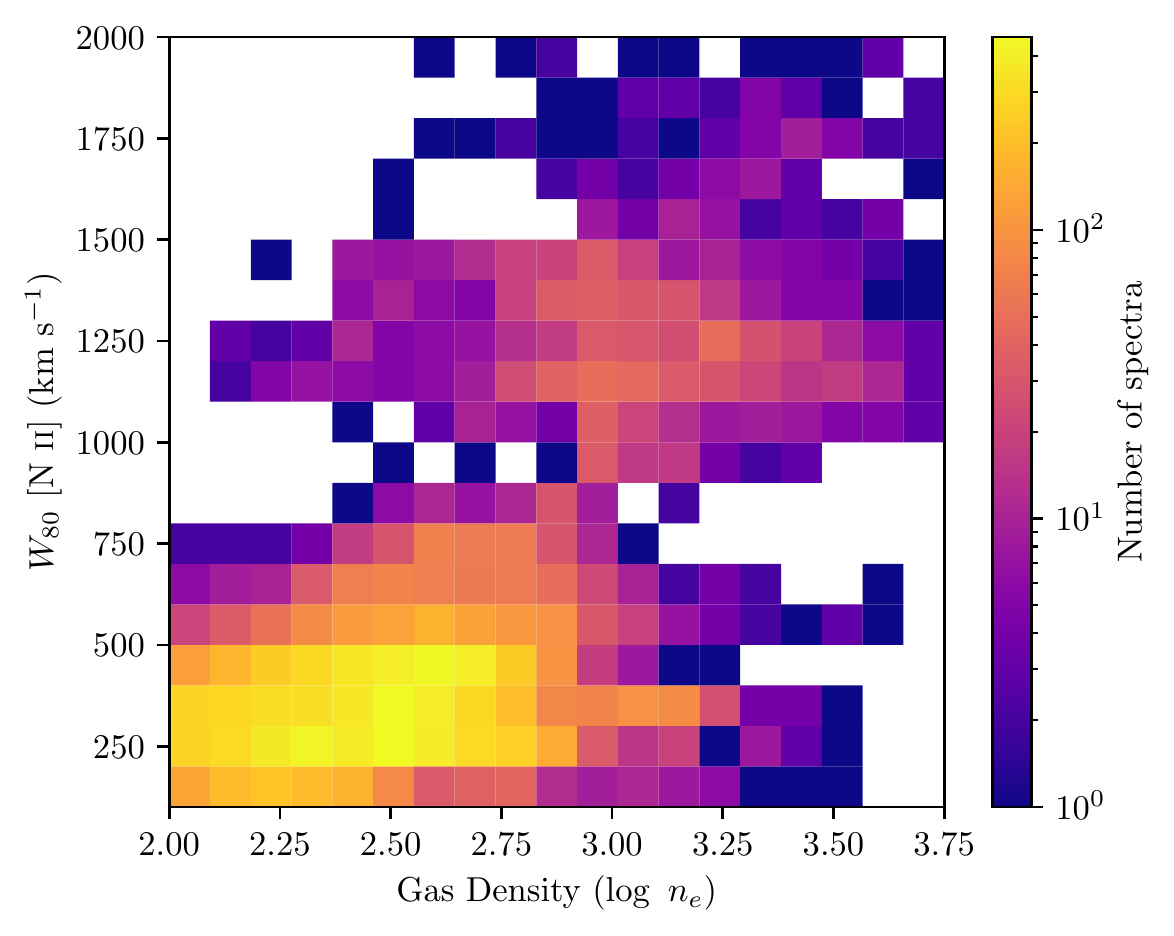}
        \caption{
            Comparison between the density and the \nii \woi.
            Each axis has been divided into 20 cells and color coded according to the number of points in each cell, totalling $\sim 1.4\times 10^{4}$ individual spectra (see text).
        }
        \label{fig:density_w80_n2}
    \end{figure}

\begin{figure}
        \centering
        \includegraphics[width=\linewidth]{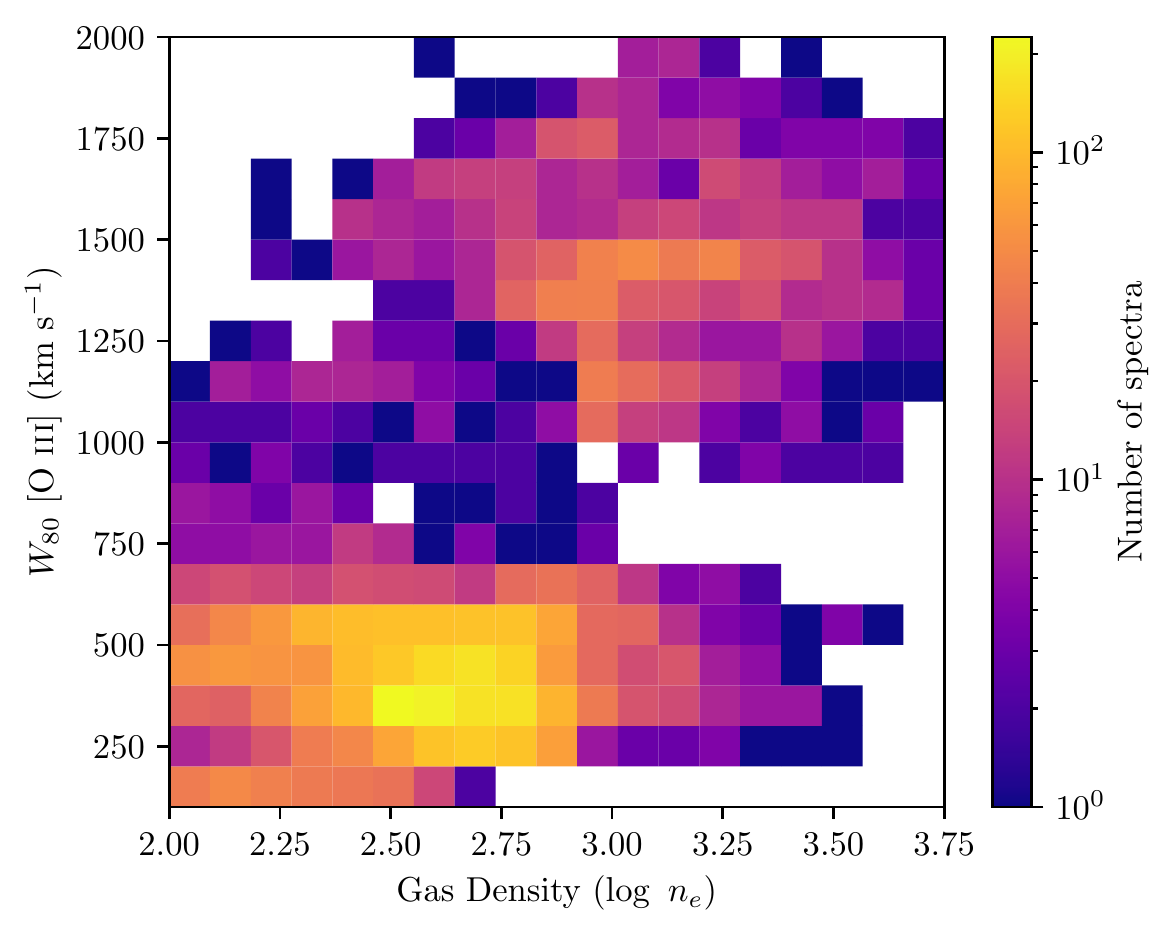}
        \caption{
            Analagous to Fig. \ref{fig:density_w80_n2} but for the \oiii line.
        }
        \label{fig:density_w80_o3}
    \end{figure}

    This same effect can also be seen in plots comparing the electron density to the \woi for each spectrum, which are shown in figures \ref{fig:density_w80_n2} and \ref{fig:density_w80_o3}.
    Only spectra satisfying the following conditions were used in these figures:
        i) SNR in the \sii lines above 3;
        ii) no bad pixel flags compromising the accuracy of \woi in the \nii or \oiii lines;
        iii) \sii unaffected by strong telluric lines.
    Adoption of these criteria reduced the number of points to about $1.4\times 10^{4}$, which represent roughly ten percent of all the spectra in the sample. 
    The two regimes identified in the histogram (\autoref{fig:density_histogram}) are now seen as two clouds: one having \woi $\sim 400 ~{\rm km~s^{-1}}$ and low densities, and the other with \woi above $800~{\rm km~s^{-1}}$ and higher densities.

    Our interpretation of this connection between velocity dispersion and electron density is that the gas which forms the outflow is encountering the galaxy's interstellar medium (ISM) and increasing the local density.
    Of course, this interpretation requires the outflow to be directed along a direction which meets the disk of the galaxy.
    Outflows perpendicular to the galaxy's disk would, therefore, produce a comparatively smaller effect on the density.
    Another possibility could be that the outflowing gas is being spread out in the direction of the line of sight, hence the increase in \woi, when it meets regions of higher density in the ISM.
    Both interpretations differ in the cause attributed to the increased \woi, which is intrinsic to the outflow in the first, and a consequence of higher density clouds in the second.

\subsection{Ionised gas mass}

The ionised gas mass was calculated for each individual spaxel of each galaxy as: 

\begin{equation}
    M = \frac{m_p \lha}{n_e \jha(T)}
    \label{gas_mass}
\end{equation}

\noindent where $m_p$ is the proton mass, $n_e$ is the number density of electrons from the \sii lines flux ratio, \lha is the \Ha luminosity and \jha is the \Ha emissivity in ${\rm erg\, cm^{3}\, s^{-1}}$ for a given temperature. 

\begin{figure}
    \centering
    \includegraphics[width=\columnwidth]{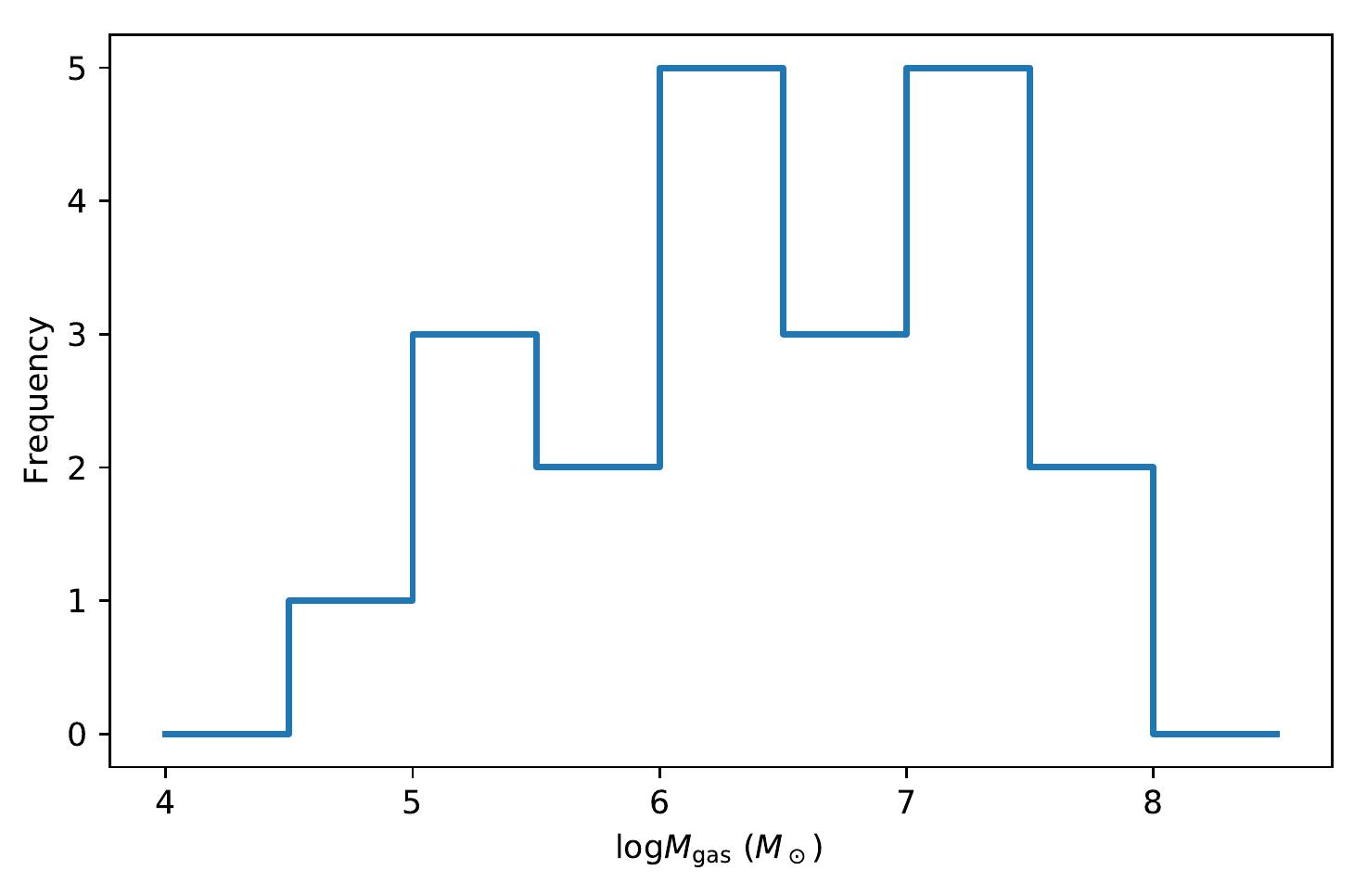}
    \caption{
        Histogram of ionised gas mass of galaxies in the sample (within the IFU's field of view), based on the H$\alpha$ luminosity and the gas density derived from the [\ion{s}{ii}] line ratio.
    }
    \label{fig:gasmass_histogram}
\end{figure}

Since we did not measure the temperature based on the nebular emission, we assume a standard value for the electron temperature of $T = 10^4\,\,{\rm K}$.
The \Ha luminosity was corrected only for foreground Galactic extinction based on the CCM \citep{Cardelli1989} extinction law, using the dust maps from IRSA \citep{Schlegel1998}.
A factor of $10^{0.4 A}$ was applied to the measured flux, where $A$ is the extinction in magnitudes for the SDSS $r'$ band.
We did not correct for extinction within the galaxy due to the fact that we do not have suitable H$\beta$ fluxes or other reddening indicators to obtain the internal attenuation.
A histogram of the total ionised gas masses for our sample, obtained by summing the masses evaluated at each spaxel within the observed field of view, is shown in Fig.\,\ref{fig:gasmass_histogram}.

\subsection{Mass outflow rates}

In order to calculate the mass outflow rates we have used the \Ha luminosity.
However, instead of using the total flux of \Ha emission, we used only the fraction of the flux which corresponds to velocities above 600\,\kms from the line centre.
This ensures that even if the \woi of \Ha is different from that of \oiii, only the emission from the outflowing gas is considered.

Mass outflow rates are given by equation \ref{outflow_rate} below. The basic assumption is that the outflow velocity $v$ we see now is approximately the average velocity of the gas since it left the vicinity of the AGN. Therefore, the time it took for the gas to reach its current distance from the central engine is $R/v$.

\begin{equation}
    \dot{M} = M \frac{v}{R} 
    \label{outflow_rate}
\end{equation}
where we have adopted 1/2\,\woi as a proxy for the outflow velocity, which should be a good approximation when considering many galaxies, and therefore, many different projections for the outflow.
The resulting mass outflow rate values for each galaxy, as derived from the kinematics of both the \oiii and \nii emission lines, are given in table \ref{tab:outflow}.

    By far the most uncertain term in \ref{outflow_rate} is the distance $R$.
    Even before any projection effects are taken into account, one first has to consider where the gas is being accelerated, either at the vicinity of the AGN or in situ \citep[i.e. ][and references therein]{kraemer2020}.
    In this work we adopt a model of outflow in the form of an expanding spherical shell, which we assume to be accelerated at the nucleus.
    We define the outflow travel distance $R$ as the largest radius, in the plane of the sky, for which we observe $\woi \ge 600 \kms$.
    This is, of course, a lower limit for the travel distance in the likely case that the outflows are not spherically symmetric.
    As a result, our estimates of mass outflow rates should be considered as upper limits.

\begin{figure}
    \centering
    \includegraphics[width=\columnwidth]{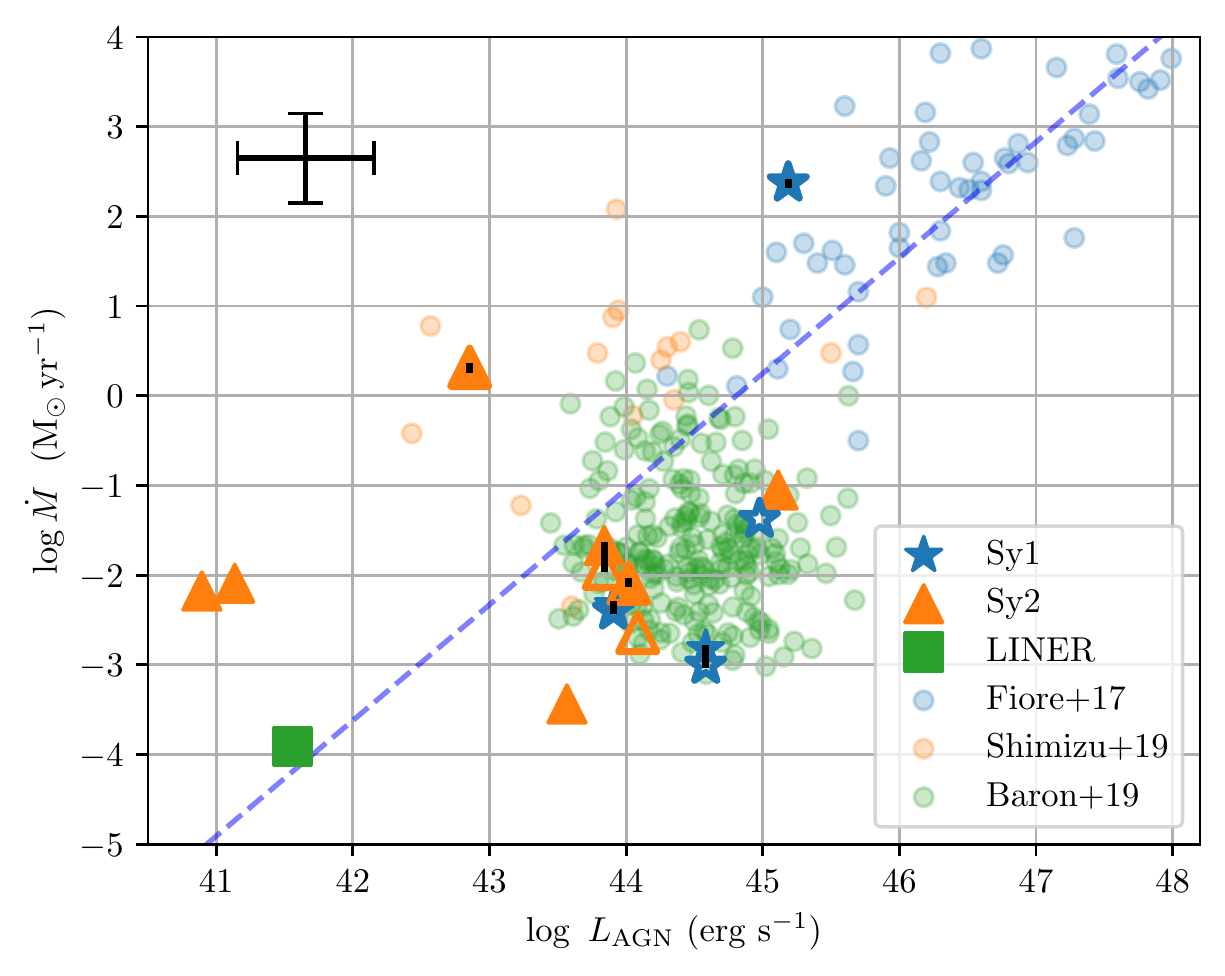}
    \caption{
        Relation between the mass outflow rate $\dot{M}$ and the AGN's bolometric luminosity \lagn. Stars, triangles and squares represent Sy\,1, Sy\,2 and LINERs, respectively.
        Filled symbols are for outflow rates based on the [\ion{N}{ii}] line, while open symbols are based on [\ion{O}{iii}].
        When estimates based on both \oiii and \nii are available, the symbols are joined by a vertical black line.
        Blue circles are from \citet{Fiore2017}, orange circles from \citet{Shimizu2019} and green circles from \citet{Baron2019b}.
        The blue dashed line is the sub-linear fit of \citet{Fiore2017}.
        Typical uncertainties are represented by the black cross at the
             upper left.
             When an outflow is detected in both \nii and \oiii, the symbols are linked by a vertical black line.
    }
    \label{fig:outflow_rate}
\end{figure}

In figure \ref{fig:outflow_rate} we show the relation between the mass outflow rates \mdot and the AGN bolometric luminosity \lagn for the 13 galaxies of our sample with outflows as traced by $\woi \ge 600 \kms$ together with previous results from the literature.
Spectral types Sy\,1 and Sy\,2 and LINERs are represented by different symbols in the figure, although we are not implying any distinction in the outflow mechanism.
In this plot we see that the values obtained for our galaxies follow the trend of the sub-linear correlation with \lagn from \citet{Fiore2017} -- the blue dashed line in the Figure.
The scatter is nevertheless large, with four points two orders of magnitude above the mean relation, one on the relation and the rest of the sample 1--2 orders of magnitude below the relation.
Most of the points below the relation occupy the same space as those from \citet{Baron2019b}.
We have also added points compiled by \citet{Shimizu2019} for reference.
Some of our galaxies are also present in these references, therefore the same galaxy might be represented by more than one point.

When comparing the galaxies with and without outflows a small trend is also observed, although the luminosities are consistent within uncertainties.
For the 13 galaxies with outflows the average AGN luminosity is $\log \lagn$ of $43.83 \pm 0.83$, while for galaxies lacking outflows the average value is $\log \lagn$ is $42.94 \pm 1.54$. 

\subsection{Outflow power}

Having the mass outflow rate, estimating the outflow power is relatively straightforward. We consider only the mechanical power of the outflow, disregarding eventual heating and expansion of the outflowing gas. The total kinetic power of the outflow is given by: 

\begin{equation}
    P = \frac{1}{2} \dot{M} v^2
    \label{outflow_power}
\end{equation}

\noindent where $v$ is the outflow velocity, taken to be 1/2\,\woi for \woi higher than 600\,\kms. The calculated values are shown in the 4$^{th}$ and 7$^{th}$ columns of Table \ref{tab:outflow}, and  range between 10$^{36.2}$\,erg\,s$^{-1}$ for the LINER in NGC\,2787 to 10$^{43.3}$\,erg\,s$^{-1}$ for the Sy\,1 in Mrk\,6, with a median value of 
$\log\, [\dot{E} / ({\rm erg\,s^{-1}})] = 38.5_{-0.9}^{+1.8}$ when considering \nii based \woi.
Similar results are found for \oiii based estimates, with a median outflow power of $\log\, [\dot{E} / ({\rm erg\,s^{-1}})] = 38.4_{-0.8}^{+2.9}$.

\input{outflow_table.tex}

\begin{figure}
    \centering
    \includegraphics[width=\columnwidth]{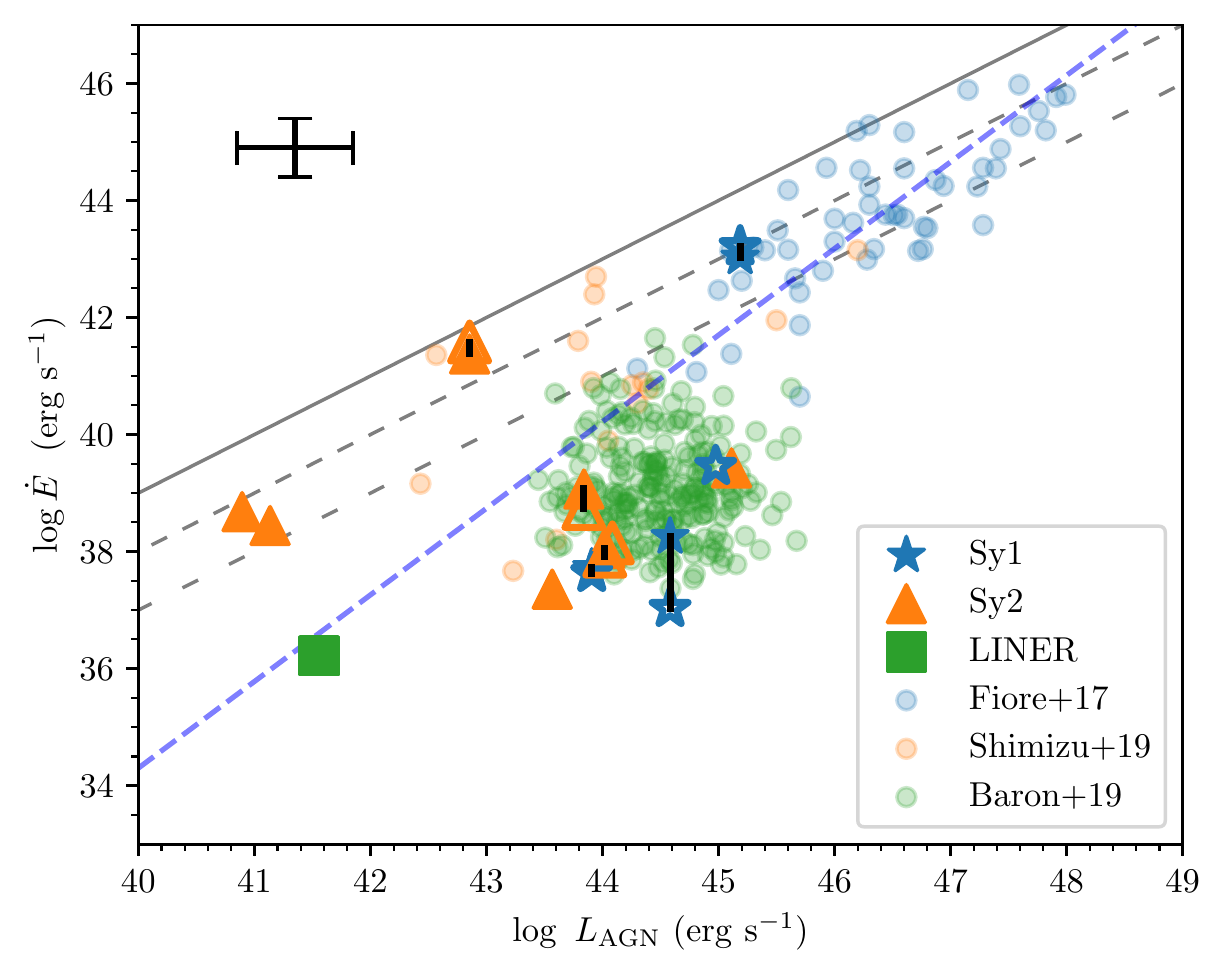}
    \caption{
        Relation between the outflow power $\dot{E}$ and the AGN bolometric luminosity \lagn for our galaxies compared to previous value from the literature. The dashed blue line shows the sub-linear fit from \citet{Fiore2017}.
        Continuous and dashed grey lines are a visual aid representing outflow powers, from top to bottom, of 0.1\,\lagn, 0.01\,\lagn and 0.001\,\lagn, respectively. Distinct AGN spectral types are represented by different symbols. Filled symbols represent values based on the \nii line while open symbols represent values based on the \oiii line.}
    \label{fig:outpower_lagn}
\end{figure}

\begin{figure}
    \centering
    \includegraphics[width=\columnwidth]{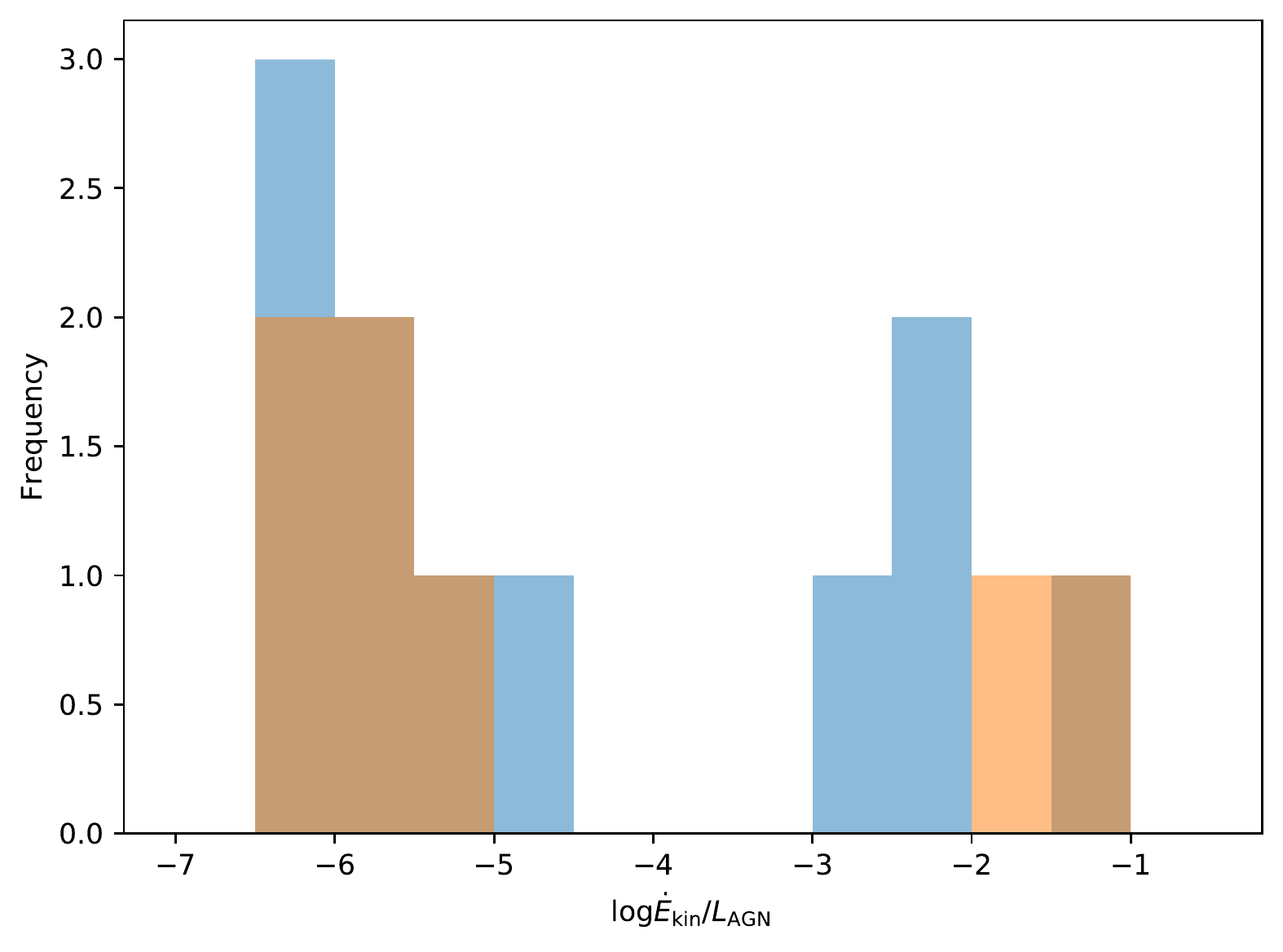}
    \caption{Histogram of the ratios between the outflow power $\dot{E}$ and the AGN bolometric luminosity \lagn. Blue bars are for estimates based on \nii and orange bars are based on \oiii. 
    }
    \label{fig:outpower_lagn_hist}
\end{figure}

Figure \ref{fig:outpower_lagn} shows the relation between the outflow kinetic power $\dot{E}$ and the AGN luminosity \lagn as compared with previous ones from the literature in the reference plot of \citet{Fiore2017}.
The blue dashed line in the Figure shows the sub-linear correlation from \citet{Fiore2017} for ionised gas outflows. 

Previous models and studies have argued that the relation between the kinetic power and the AGN luminosity should be $0.05\lagn\ge\dot{E}\ge0.5\lagn$ for a significant impact of the AGN on the evolution of the host galaxy \citep{dimatteo2005,hopkins2010,Zubovas2012}.
As a reminder, we are only addressing one manifestation of outflow -- the ionised gas phase -- which is linked to radiation pressure driven winds.
Figure \ref{fig:outpower_lagn} and the histogram of the ratios $\dot{E}/\lagn$ shown in Fig. \ref{fig:outpower_lagn_hist} reveal that only for two galaxies of our sample -- NGC\,1068 and Mrk~6 -- the outflows have such power.
The remaining outflows have powers below $10^{-2}$\,\lagn.
When compared to previous results from the literature, our 4 most powerful outflows follow the trend of the higher luminosity sources of \citet{Fiore2017} and \citet{Shimizu2019}, while the other sources are much less powerful, showing a behaviour more similar to the lower-luminosity sources  of \citet{Baron2019b}.

\section{Conclusions}
\label{sec:conclusions}

We present an analysis of GMOS-IFU optical datacubes of the inner kpc of 30 nearby AGN (mostly at $z\le0.01)$ that our research group AGNIFS has collected over the years, and have measured the gas excitation and kinematics via the fit of Gauss-Hermite polynomials to the emission lines.
We have obtained maps of the gas emission-line flux distributions, line-ratios and kinematics with spatial resolutions in the range 50--300\,pc and velocity resolution of $\approx$\,50\,\kms\, and used the parameter \woi as an indicator of outflows.
Twenty one of the 30 galaxies of our sample are Swift/BAT AGN sources.
    We determined the orientation of the ionisation axis by finding the direction of the peak fluxes in the polar flux distributions of \nii and \oiii.
    We fitted the \nii and \oiii velocity fields with a simple symmetric model
to measure the
gaseous systemic velocities, the P.A. of the corresponding kinematic major axes, and to reveal the presence of outflows.

The main conclusions we have reached on the basis of the measurements outlined above are:
\begin{itemize}

\item {\it Gas excitation and ionisation axis:} Emission line ratios characteristic of AGN excitation are observed over most of the FoV and the gas emission clearly extends beyond the inner kpc; the ionisation axis shows a random orientation relative to the photometric major axis of the galaxy indicating no preferred orientation of the AGN relative to the galaxy; 

\item {\it Gas kinematics:} Disk rotation is found in most cases, as expected given the spiral morphology of the host galaxies, though disturbances due to inflows and outflows are also seen.
The gas velocity dispersion and \woi are usually enhanced at the nucleus and/or surrounding outflows,
where the highest gas densities are frequently also seen;

\item {\it Outflows:} Outflows have been found in 21 sources.
    Most of them are oriented along the ionisation axis and associated with an increase in the \woi.
    In 7 sources, increased \woi values occur in a band crossing the nucleus perpendicularly to the ionisation axis which we attribute to the passage of an outflow or radio jet pushing the ambient gas sideways.
    Of the 21 sources with outflows, only 13 show $\woi\ge 600~\kms$; mass outflow rates and powers are only calculated for the latter;

\item {\it Impact of outflows via \woi:} We have employed the \woi index as an indicator of outflows, using a lower limit of 600\,{\kms} to isolate regions dominated by outflows (as lower profile widths can be associated to rotation).
    We find such signature in 13 of the 30 AGNs of our sample.
Seven additional galaxies show other signatures of outflows, although presenting $\woi < 600$\,\kms; we attribute these low values to the fact that such outflows have a low impact on the surrounding gas;

\item {\it Gas densities $n_e$ in the outflows:} We found that the gas densities -- determined via the [S{\sc ii}] line ratio -- tend to be higher in the regions with outflows ($\woi \ge 600 ~{\rm km~s^{-1}}$) than in those without, with median values of of $n_e \approx$\,800\,cm$^{-3}$ and $n_e \approx$\,300\,cm$^{-3}$, respectively;

\item {\it Ionised gas mass within the inner kpc:} Total ionised gas masses within the $\approx$ inner kiloparsec are in the range 10$^{4.5}-10^{8}$\,M$_\odot$; 

\item {\it Mass outflow rate $\dot{M}$:} Using \woi as a proxy for the outflow gas velocity $v=1/2\,\woi$ and a distance of the outflow corresponding to the largest observed radial distance from the nucleus at which $\woi \ge 600~\kms$, we obtain mass outflow rates in the range
$\log\,[\dot{M} / ({\rm M_\odot\, yr^{-1}})] = -3.91$ to $\log\,[\dot{M} / ({\rm M_\odot\, yr^{-1}})] = 2.38$,
with a median value of
$\log\,[\dot{M} / ({\rm M_\odot\, yr^{-1}})] = -2.1_{-1.0}^{+1.6}$,
where the upper and lower limits represent the 16 and 84 percentiles.

\item {\it Outflow power $\dot{E}$:} Outflow powers are in the range $\sim 10^{37}\,\,{\rm erg\,s^{-1}}$ to $\sim 10^{43}\,\,{\rm erg\,s^{-1}}$.
When compared to the AGN bolometric luminosities \lagn, log($\dot{E}$)$\ge$0.01\,\lagn only for 2 sources, for two others 0.001\,$\ge$\,log($\dot{E}$)/\lagn\,$\ge$0.01, while for the remainder 10 galaxies, the powers are lower than 0.001\,\lagn.

\item {\it Relation $\dot{M}$ vs. \lagn:} The mass outflow rate shows some correlation with \lagn, with a large scatter; when compared to the
previous results from the literature \citep[e.g.][]{Fiore2017}, they seem to approximately follow the same relation, on average, with three points above, one on the relation and most of them below the relation;

\item {\it Relation $\dot{E}$ vs. \lagn:} There is also some correlation between the outflow powers and \lagn, also with a large scatter; when added to the previous relation of \citet{Fiore2017}, the 4 most powerful outflows in our sample follow the trend of the being in the range 0.001\,\lagn$\ge$\,log($\dot{E}$)\,$\ge$0.1\,\lagn, while the outflows in the remaining galaxies of our sample populate a region well below the previous relation.

\end{itemize}

\section*{Ackowledgments}

TSB would like to thank the Gemini Brazilian National Time Allocation Committee NTAC, the International Time Allocation Committee ITAC and the Gemini Observatory for the support to all these observations, collected over approximately 10 years. TSB would like to thank as well all the Institutes where the proposals and the analyses of the observations have been carried out: Instituto de F\'isica, Universidade Federal do Rio Grande do Sul, Universidade Federal de Santa Catarina, Universidade Federal de Santa Maria, Rochester Institute of Technology and Harvard-Smithsonian Center for Astrophysics, as well as CAPES, CNPq and FAPERGS for financial support.
DRD would like to thank CAPES and CNPq for financial support during this research.
RAR thanks partial financial support from Conselho Nacional de Desenvolvimento Cient\'ifico e Tecnol\'ogico (202582/2018-3 and 302280/2019-7) and Funda\c c\~ao de Amparo \`a pesquisa do Estado do Rio Grande do Sul (17/2551-0001144-9 and 16/2551-0000251-7).
NN acknowledges support from Conicyt (PIA ACT172033, Fondecyt 1171506, and BASAL AFB-170002).
This work was  supported in part by the National Science Foundation under Grant No. AST-1108786

 Based on observations obtained at the Gemini Observatory, which is operated by the Association of Universities for Research in Astronomy, Inc., under a cooperative agreement with the NSF on behalf of the Gemini partnership: the National Science Foundation (United States), National Research Council (Canada), CONICYT (Chile), Ministerio de Ciencia, Tecnolog\'{i}a e Innovaci\'{o}n Productiva (Argentina), Minist\'{e}rio da Ci\^{e}ncia, Tecnologia e Inova\c{c}\~{a}o (Brazil), and Korea Astronomy and Space Science Institute (Republic of Korea).

\section*{Data availability}
The data underlying this article will be shared on reasonable request to the corresponding author.
The following additional information and figures are available as online supplementary material:
comments on individual galaxies (Appendix A);
acquisition images with the overlaid GMOS field-of-view and sample spectrum (Appendix B);
data plots for all galaxies (Appendix C);
results of Gauss-Hermite higher order moments (Appendix D);
WHAN diagnostic diagrams (Appendix E).


\bibliographystyle{mnras}
\bibliography{references}

\label{lastpage}



\input{all_appendices.tex}

\end{document}

%% file: sample_table.tex
\begin{table}
\label{tab:sample}
\caption{Sample properties: activity type, morphology, distance and projected scale.}
\begin{tabular}{llrcc}
\hline
\hline
Galaxy & Act. & Morph. & Distance & Scale \\
 &  &  & Mpc & pc/$^{\prime\prime}$ \\
\hline
Mrk~348 & S2 & SA(s)0/a: & 21.5 & 104 \\
NGC~1068 & S2 & (R)SA(rs)b & 12.6 & 61 \\
Mrk~1058* & S2 & S? & 72.8 & 346 \\
Mrk~607 & S2 & Sa & 37.7 & 181 \\
NGC~1365 & S1 & SB(s)b & 18.1 & 87 \\
NGC~1358* & S2 & SAB0/a(r) & 56.0 & 267 \\
NGC~1386* & S2 & SB0\^{}+(s) & 15.9 & 77 \\
NGC~1566 & S1 & SAB(s)bc & 6.6 & 32 \\
NGC~1667* & S2 & SAB(r)c & 42.8 & 205 \\
NGC~2110 & S2 & SAB0\^{}- & 36.9 & 177 \\
Mrk~6 & S1 & SAB0\^{}+: & 95.0 & 449 \\
Mrk~79 & S1 & SBb & 90.5 & 428 \\
NGC~2787* & L2 & SB0\^{}+(r) & 7.4 & 36 \\
MCG~-05-23-016 & S1 & S0? & 41.3 & 198 \\
NGC~3081 & S2 & (R)SAB0/a(r) & 32.5 & 156 \\
NGC~3227 & S1 & SAB(s)a pec & 20.6 & 99 \\
NGC~3516 & S1 & (R)SB0\^{}0?(s) & 38.9 & 187 \\
NGC~3783 & S1 & (R')SB(r)ab & 38.5 & 185 \\
NGC~3786 & S2 & SAB(rs)a pec & 47.0 & 225 \\
NGC~3982* & S2 & SAB(r)b? & 21.6 & 104 \\
NGC~4180 & S2 & Sab? & 37.5 & 180 \\
NGC~4450* & L1 & SA(s)ab & 15.3 & 74 \\
NGC~4501* & S2 & SA(rs)b & 19.7 & 95 \\
NGC~4593 & S1 & (R)SB(rs)b & 25.6 & 123 \\
MCG~-06-30-015 & S1 & S? & 37.4 & 180 \\
NGC~5728 & S2 & SAB(r)a? & 26.4 & 127 \\
NGC~5899 & S2 & SAB(rs)c & 45.3 & 217 \\
NGC~6300 & S2 & SB(rs)b & 14.0 & 68 \\
NGC~6814 & S1 & SAB(rs)bc & 22.8 & 110 \\
NGC~7213 & S1 & SA(s)a? & 22.0 & 106 \\
\hline
\end{tabular}

$^*$ Galaxies with an X-Ray counterpart in Swift/BAT 105 month catalog \citep{Oh2018}.
\end{table}

%% file: obslog_table.tex
\begin{table}
\caption{\label{tab:obslog}Observation log -- Columns are: (1) galaxy name; (2) Gemini program ID; (3) on-source exposure time in seconds; (4) field-of-view in arcsec.}
\begin{tabular}{llcc}
\hline
\hline
Galaxy & Program ID & Exp. Time & FoV \\
\hline
Mrk~348$^\dagger$ & GN-2014B-Q-87 & 4860 & $1.7 \times 2.5$ \\
NGC~1068$^\dagger$ & GS-2010B-Q-81 & 2491 & $3.3 \times 4.9$ \\
Mrk~1058$^\dagger$ & GN-2014B-Q-87 & 4860 & $1.7 \times 2.5$ \\
Mrk~607$^\dagger$ & GN-2014B-Q-87 & 4860 & $1.7 \times 2.5$ \\
NGC~1365$^\dagger$ & GS-2014B-Q-30 & 1890 & $3.3 \times 5.3$ \\
NGC~1358 & GS-2010B-Q-19 & 4206 & $7.4 \times 9.5$ \\
NGC~1386 & GS-2011B-Q-23 & 4506 & $7.4 \times 9.5$ \\
NGC~1566 & GS-2011B-Q-23 & 4506 & $7.4 \times 9.5$ \\
NGC~1667 & GS-2010B-Q-19 & 4206 & $7.4 \times 9.5$ \\
NGC~2110 & GS-2010B-Q-19 & 4206 & $7.4 \times 9.6$ \\
Mrk~6$^\dagger$ & GN-2014B-Q-87 & 5670 & $3.4 \times 5.1$ \\
Mrk~79$^\dagger$ & GN-2014B-Q-87 & 4860 & $1.7 \times 2.5$ \\
NGC~2787 & GN-2011A-Q-85 & 4920 & $7.1 \times 9.2$ \\
MCG~-05-23-016$^\dagger$ & GS-2014A-Q-78 & 2400 & $6.3 \times 5.5$ \\
NGC~3081 & GN-2011A-Q-85 & 4920 & $7.1 \times 9.2$ \\
NGC~3227$^\dagger$ & GN-2013A-Q-61 & 2400 & $6.3 \times 4.9$ \\
NGC~3516$^\dagger$ & GN-2013A-Q-61 & 7201 & $9.3 \times 4.9$ \\
NGC~3783$^\dagger$ & GS-2014A-Q-78 & 4802 & $6.3 \times 5.8$ \\
NGC~3786$^\dagger$ & GN-2013A-Q-61 & 4800 & $6.3 \times 4.9$ \\
NGC~3982 & GN-2006B-Q-94 & 4680 & $7.4 \times 15.5$ \\
NGC~4180 & GN-2014A-Q-90 & 4800 & $6.3 \times 5.8$ \\
NGC~4450 & GN-2006B-Q-94 & 4682 & $21.4 \times 5.5$ \\
NGC~4501 & GN-2008A-Q-8 & 6000 & $7.4 \times 15.5$ \\
NGC~4593$^\dagger$ & GN-2013A-Q-61 & 4800 & $6.3 \times 4.9$ \\
MCG~-06-30-015$^\dagger$ & GS-2014A-Q-78 & 4802 & $6.3 \times 5.8$ \\
NGC~5728$^\dagger$ & GS-2013A-Q-56 & 4802 & $6.3 \times 4.9$ \\
NGC~5899$^\dagger$ & GN-2013A-Q-61 & 6818 & $6.7 \times 5.0$ \\
NGC~6300$^\dagger$ & GS-2014A-Q-78 & 8823 & $6.8 \times 5.8$ \\
NGC~6814$^\dagger$ & GS-2014A-Q-78 & 2400 & $6.3 \times 5.6$ \\
NGC~7213 & GS-2011B-Q-23 & 4506 & $7.4 \times 9.5$ \\
\hline
\end{tabular}

$^\dagger$ Galaxies observed in single slit mode.
\end{table}

%% file: vsys_table.tex
\begin{table*}
{\footnotesize
\caption{\label{tab:vsys}Velocities, kinematic and photometric major axis  Columns are: (1) name of the galaxy; (2) systemic velocity of the \nii velocity field; (3) kinematic major axis of the \nii velocity field; (4) \nii ionisation axis; (5) systemic velocity of the \oiii velocity field; (6) kinematic major axis of the \oiii velocity field; (7) \oiii ionisation axis; (8) Photometric major axis; }
\begin{tabular}{lccccccc}
\hline
\hline
Galaxy & [N~{\sc ii}] $v_{\rm sys}$ & KMA [N~{\sc ii}] & [N~{\sc ii}] IA & [O~{\sc iii}] $v_{\rm sys}$ & KMA [O~{\sc iii}] & [O~{\sc iii}] IA & PMA \\
 & $\mathrm{km\,s^{-1}}$ & $\mathrm{{}^{\circ}}$ &  & $\mathrm{km\,s^{-1}}$ & $\mathrm{{}^{\circ}}$ &  & $\mathrm{{}^{\circ}}$ \\
\hline
Mrk~348 & 4531 & 25 & 12 & 4529 & 25 & 9 & 160 \\
NGC~1068 & 1009 & 23 & 9 & 984 & 28 & 18 & 82$^\dagger$ \\
Mrk~1058 & 5102 & 127 & 27 & 5089 & 167 & 18 & 115 \\
Mrk~607 & 2777 & 133 & 139 & 2750 & 131 & 140 & 140 \\
NGC~1365 & 1617 & 60 & 153 & -- & -- & -- & 49 \\
NGC~1358 & 4083 & 84 & 124 & -- & -- & -- & 15 \\
NGC~1386 & 822 & 22 & 7 & -- & -- & -- & 25 \\
NGC~1566 & 1469 & 33 & 30 & -- & -- & -- & 40$^\dagger$ \\
NGC~1667 & 4641 & 138 & 159 & -- & -- & -- & 165 \\
NGC~2110 & 2344 & 172 & 167 & -- & -- & -- & 165 \\
Mrk~6 & 5526 & 166 & 8 & 5628 & 171 & 36 & 130 \\
Mrk~79 & 6625 & 147 & 8 & 6618 & 31 & 5 & 65 \\
NGC~2787 & 672 & 68 & 75 & -- & -- & -- & 110 \\
MCG~-05-23-016 & 2537 & 66 & 66 & 2549 & 72 & 157 & 50 \\
NGC~3081 & 2412 & 75 & 143 & -- & -- & -- & 70 \\
NGC~3227 & 1122 & 0 & 171 & 1041 & 14 & 175 & 153 \\
NGC~3516 & 2608 & 36 & 10 & 2621 & 24 & 13 & 27 \\
NGC~3783 & 2969 & 75 & 15 & -- & -- & -- & 100 \\
NGC~3786 & 2682 & 74 & 82 & 2685 & 75 & 90 & 70 \\
NGC~3982 & 1107 & 25 & 34 & -- & -- & -- & 15 \\
NGC~4180 & 2042 & 177 & 52 & -- & -- & -- & 20 \\
NGC~4450 & 1920 & 36 & 84 & -- & -- & -- & 0 \\
NGC~4501 & 2238 & 124 & 160 & -- & -- & -- & 140 \\
NGC~4593 & 2485 & 95 & 115 & 2489 & 83 & 125 & 120$^\dagger$ \\
MCG~-06-30-015 & 2325 & 117 & 112 & 2323 & 111 & 109 & 115 \\
NGC~5728 & 2772 & 173 & 143 & 2812 & 172 & 148 & 2$^\dagger$ \\
NGC~5899 & 2671 & 49 & 177 & 2672 & 9 & 176 & 20 \\
NGC~6300 & 1111 & 161 & 58 & 1089 & 24 & 61 & 118 \\
NGC~6814 & 1637 & 170 & 135 & 1667 & 144 & 148 & 65 \\
NGC~7213 & 1859 & 128 & 159 & -- & -- & -- & 70 \\
\hline
\end{tabular}

}
\begin{flushleft}
Photometric major axis for most galaxies is taken from the 2MASS Extended Source Catalog (XSC), 
with the exception of those marked with $^\dagger$. For NGC\,5729 the PA was taken from 
\citep{Erwin2004}, and for the remaining three the PA was visually estimated from the DSS images.
\end{flushleft}
\end{table*}

%% file: outflow_table.tex
\begin{table*}
\caption{Outflow rate, power and bolometric luminosity -- Columns are: (1) name of the galaxy; (2 and 5) distance in the plane of the sky from the galactic centre to the last spaxel identified as outflow dominated; (3 and 6) mass outflow rates; (4 and 7) outflow kinetic power; (8) outflow position angle; (9) AGN bolometric luminosity.}
\label{tab:outflow}
\begin{tabular}{lrrrrrrrr}
\hline\hline
Galaxy & $R_{\rm max}\,\,\mbox{[N {\sc ii}]}$ & $\log \dot{M}\,\,\mbox{[N {\sc ii}]}$ & $\log \dot{E}\,\,\mbox{[N {\sc ii}]}$ & $R_{\rm max}\,\,\mbox{[O {\sc iii}]}$ & $\log \dot{M}\,\,\mbox{[O {\sc iii}]}$ & $\log \dot{E}\,\,\mbox{[O {\sc iii}]}$ & Outflow PA & $\log L_{\rm AGN}$ \\
 & $\mathrm{pc}$ & $\mathrm{M_{\odot}\,yr^{-1}}$ & $\mathrm{erg\,s^{-1}}$ & $\mathrm{pc}$ & $\mathrm{M_{\odot}\,yr^{-1}}$ & $\mathrm{erg\,s^{-1}}$ & $\mathrm{{}^{\circ}}$ & $\mathrm{erg\,s^{-1}}$ \\
\hline
Mrk\,348 &  &  &  & 158 & -2.64 & 38.15 & 19 & 44.08 \\
NGC\,1068 & 171 & 0.30 & 41.38 & 171 & 0.32 & 41.58 & 30 & 42.85 \\
Mrk\,1058 &  &  &  &  &  &  & 44 &  \\
Mrk\,607$^\dagger$ &  &  &  &  &  &  & 45 & 43.42 \\
NGC\,1365 &  &  &  &  &  &  & 118 & 43.48 \\
NGC\,1358 &  &  &  &  &  &  & 120 & 42.05 \\
NGC\,1386$^\dagger$ & 140 & -2.18 & 38.68 &  &  &  & 110 & 40.89 \\
NGC\,1566 &  &  &  &  &  &  &  & 41.92 \\
NGC\,1667 & 235 & -2.09 & 38.45 &  &  &  & 164 & 41.14 \\
NGC\,2110 & 336 & -1.05 & 39.43 &  &  &  & 54 & 45.11 \\
Mrk\,6$^\dagger$ & 1217 & 2.35 & 43.03 & 1217 & 2.38 & 43.22 & 0 & 45.19 \\
Mrk\,79$^\dagger$ &  &  &  & 1068 & -1.38 & 39.46 & 10 & 44.98 \\
NGC\,2787 & 30 & -3.91 & 36.22 &  &  &  & 158 & 41.56 \\
MCG\,-05-23-016 &  &  &  &  &  &  &  & 44.99 \\
NGC\,3081 &  &  &  &  &  &  & 0 & 44.22 \\
NGC\,3227 & 352 & -2.33 & 37.73 & 269 & -2.40 & 37.63 & 27 & 43.91 \\
NGC\,3516 & 501 & -2.82 & 38.26 & 577 & -3.00 & 37.03 & 15 & 44.58 \\
NGC\,3783 &  &  &  &  &  &  &  & 44.81 \\
NGC\,3786 &  &  &  &  &  &  &  & 43.70 \\
NGC\,3982 &  &  &  &  &  &  &  & 40.86 \\
NGC\,4180 & 115 & -3.43 & 37.36 &  &  &  &  & 43.57 \\
NGC\,4450 &  &  &  &  &  &  &  & 41.04 \\
NGC\,4501 &  &  &  &  &  &  & 45 & 40.57 \\
NGC\,4593 &  &  &  &  &  &  &  & 44.01 \\
MCG\,-06-30-015 &  &  &  &  &  &  &  & 44.20 \\
NGC\,5728 & 238 & -2.07 & 38.06 & 333 & -2.09 & 37.92 & 135 & 44.02 \\
NGC\,5899$^\dagger$ & 202 & -1.67 & 39.07 & 245 & -1.93 & 38.74 & 0 & 43.84 \\
NGC\,6300 &  &  &  &  &  &  & 26 & 43.43 \\
NGC\,6814 &  &  &  &  &  &  & 151 & 43.72 \\
NGC\,7213 &  &  &  &  &  &  &  & 43.43 \\
\hline
\end{tabular}

\begin{flushleft}
$^\dagger$ Galaxies with equatorial outflow.
\end{flushleft}
\end{table*}

%% file: all_appendices.tex
\newpage

\appendix

\section{Comments on individual galaxies}
\label{sec:individual}

\noindent{\bf Mrk\,348 (Fig.\,\ref{fig:data_mrk348}):} This Seyfert 2 galaxy shows 
the SER (see section \ref{sec:results} for a definition) more extended along NE-SW, at an angle of $\approx$\,40$^\circ$ with the PMA. We identify this direction as that of the ionisation axis of the AGN, along which the gas velocity map shows two compact regions (diameters of $\sim$\,200\,pc) at opposite sides of the nucleus, one blueshifted by $\approx-100$\kms at 1.4\,arcsec (420\,pc) to the NE and the other redshifted to the SW (with similar velocity and distance from the nucleus). These properties can be interpreted as due to {\it outflows} from the nucleus. These outflows are associated to an increase in the \woi values, mostly to the North, suggesting an interaction 
between the outflow and the local ISM (even though \woi is lower than the 600\kms\ threshold we have adopted for outflows over the whole FoV). The electronic density map shows an increase towards the SW, in the direction of the compact redshifted region. The data for this galaxy has been previously analysed by \citet{Freitas2018}. 

\noindent {\bf NGC\,1068 (Fig.\,\ref{fig:data_ngc1068}):} This well studied Sy 2 galaxy shows the peak of gas emission displaced by $0\farcs5$ (30\,pc) towards the NE from the centre, with the SER extending along this direction, that makes an angle of $\approx$\,50$^\circ$ with the PMA.
Along this direction -- that can be identified with the AGN ionisation axis -- blueshifts are observed to the NE and redshifts to the SW reaching  $\approx$\,800\,km\,s$^{-1}$ in \oiii at $\approx 2\farcs5$ from the nucleus and somewhat lower values ($\approx$600\,\kms) in the \nii lines.
Beyond the high blueshifts, redshifts are observed to the NE, evidencing the presence of more than one kinematic component and that the high blue and redshifts are due to \emph{nuclear outflows}.
W$_{80}$ is enhanced to values as high as 2500\,km s$^{-1}$ surrounding the regions with the highest velocities.

\noindent {\bf Mrk\,1058 (Fig.\,\ref{fig:data_mrk1058}):} The largest extent of the SER of this Seyfert\,2 galaxy is observed  towards the SW, that can be identified with the orientation of the ionisation axis. This direction is almost perpendicular to the PMA.
Excess blueshifts are observed in the gas relative {\bfseries to the symmetric \nii velocity} field everywhere, as previously shown by \citet{Freitas2018}.
The blueshifts are more conspicuous in the \oiii velocity field along the ionisation axis and are associated with an increase in the \nii \woi, suggesting the presence of an {\it outflow} impacting the ISM there, even though \woi$\le$600\kms everywhere.

\noindent {\bf Mrk\,607 (Fig.\,\ref{fig:data_mrk607}):} The SER in this Seyfert 2 galaxy is most extended to the NW along the PMA which also coincides with the kinematic major axis (KMA). \citet{Freitas2018} have shown that the gas in this galaxy counter-rotates relative to the stars, suggesting an external origin. An increase in \woi is observed 
in a $\sim$\,200\,pc wide band crossing the nucleus
perpendicularly to the PMA.
This increase is approximately co-spatial with the increase in the gas density and in the gas velocity dispersion as previously reported by \cite{Freitas2018} and tentatively attributed to an {\it equatorial outflow} related to the AGN and its radio emission, as the ionisation axis seems to be along the PMA.
\citet{schoenell2019} reach similar conclusions, reporting an equatorial outflow of molecular gas, traced by the H$_2$ emission.

\noindent {\bf NGC\,1365 (Fig.\,\ref{fig:data_ngc1365}):} The FoV of our observations of this nearby Sy\,1 galaxy covers only $\approx300\times400$\,pc at the galaxy, the SER being extended beyond the FoV. An increase in the electronic density is observed towards the SE, in agreement with the measurements by \citet{lena2016}  who have also obtained GMOS-IFU spectroscopy of this galaxy over a somewhat larger
FoV. Over an even larger FoV, a cone-shaped structure has been observed extending by 1.5\,kpc to the SE in \oiii by \citet{Storchi-Bergmann1991} and associated {\it outflows} have been observed in previous studies \citep{Sharp2010}.

\noindent {\bf NGC\,1358 (Fig.\,\ref{fig:data_ngc1358}):} This Sy\,2 galaxy shows emission that peaks at the nucleus,
with the SER, observed in [N\,{\sc ii}], showing an S-shape structure, being most extended approximately perpendicularly to the PMA. The kinematics show a distorted rotation pattern with the KMA approximately along the S-shape structure, almost
perpendicular to the PMA. Two emission blobs at 2\,arcsec ($\approx$\,500\,pc) from the nucleus at the end of the {\it S} are
associated with an increase in W$_{80}$ and can be attributed to {\it outflows} from the nucleus (as previously
discussed in \citet{SchnorrMuller2017a}).

\noindent {\bf NGC\,1386 (Fig.\,\ref{fig:data_ngc1386}):} This Sy\,2 galaxy shows emission peaking at the nucleus and an ionisation axis making a small angle ($\approx$\,20$^\circ$) with the PMA, the latter of which is in approximate agreement with the KMA.
\citet{Lena2015}) pointed out the presence of a compact outflow along the ionisation axis that is not resolved by our observations.
The kinematics show distorted rotation and the most conspicuous feature is an increase in \woi perpendicularly to the ionisation axis, attributed to gas in {\it equatorial rotation and outflow} (as previously discussed in \citet{Lena2015}).
This increase in \woi is associated with an increase in the gas density.
Based on coronal lines seen in the near infrared spectrum of this galaxy, \citet{rodriguez-ardila2017b} estimate a mass outflow rate of $11\,\,{\rm M_\odot\,yr^{-1}}$.

\noindent {\bf NGC\,1566 (Fig.\,\ref{fig:data_ngc1566}):} This Sy\,1 galaxy shows emission peaking at the nucleus and a very compact SER; at lower flux levels it is more extended along the PMA that seems to coincide with the KMA.
A large HII region is observed at $\approx$ 2\,arcsec ($\approx$\,60\,pc) SW of the nucleus.
A distorted rotation pattern seems to be due to motions along nuclear spiral arms that could be associated to inflows seen in cold molecular gas kinematics in observations with the Atacama Large Milimetric Array ALMA \citep{Combes2014}.
Combining data from ALMA and GMOS \citet{Slater2019} argue for the presence of molecular and ionised gas outflows in the inner kiloparsec of this galaxy.

\noindent {\bf NGC\,1667 (Fig.\,\ref{fig:data_ngc1667}):}  This Sy\,2 galaxy shows the SER slightly more extended towards the SW.
The kinematics show a distorted rotation pattern with an KMA apparently tilted by $\approx30^\circ$ relative to the PMA. This distortion has been attributed to inflows along nuclear spiral arms \citep{SchnorrMuller2017a}. \nii \woi shows the highest values in an arc structure just to the north of the nucleus. As \woi$\ge$600\kms, we have considered the presence of an {\it outflow} there.

\noindent {\bf NGC\,2110 (Fig. \ref{fig:data_ngc2110}):} This Sy\,2 galaxy shows the SER more elongated towards the N-NW, that can be identified with the orientation of the ionisation axis, which is approximately also the orientation of the PMA and KMA.
The kinematics shows a distorted rotation pattern -- previously analysed by \citet{SchnorrMuller2014} with optical data and \citet{Diniz2015} in the near-IR -- with enhanced \woi along a ``band" crossing the nucleus almost perpendicular to the ionisation axis, attributed to an {\it equatorial outflow} as in the case of NGC\,1386. The increased \woi seems to be associated to an increase also in the gas density.

\noindent {\bf Mrk\,6 (Fig.\,\ref{fig:data_mrk6}):} This Seyfert\,1 galaxy shows the SER more extended along the N-S direction and a distorted rotation pattern in the emitting gas, as previously pointed out by \citet{Freitas2018}. The rotation component seems to have a KMA following the orientation of the PMA, more clearly seen in \nii. In \oiii, the increase in the blueshifts to the North and redshifts to the South, indicate the presence of an {\it outflow} making an angle of $\approx50^\circ$ with the PMA. This N-S outflow is approximately co-spatial with a radio structure seen in a 3.6\,cm image \citep{Freitas2018}. High values of \woi are observed over most of the FoV, and being highest in two regions aligned almost perpendicularly to the radio-structure and outflow, as in NGC\,2110 and NGC\,1386, probably due again to an {\it equatorial outflow} produced by the passage of the radio jet.

\noindent {\bf Mrk\,79 (Fig.\,\ref{fig:data_mrk79}):} This Sy\,1 galaxy shows the SER elongated towards the S-SW, at an angle of 55$^\circ$ relative to the PMA, extending beyond the border of the FoV. This direction (P.A.$\approx$190$^\circ$) can be identified with the ionisation axis of the AGN, as evidenced in the narrow-band image of \citet{Schmitt2003}, which is also the orientation of the jet-like 3.6\,cm radio continuum image \citep{Schmitt2001}. In \oiii, we find blueshifts to the South that we attribute to an {\it outflow} along the ionisation axis, while in \nii lower velocity blueshifts are also observed there. Blueshifts are also observed to the NW, but their origin is not clear, seeming to be a counterpart to a redshifted region observed to the SE in the \nii emission. \woi is enhanced approximately perpendicularly to the ionisation axis, this can again be attributed to an {\it equatorial outflow} due to the passage of the radio jet. The gaseous kinematics of this galaxy has been previously studied in the near-IR by \citet{Riffel2013}, who also found strongest gas emission to the south, and similar kinematics to ours in the Pa\,$\beta$ and [Fe\,{\sc ii}] emission lines. \citet{Freitas2018} has analysed these data in a previous study but the SER elongation was mistakenly identified as being oriented towards the North in that paper.

\noindent {\bf NGC\,2787 (Fig.\,\ref{fig:data_ngc2787}):} This galaxy has a LINER nucleus, very compact SER ($\approx$40\,pc) that peaks at the nucleus and extended emission that shows a rotation pattern with KMA tilted by $\approx34^\circ$ relative to the PMA. Its kinematics and excitation have been investigated in \citet{brum2017}. The [N\,{\sc ii}] \woi is high at the nucleus and its distribution is elongated to the NW approximately along the kinematic minor axis, which could be due to an {\it outflow} along this direction.

\noindent {\bf MCG-5-23-16 (Fig.\,\ref{fig:data_mcg0523}):} This Seyfert\,1 galaxy shows the peak of the [O\,{\sc iii}] emission off-centred by $0\farcs5$ to the E of the nucleus and the SER in [O\,{\sc iii}] being most extended approximately along the PMA, that seems to coincide with the KMA. The [N\,{\sc ii}] kinematics is dominated by rotation, with a steeper gradient than that of the [O\,{\sc iii}] kinematics, suggesting the presence of another, non-rotating component in the latter. 

\noindent {\bf NGC\,3081 (Fig.\,\ref{fig:data_ngc3081}):} This Sy\,2 galaxy shows gas emission peaking at the nucleus, a SER elongated to the N-NW, approximately perpendicularly to the PMA. A nuclear bar is observed along the PMA, whose orientation is in approximate agreement with that of the KMA. A distorted rotation pattern, observed in [N\,{\sc ii}] has been associated to inflows along the bar, combined with a compact nuclear {\it outflow} almost perpendicular to the bar -- as previously discussed in \citet{SchnorrMuller2016}. The outflow is approximately oriented along the SER, and is associated with an increase in the gas density. An increase in \woi is observed surrounding the SER, probably due again to an {\it equatorial outflow}, or lateral displacement of the ambient gas by the outflow. Both outflows lead to \woi below the 600\,\kms threshold.

\noindent {\bf NGC\,3227 (Fig.\,\ref{fig:data_ngc3227}):} The SER in this Sy\,1 galaxy is very compact, but at fainter levels of gas emission is most extended towards the NE, making an angle of $\approx$\,54$^\circ$ with the PMA.
The highest blueshifts -- seeming to extend beyond our FoV are also observed in this direction. \woi increases to values of the order of 1000\,km\,s$^{-1}$ towards the NE (P.A.$\approx27^\circ$), where blueshifts are observed and can be attributed to an {\it outflow}, which is also supported by [Fe\,{\sc ii}] emission in the near-IR \citep{schoenell2019}.
The velocity fields are distorted, dominated by blueshifts in the case of [O\,{\sc iii}] and showing also a rotation component in the [N\,{\sc ii}] kinematics, with a KMA approximately along the PMA. 

\noindent {\bf NGC\,3516 (Fig.\,\ref{fig:data_ngc3516}):} In this Sy\,1 galaxy, the SER in [N\,{\sc ii}] is elongated to the N-NE border of FoV at 3\farcs5 ($\approx$\,1\,kpc) from the nucleus, similarly observed also in [O\,{\sc iii}] but at lower flux levels. This orientation is approximately that of the PMA (also similar to that of the KMA). The kinematics suggest again a distorted rotation pattern, with the highest blueshifts ($\approx$-300\,km\,s$^{-1}$) being observed closer to the nucleus -- at $\approx$2\,arcsec370\,pc N-NE, than the highest redshifts at $\approx$\,1\,kpc to the S-SW. The highest redshifts seem to occur beyond the limits of the FoV and are probably due to rotation. The region with the highest blueshifts is surrounded by very high \woi values ($\approx$1000\,km\,s$^{-1}$ and gas densities, which suggests that these blueshifts are due to an {\it outflow} that is pushing the surrounding gas.

\noindent {\bf NGC\,3783 (Fig.\,\ref{fig:data_ngc3783}):} This Sy\,1 galaxy shows a round SER, with a velocity field showing a low amplitude ($\le$40\,km\,s$^{-1}$) distorted rotation with a KMA apparently tilted by $\approx45^\circ$ relative to the PMA. There is some elongation to the East in the [N\,{\sc ii}] gas emission flux, \woi and gas density distributions following the PMA.

\noindent {\bf NGC\,3786 (Fig.\,\ref{fig:data_ngc3786}):} This Sy\,2 galaxy shows a compact SER with the gas velocity fields showing a rotation pattern with a KMA approximately along the PMA. The highest \woi values are observed in an elongated structure reaching $\sim 1\farcs4$ ($\approx$\,300\,pc) from the nucleus to the SE with some correspondence in the gas density map, although no signature of outflows is seen in the velocity fields. 

\noindent {\bf NGC\,3982 (Fig.\,\ref{fig:data_ngc3982}):} The \nii SER in this Sy\,2 galaxy is round and compact, reaching $\approx$\,100\,pc from the nucleus, where the highest \woi and gas densities are observed. 
\citet{brum2017} has studied the gas kinematics and excitation of this galaxy showing deviations from pure rotation associated with nuclear spiral structure.

\noindent {\bf NGC\,4180 (Fig.\,\ref{fig:data_ngc4180}):} In this Sy\,2 galaxy, the gas emission peaks $\approx\,0\farcs2$ (40\,pc) E of the nucleus and the SER is more extended approximately along the PMA. The gas kinematics shows again a distorted rotation pattern with a KMA apparently tilted 
relative to the PMA. \woi is largest just to the West of the nucleus, where it reaches $\ge$600\kms that we  attribute to an {\it outflow} whose orientation is not clear in the the kinematic maps.

\noindent {\bf NGC\,4450 (Fig.\,\ref{fig:data_ngc4450}):} This galaxy has a nucleus classified as LINER 1. The long FoV reaches HII regions 5$^{\prime\prime}$ from nucleus. The gas kinematics shows again a distorted rotation pattern whose KMA seems to approximately coincide with the PMA. A blueshifted knot of gas emission at $\approx\,1\farcs5$ (100\,pc) E of the nucleus coincides with an increase in the velocity dispersion that suggests it is gas in {\it outflow}, as previously discussed in \citet{brum2017}, but with \woi below the 600\kms threshold.

\noindent {\bf NGC\,4501 (Fig.\,\ref{fig:data_ngc4501}):} This Sy\,2 galaxy has the \nii SER extended along the PMA.
The kinematics shows a distorted rotation pattern with a KMA with similar orientation to that of the PMA.
The distortion along the minor axis seems to be associated with an increase in \woi that suggests a small {\it outflow} along the minor axis, in particular to the SW, as seen also in the previous study by \citep{brum2017}.
The \woi is nevertheless below the 600\kms threshold.

\noindent {\bf NGC\,4593 (Fig.\,\ref{fig:data_ngc4593}):} This Sy\,1 galaxy has the \oiii SER most extended along the PMA, with the peak emission displaced by $\approx 0\farcs3$ (40\,pc) to the NE of the nucleus. The gas kinematics show a rotation pattern with a tilted KMA relative to the PMA by $\approx\,30^\circ$. The highest \woi and gas densities are observed to the E-SE, what could indicate the presence of an outflow there. This seems not be supported by the velocity fields even though the \oiii one is shallower than that in \nii suggesting the presence of another kinematic component in \oiii. 

\noindent {\bf MCG-6-30-15 (Fig.\,\ref{fig:data_mcg0630}):} This is a Sy\,1.2 galaxy close to edge-on, with emission
peaking at the nucleus, SER most extended along the PMA, that again coincides with the KMA. Both  [N\,{\sc ii}] and [O\,{\sc iii}] kinematics are dominated by rotation (although again showing distortions) with W$_{80}$[N\,{\sc ii}] and gas density showing a small increase at the nucleus.

\noindent {\bf NGC\,5728 (Fig.\,\ref{fig:data_ngc5728}):} This Sy\,2 galaxy has elongated \nii and \oiii SER's reaching the borders of the FoV to the NW of the nucleus, making an angle of $\approx$\,45$^\circ$ with the PMA that is along N-S. The peak emission is displaced by $\approx 0\farcs5$ (60\,pc) to the NW of the nucleus, where high redshifts (up to 350\,\kms) are observed.
Another peak of emission is observed further to the NW at $\approx$ 3\,arcsec (400\,pc) from the nucleus, where the highest blueshifts (up to $-$300\,\kms) are observed. There is a steep velocity gradient between these two regions, that is co-spatial with a region of very high values of \woi. A recent study of this galaxy by \citet{Shimizu2019} has confirmed the presence of a {\it bipolar outflow} associated with the above two regions. Enhanced \woi is observed perpendicularly to the ionization and outflow axis.

\noindent {\bf NGC\,5899 (Fig.\,\ref{fig:data_ngc5899}):} This Sy\,2 galaxy shows the SER elongated in the direction N-S, tilted by $\approx$\,20$^\circ$ relative to the PMA. A hint of a rotation pattern with the KMA similar to the PMA is observed towards the borders of the FoV, but opposite blueshifts and redshifts are observed internal to the these regions (within the inner $\approx$200\,pc radius) due to a compact outflow, with blueshifts to the south and redshifts to the north, in agreement with near-IR studies of the same galaxy \citep{schoenell2019}. Farther out, blueshifts and redshifts are also observed perpendicularly to the major axis, where enhanced \woi\ is observed.

\noindent {\bf NGC\,6300 (Fig.\,\ref{fig:data_ngc6300}):} This Sy\,2 shows the \nii and \oiii SER more extended to the NE, approximately perpendicularly to the PMA, where blueshifts are observed in both emission lines. Blueshifts are observed also at the nucleus, while to the SW, both [N\,{\sc ii}] and [O\,{\sc iii}] show redshifts. The highest values of \woi and gas density are observed to the SW. As these structures are observed perpendicularly to the PMA, we interpret them as due to {\it outflows}, although the \woi value remain below the 600\kms threshold.

\noindent {\bf NGC\,6814 (Fig.\,\ref{fig:data_ngc6814}):} This Sy\,1 galaxy shows the SER more extended along SE-NW, with blueshifts observed to the SE and redshifts to the NW along P.A. $\approx$151$^\circ$. As this direction is $\approx$ perpendicular to the PMA, and the velocities decrease beyond $\approx 1\farcs3$ (140\,pc) from the nucleus, the kinematics suggest a compact {\it bipolar outflow}, what is also supported by high gas density in the region and increased \woi values, although \woi does not reach the 600\,km\,s$^{-1}$ threshold.

\noindent {\bf NGC\,7213 (Fig.\,\ref{fig:data_ngc7213}):} This LINER/Sy\,1 galaxy shows a very compact SER with enhanced emission beyond this region delineating nuclear spiral arms. A distorted rotation pattern that seems to show a KMA almost perpendicular to the PMA has been found to be due to inflows along the nuclear spiral \citep{SchnorrMuller2014a}. 

\newpage
\clearpage

\section{Acquisition and spectrum}
\label{sec:acquisition_plots}

\input{acquisition.tex}

\newpage
\clearpage

\section{Galaxy by galaxy data}
\label{sec:data_plots}

\input{individual_results.tex}

\newpage
\clearpage

\section{Gauss-Hermite higher order moments}
\label{sec:hermite_plots}

\input{hermite_results.tex}

\newpage
\clearpage

\section{WHAN diagrams}
\label{sec:whan_diagrams}

\input{whan.tex}

%% file: acquisition.tex

\begin{figure*}
  \centering
  \includegraphics[page=1, width=\textwidth]{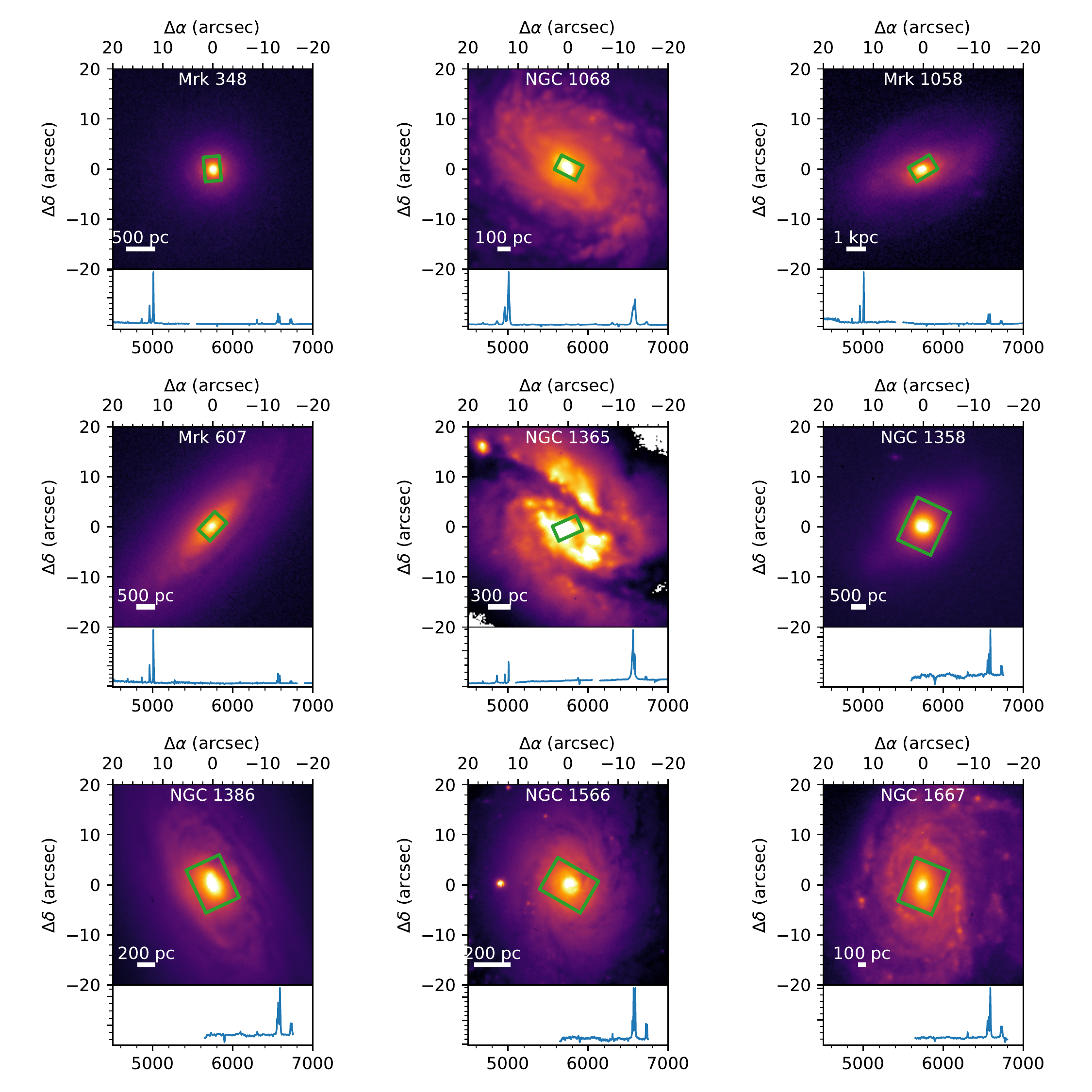}
  \caption{
      Acquisition images and central spectrum for each galaxy in the sample.
      All images are in the GMOS R filter, which is very similar to the SDSS r' filter., and all colour maps are in logarithmic scale.
      The green dashed rectangle represents the limits of the IFU field of view.
  }
  \label{fig:acqspec1}
\end{figure*}

\clearpage

\begin{figure*}
  \centering
  \includegraphics[page=2, width=\textwidth]{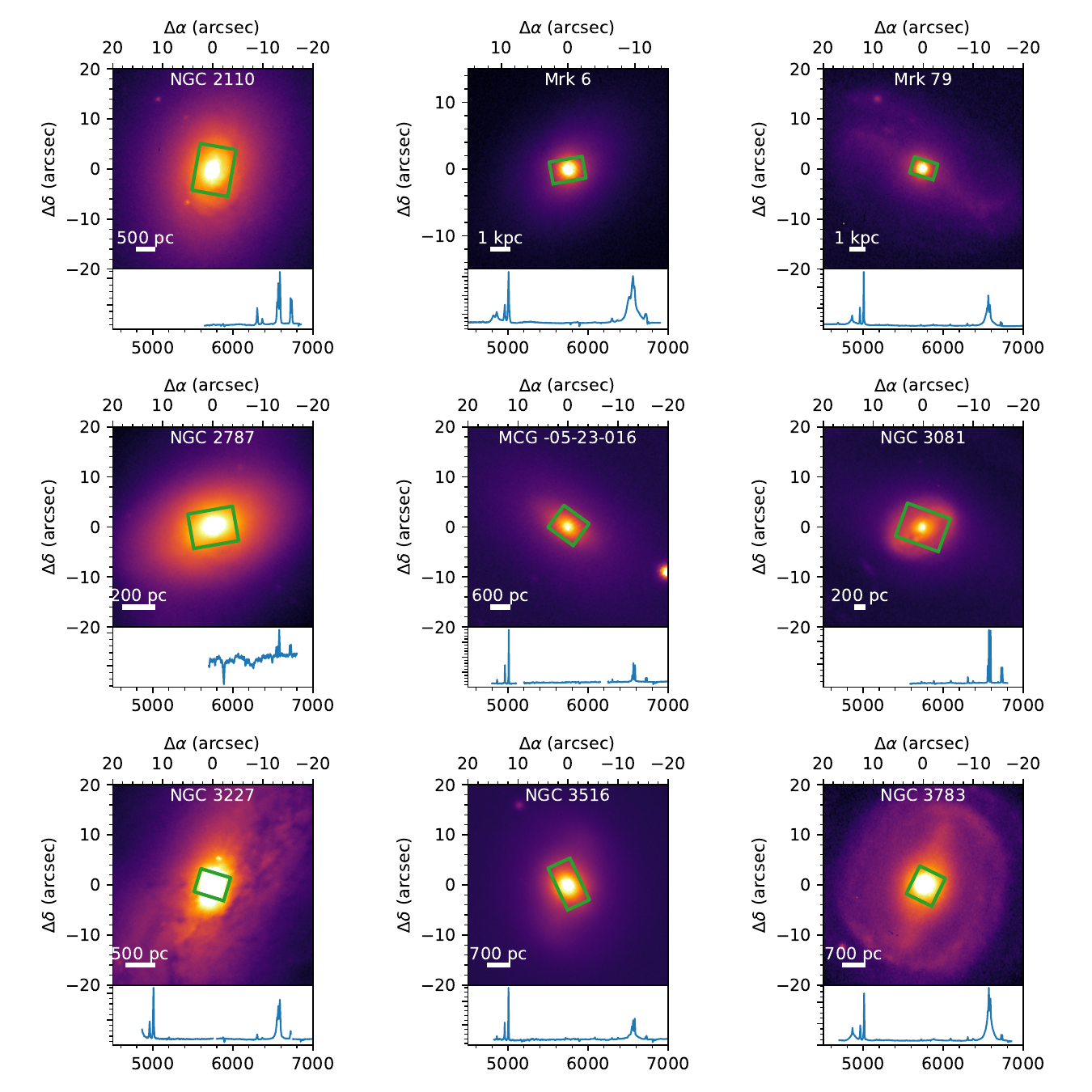}
  \caption{Continuation of \autoref{fig:acqspec1}.}
  \label{fig:acqspec2}
\end{figure*}
\newpage
\clearpage

\begin{figure*}
  \centering
  \includegraphics[page=3, width=\textwidth]{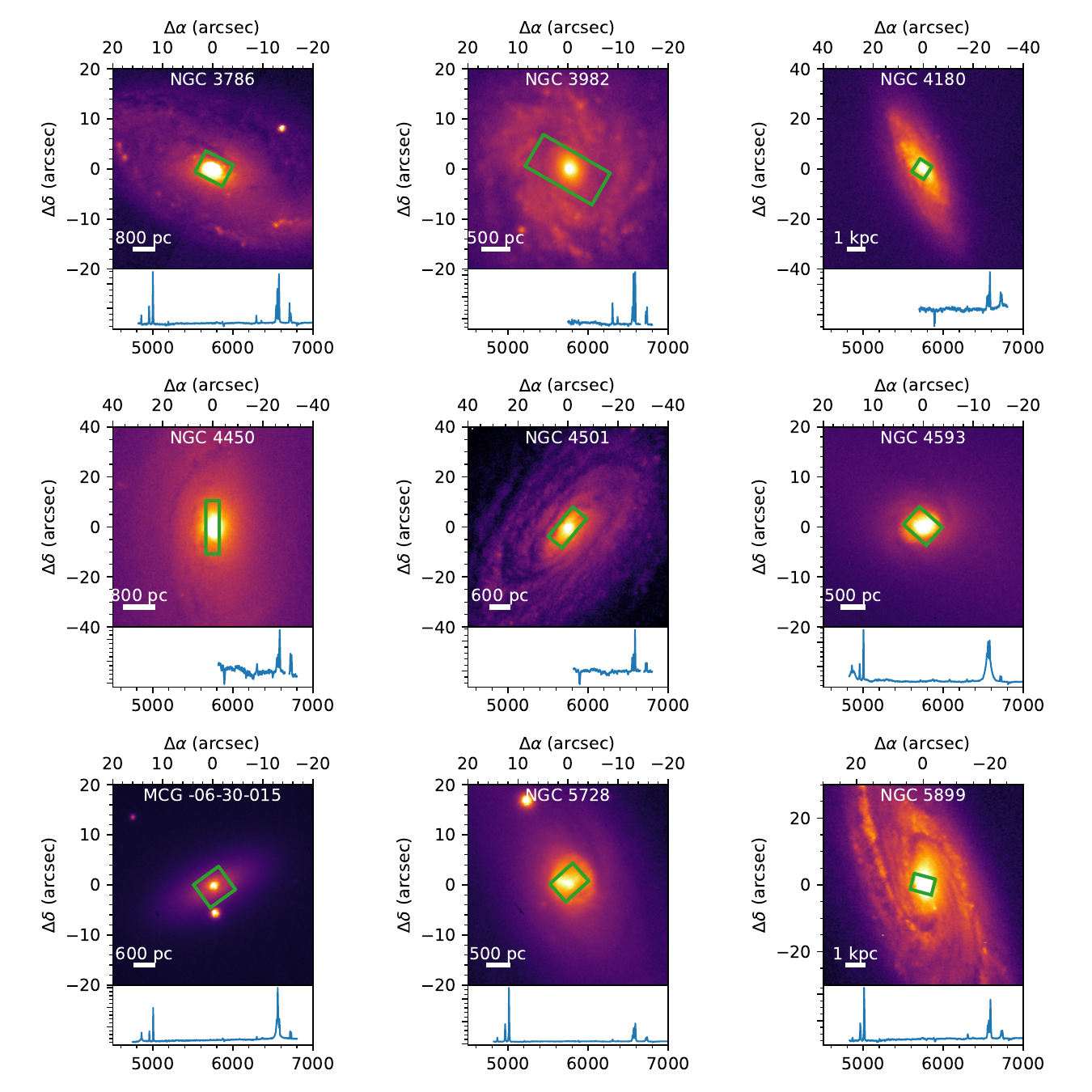}
  \caption{Continuation of \autoref{fig:acqspec1}.}
  \label{fig:acqspec3}
\end{figure*}
\newpage
\clearpage

\begin{figure*}
  \centering
  \includegraphics[page=4, width=\textwidth]{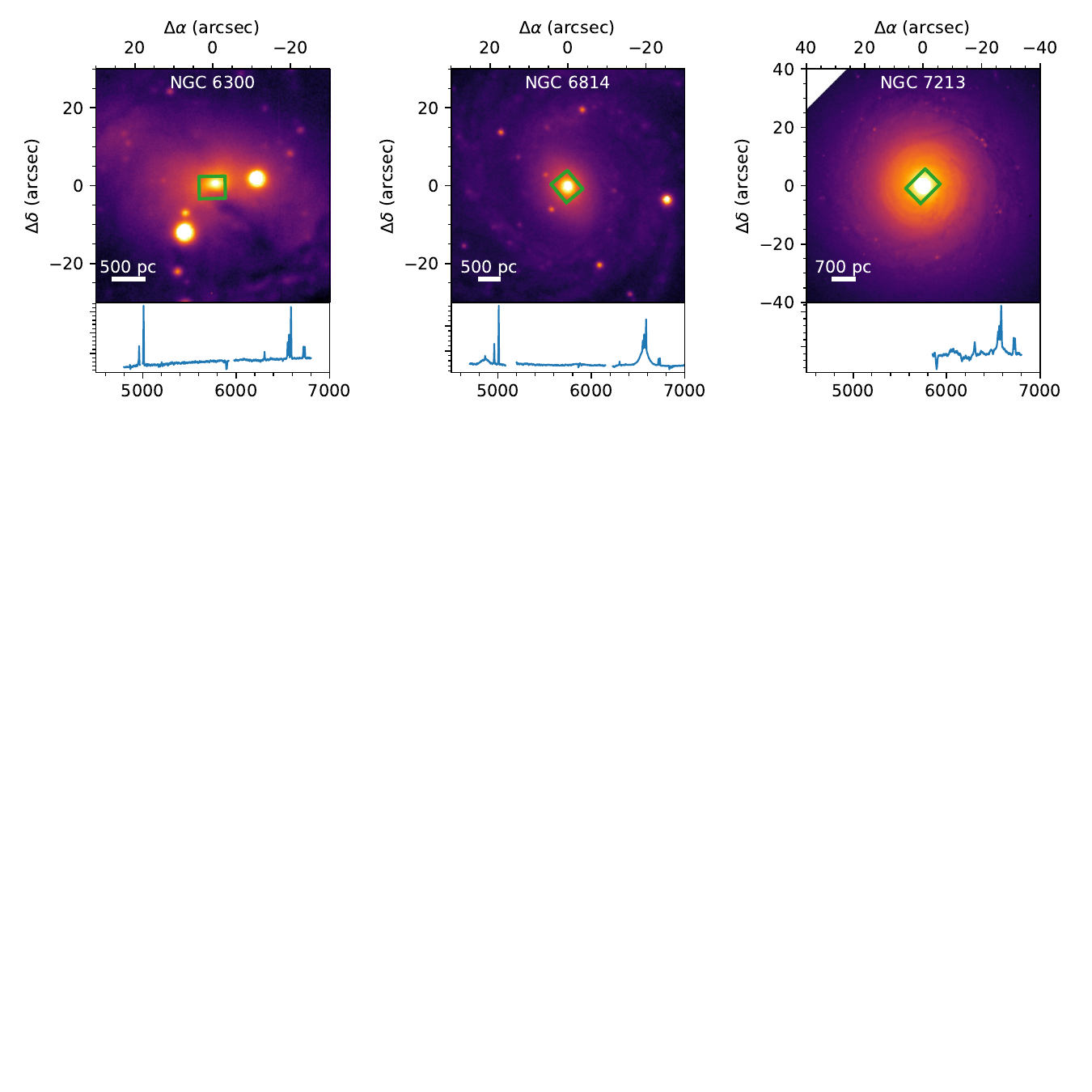}
  \caption{Continuation of \autoref{fig:acqspec1}.}
  \label{fig:acqspec4}
\end{figure*}

%% file: individual_results.tex
\begin{figure*}
    \centering
    \includegraphics[page=1,width=\textwidth]{figs/data_plots.pdf}
    \caption{
        Emission line fitting results for the galaxy Mrk~348.
        The first line of panels show
        the logarithm of the flux, in units of ${\rm 10^{-15}\,\,erg\,s^{-1}\,cm^{-2}}$,
        the radial velocity and the \woi index of the [\ion{O}{iii}]~5007{\AA} line;
        the middle line shows the same quantities for the [\ion{N}{ii}]~6583{\AA} line;
        and finally the line ratios [\ion{n}{ii}]~6583{\AA} / H$\alpha$ and  [\ion{o}{iii}]~5007{\AA} / H$\beta$,
        and the logarithm of the electron density, in units of ${\rm cm^{-3}}$ are shown in the bottom line.
        \emph{red dashed line}: photometric major axis;
        \emph{green dashed line}: ionisation axis;
        \emph{purple dashed line}: direction of the \woi enhanced region;
        \emph{blue solid line}: kinematic major axis based on \nii emission, except for the \oiii velocity field panel;
        \emph{blue dashed circle}: maximum radius of the outflow, based on the \woi threshold;
    }
    \label{fig:data_mrk348}
\end{figure*}

\begin{figure*}
    \centering
    \includegraphics[page=2,width=\textwidth]{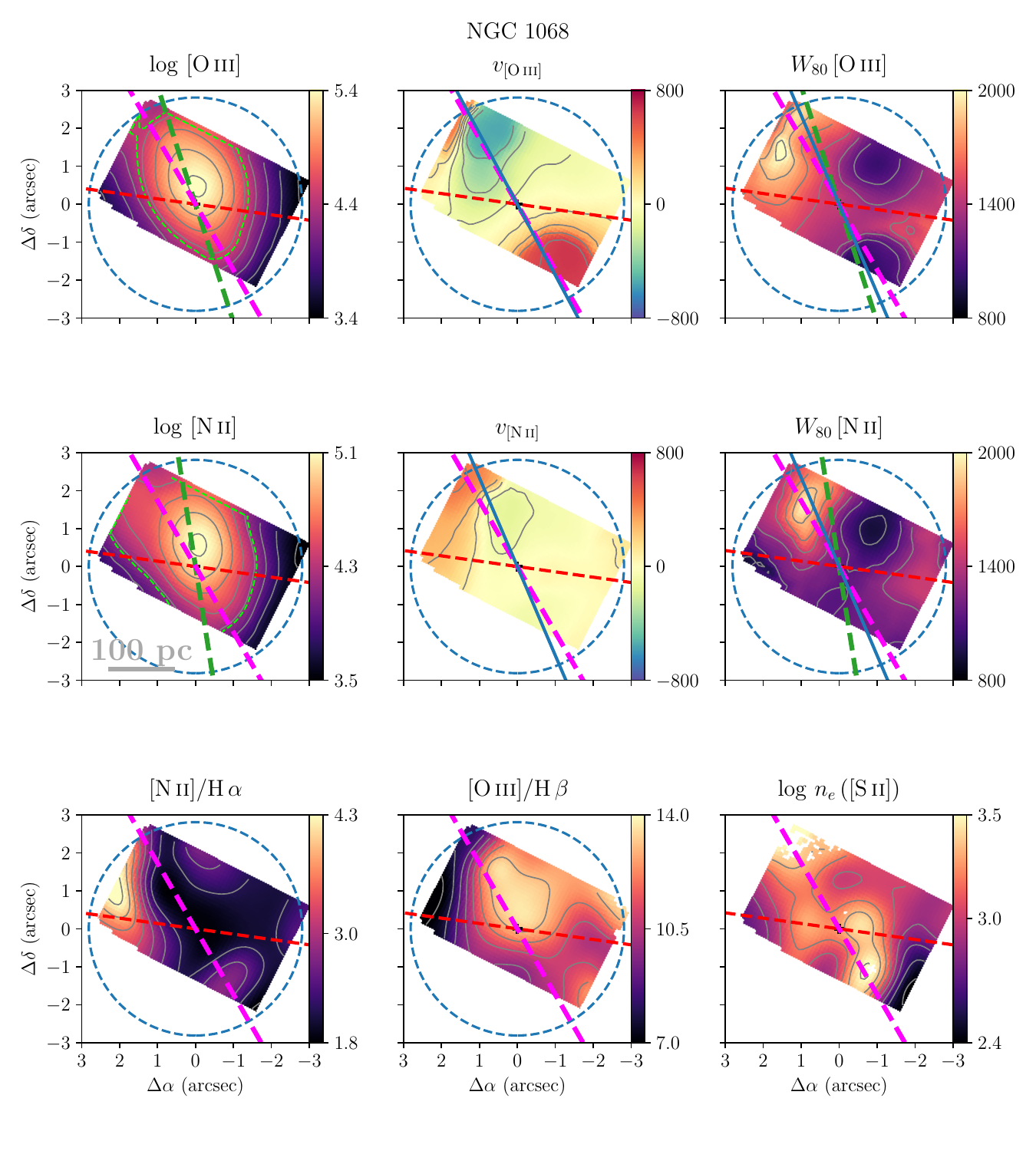}
    \caption{Continuation of figure \ref{fig:data_mrk348}.}
    \label{fig:data_ngc1068}
\end{figure*}

\begin{figure*}
    \centering
    \includegraphics[page=3,width=\textwidth]{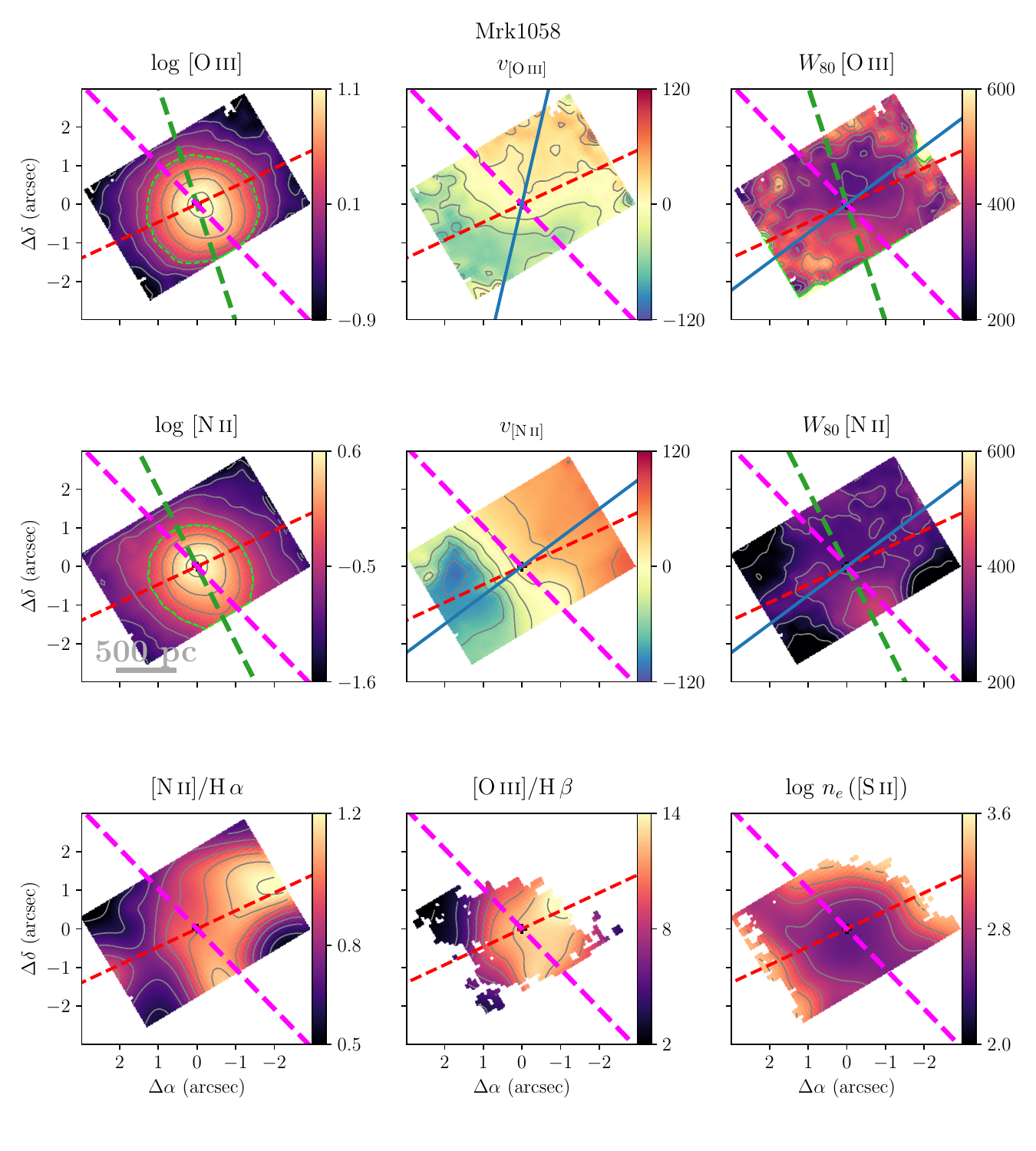}
    \caption{Continuation of figure \ref{fig:data_mrk348}.}
    \label{fig:data_mrk1058}
\end{figure*}

\begin{figure*}
    \centering
    \includegraphics[page=4,width=\textwidth]{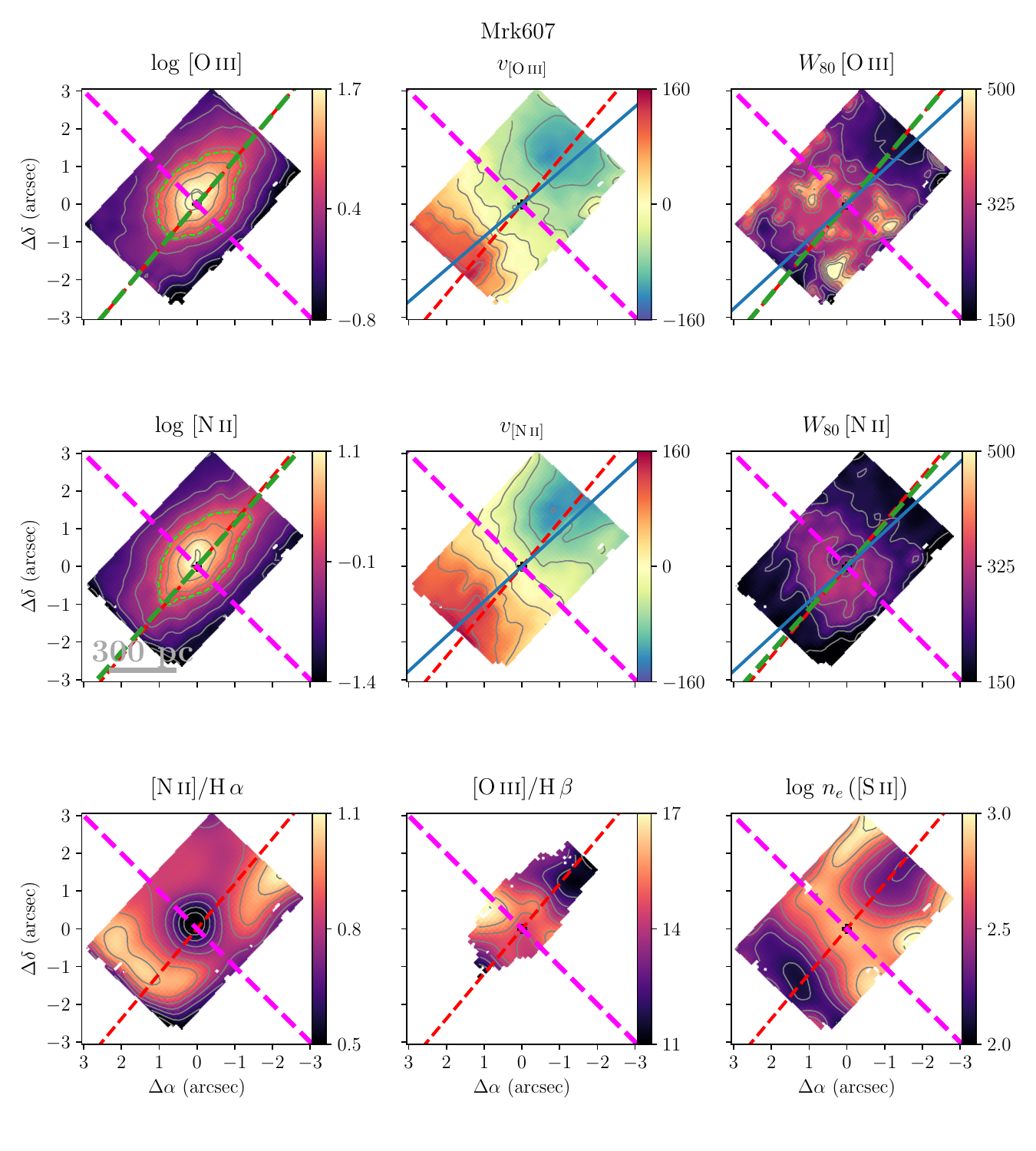}
    \caption{Continuation of figure \ref{fig:data_mrk348}.}
    \label{fig:data_mrk607}
\end{figure*}

\begin{figure*}
    \centering
    \includegraphics[page=5,width=\textwidth]{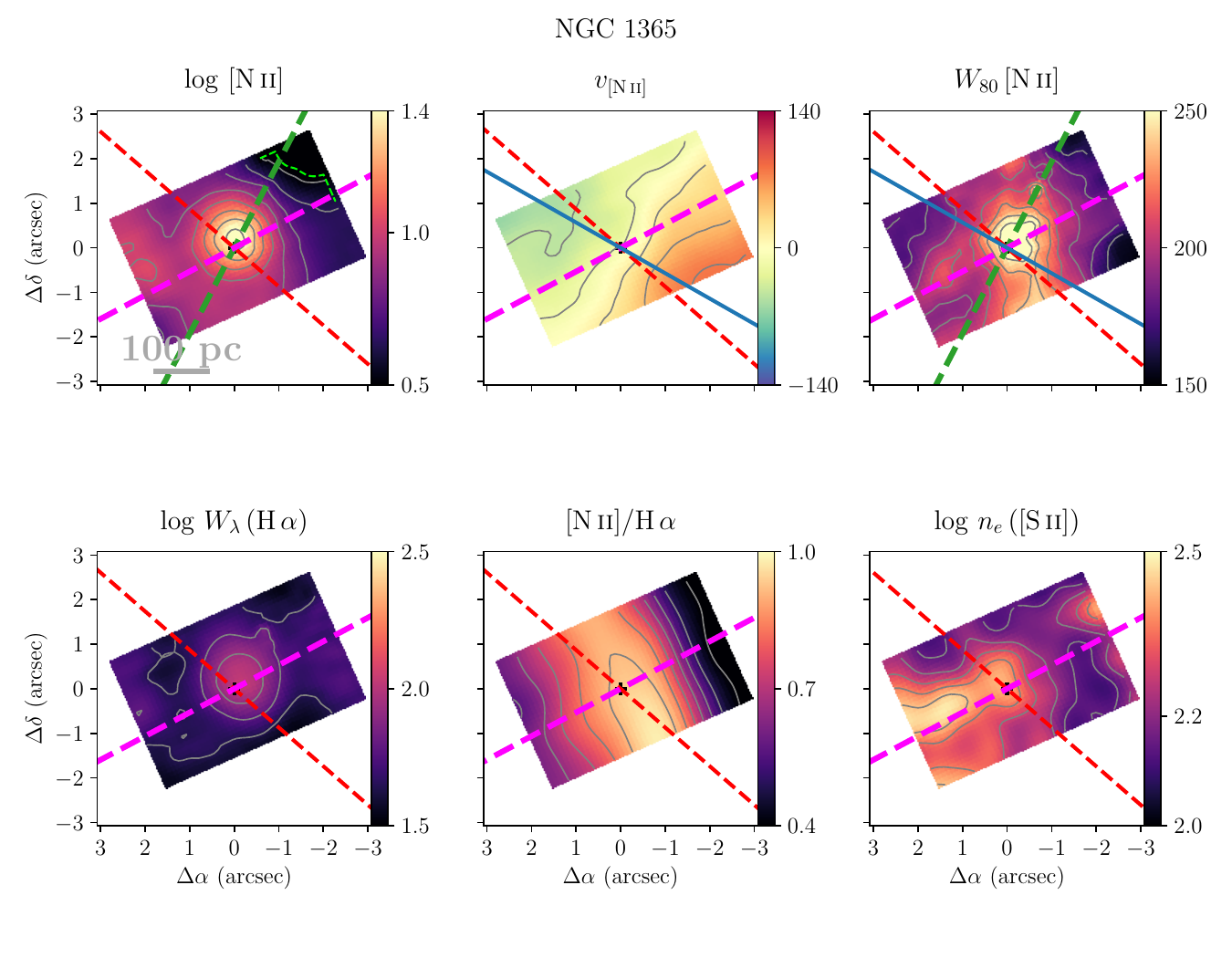}
    \caption{Continuation of figure \ref{fig:data_mrk348}.}
    \label{fig:data_ngc1365}
\end{figure*}

\begin{figure*}
    \centering
    \includegraphics[page=6,width=\textwidth]{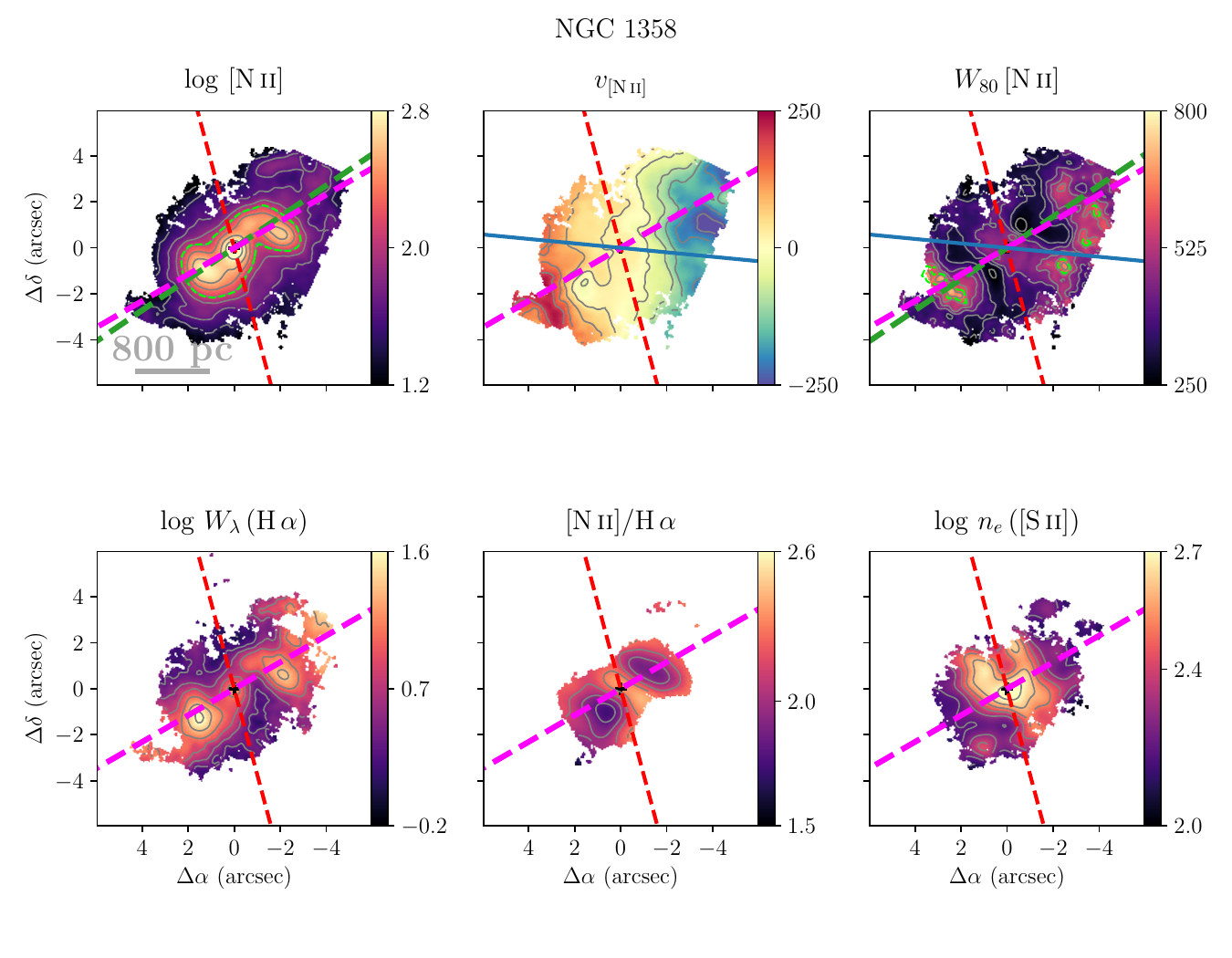}
    \caption{Continuation of figure \ref{fig:data_mrk348}.}
    \label{fig:data_ngc1358}
\end{figure*}

\begin{figure*}
    \centering
    \includegraphics[page=7,width=\textwidth]{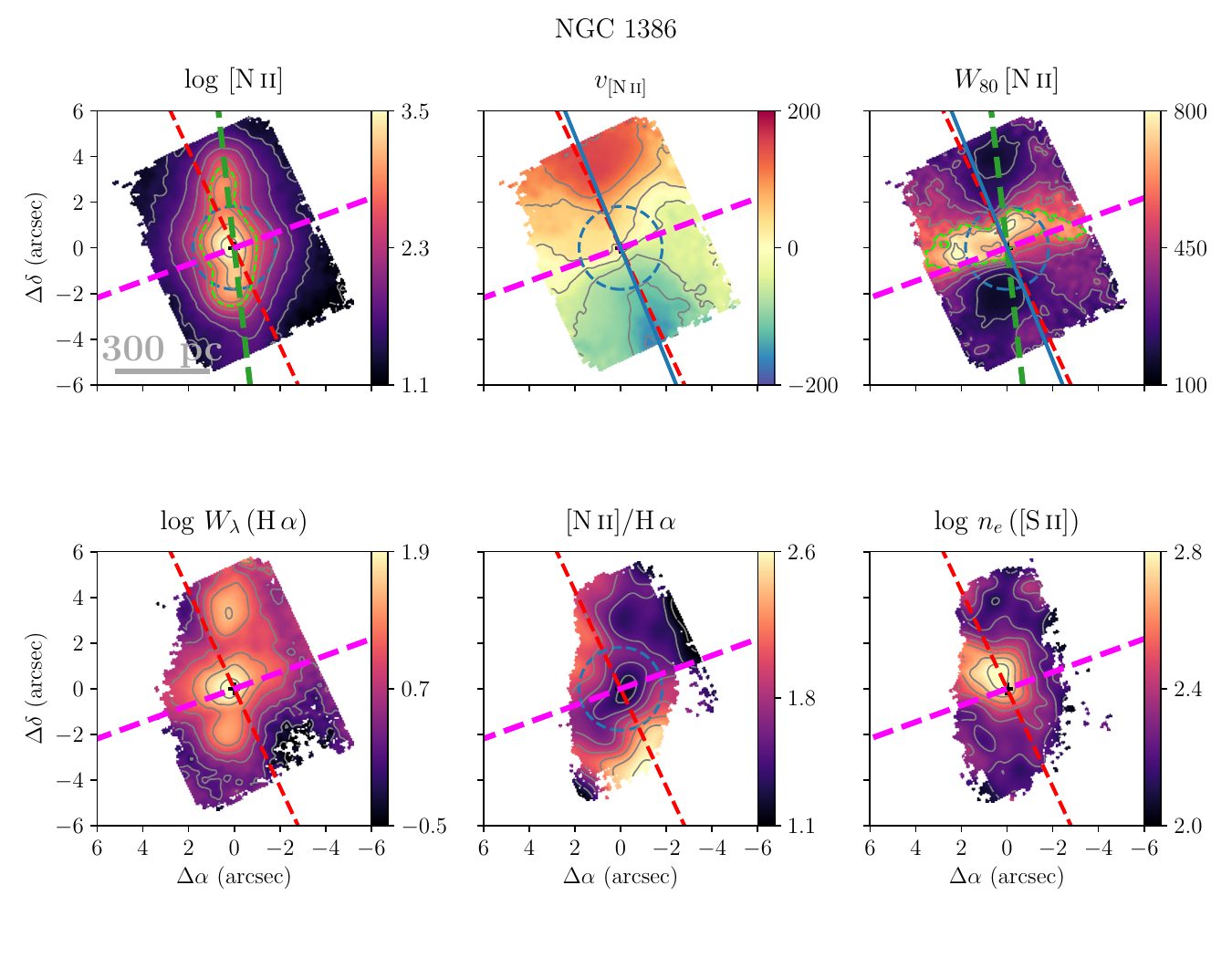}
    \caption{Continuation of figure \ref{fig:data_mrk348}.}
    \label{fig:data_ngc1386}
\end{figure*}

\begin{figure*}
    \centering
    \includegraphics[page=8,width=\textwidth]{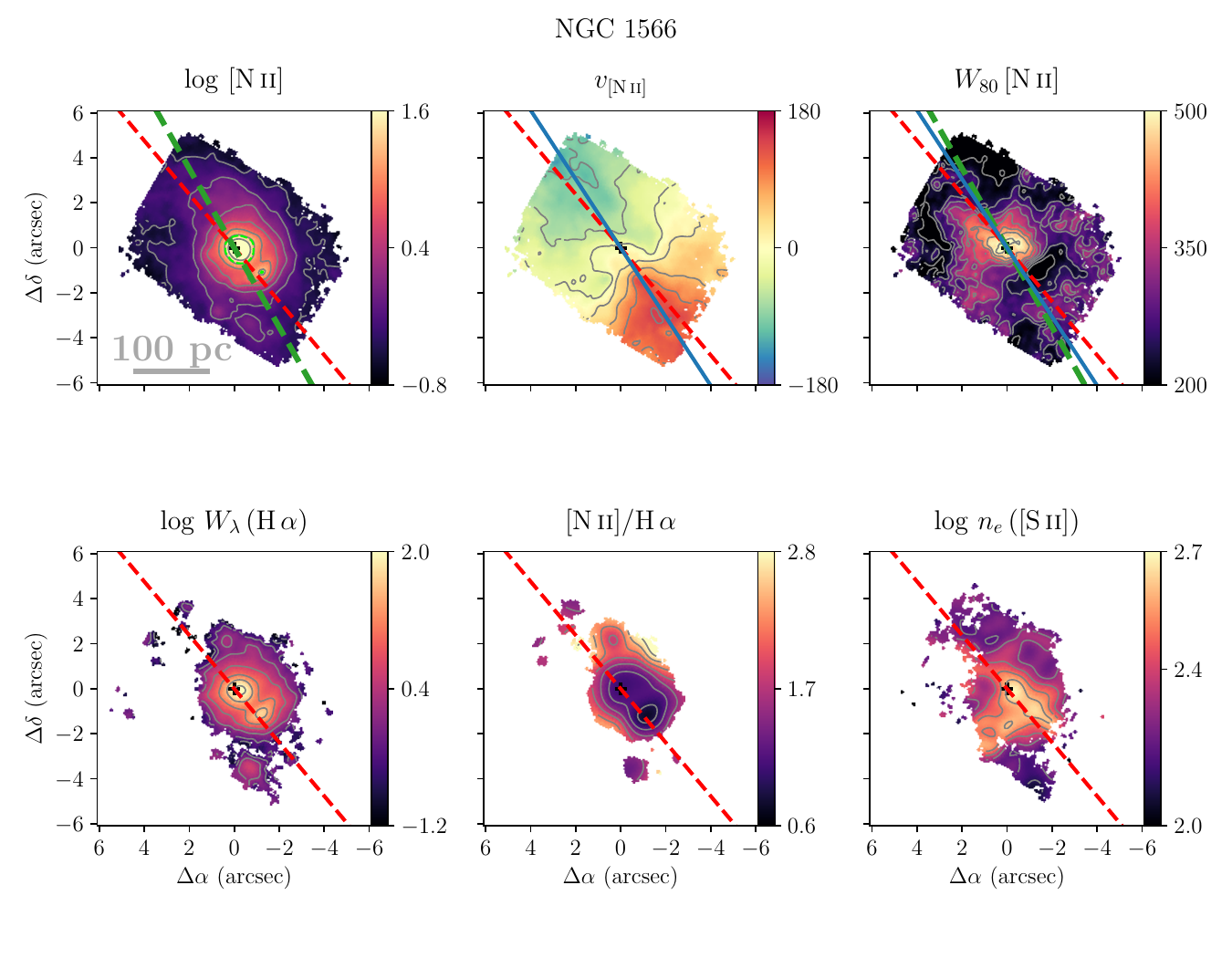}
    \caption{Continuation of figure \ref{fig:data_mrk348}.}
    \label{fig:data_ngc1566}
\end{figure*}

\begin{figure*}
    \centering
    \includegraphics[page=9,width=\textwidth]{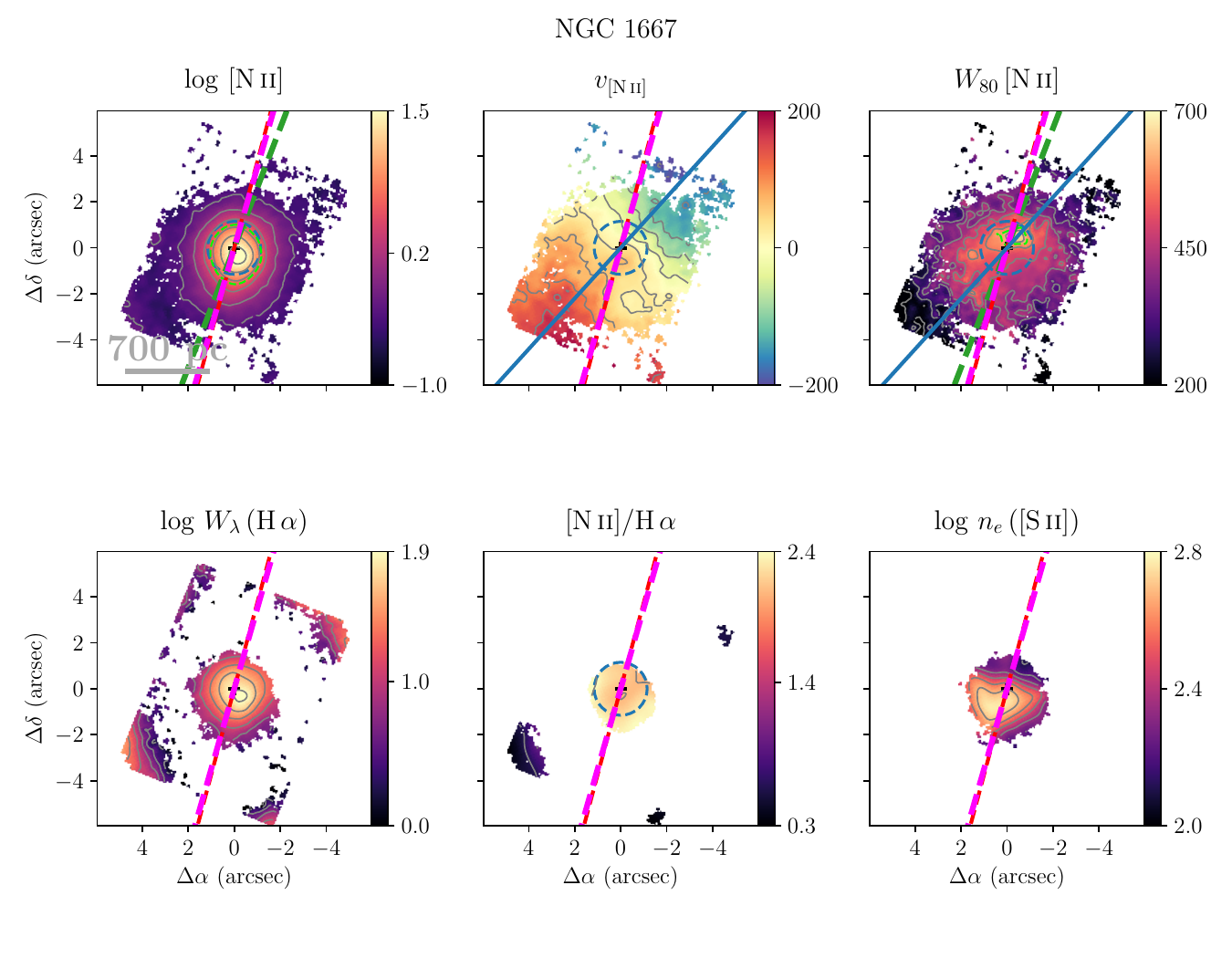}
    \caption{Continuation of figure \ref{fig:data_mrk348}.}
    \label{fig:data_ngc1667}
\end{figure*}

\begin{figure*}
    \centering
    \includegraphics[page=10,width=\textwidth]{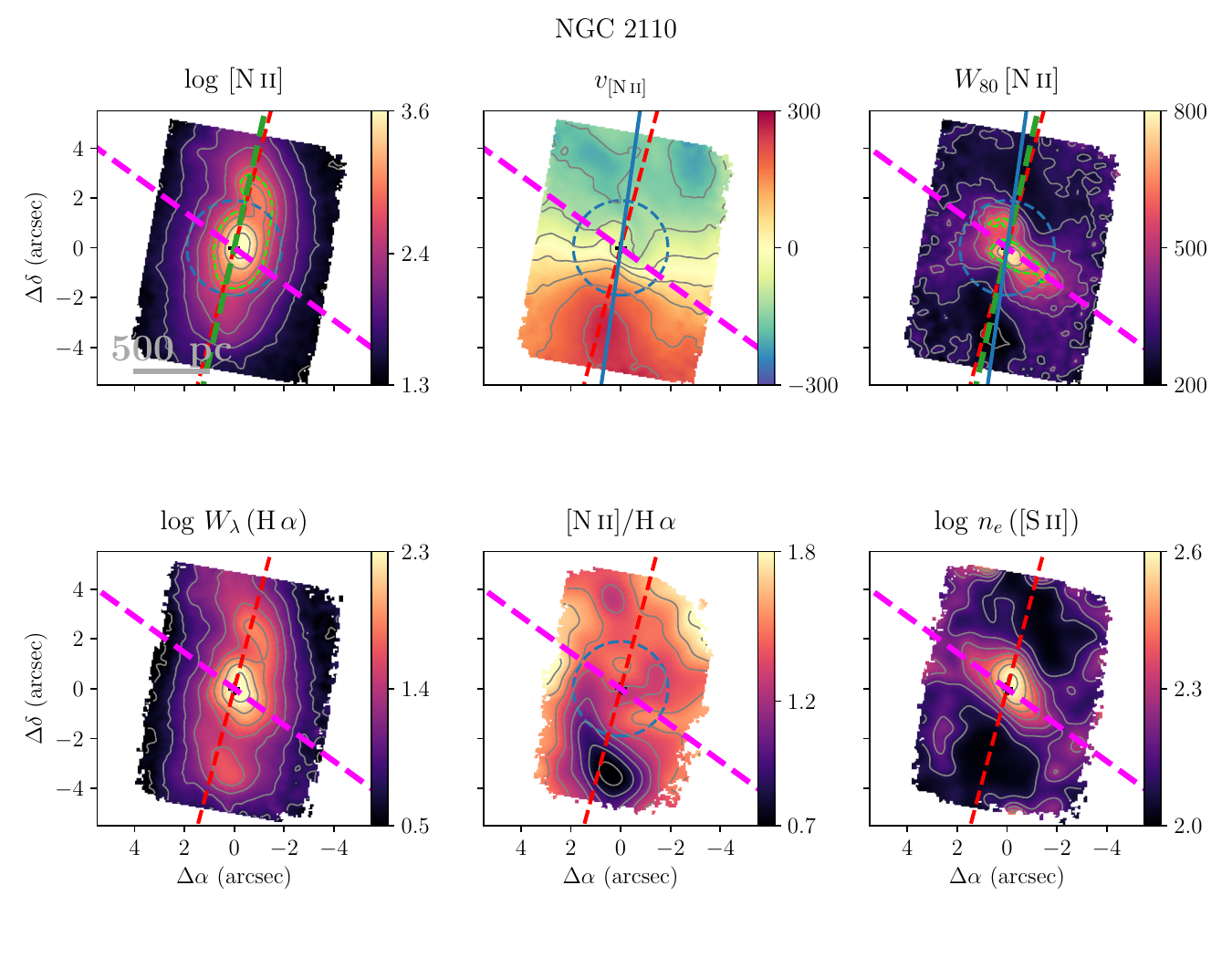}
    \caption{Continuation of figure \ref{fig:data_mrk348}.}
    \label{fig:data_ngc2110}
\end{figure*}

\begin{figure*}
    \centering
    \includegraphics[page=11,width=\textwidth]{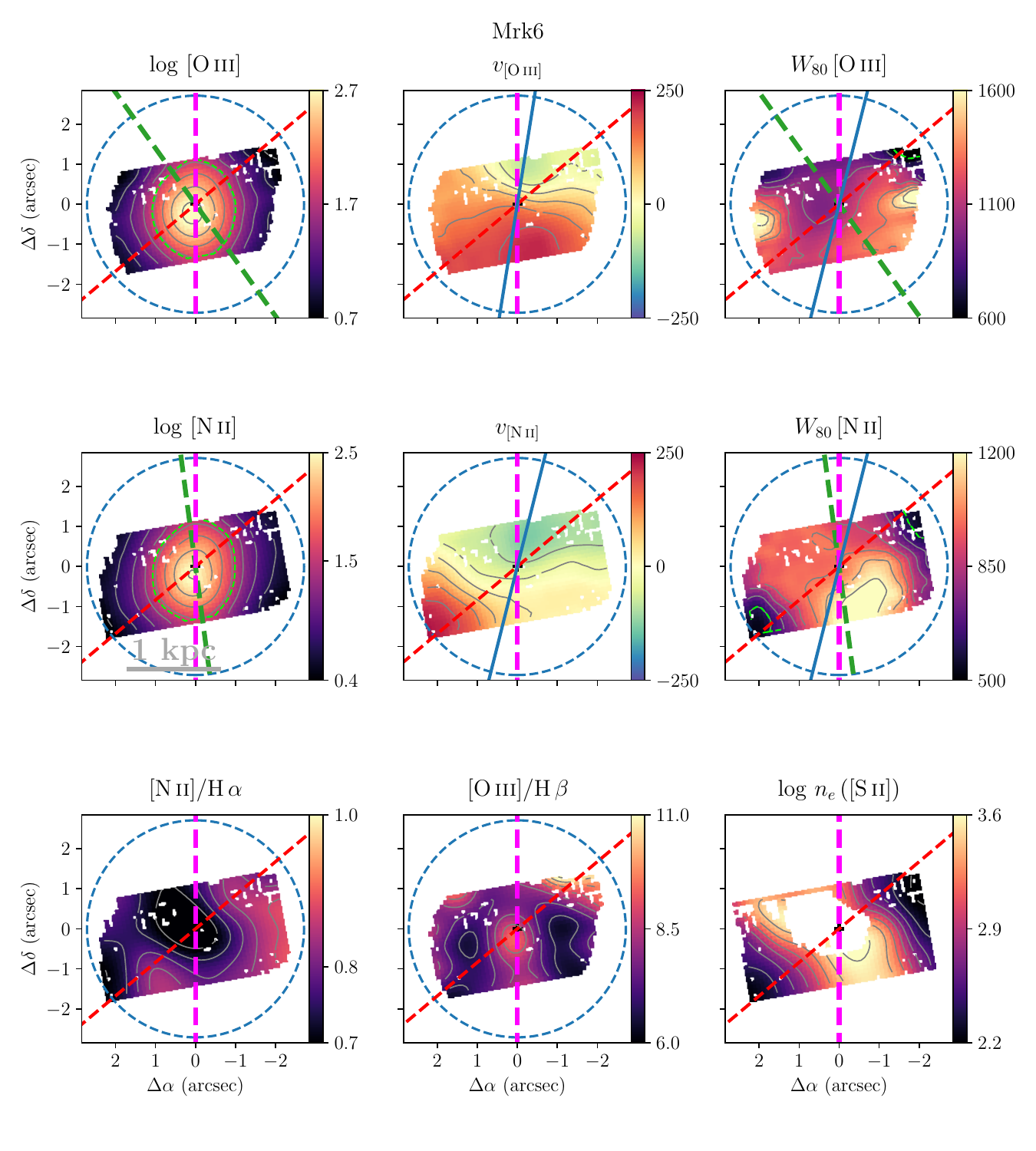}
    \caption{Continuation of figure \ref{fig:data_mrk348}.}
    \label{fig:data_mrk6}
\end{figure*}

\begin{figure*}
    \centering
    \includegraphics[page=12,width=\textwidth]{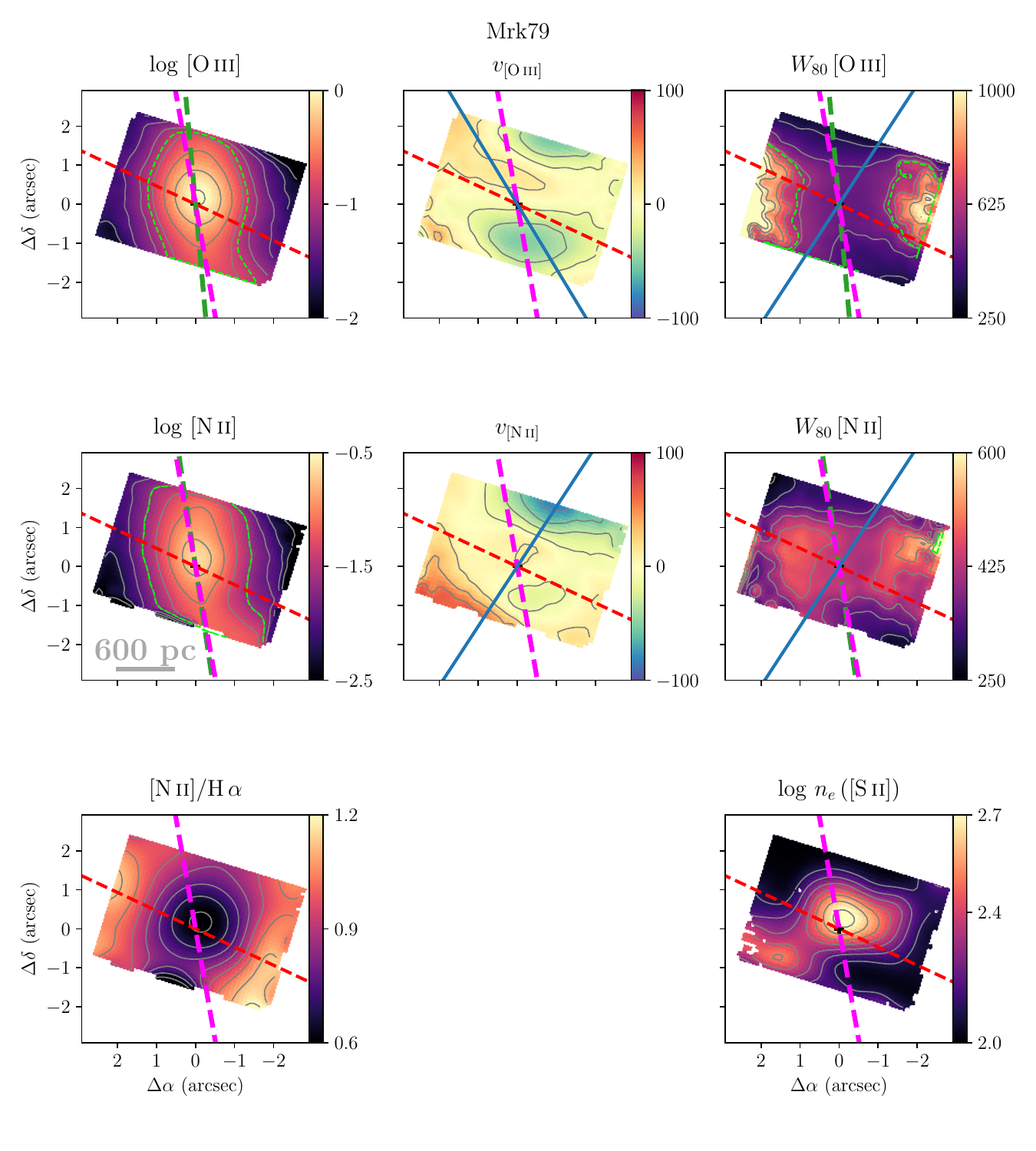}
    \caption{Continuation of figure \ref{fig:data_mrk348}.}
    \label{fig:data_mrk79}
\end{figure*}

\begin{figure*}
    \centering
    \includegraphics[page=13,width=\textwidth]{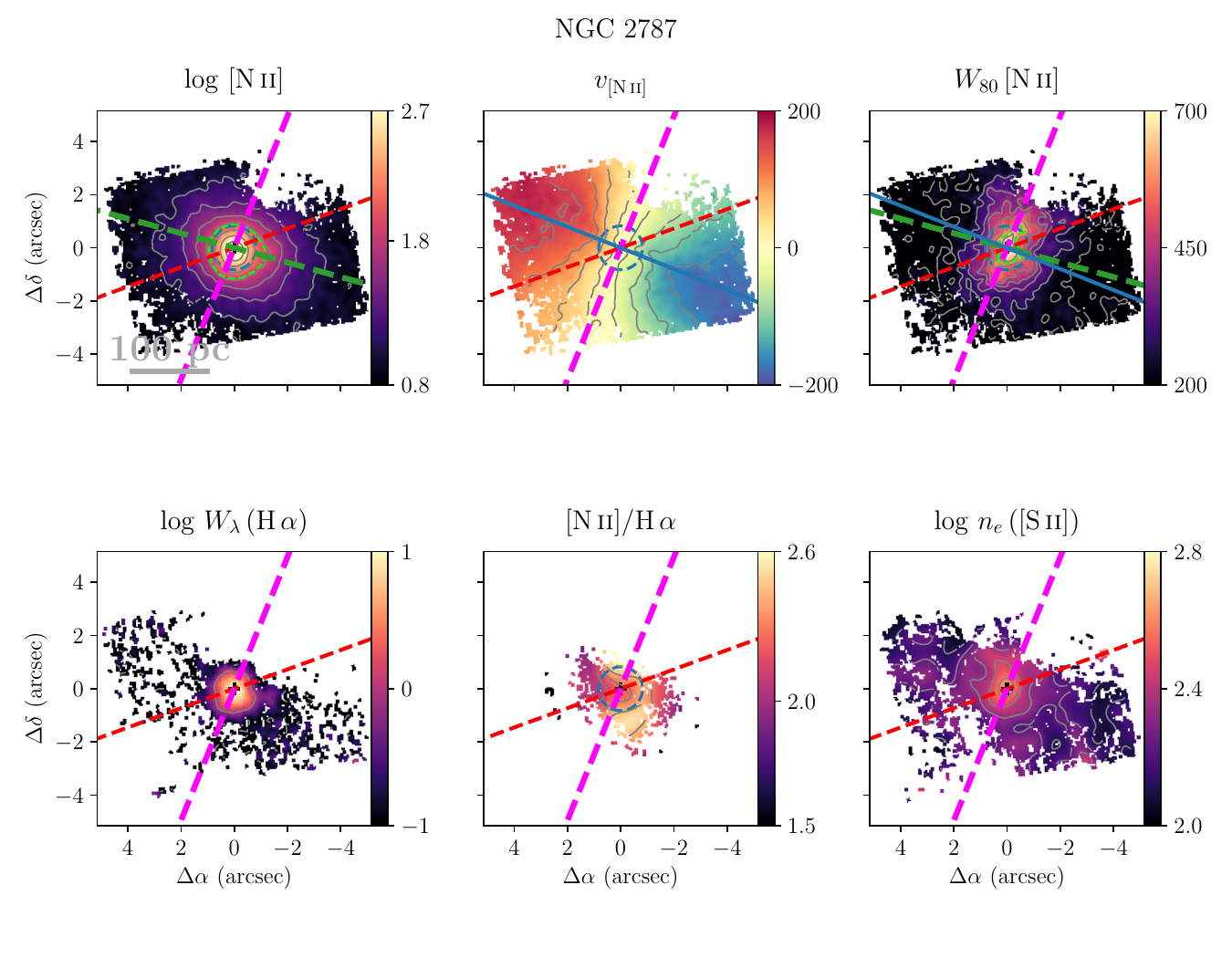}
    \caption{Continuation of figure \ref{fig:data_mrk348}.}
    \label{fig:data_ngc2787}
\end{figure*}

\begin{figure*}
    \centering
    \includegraphics[page=14,width=\textwidth]{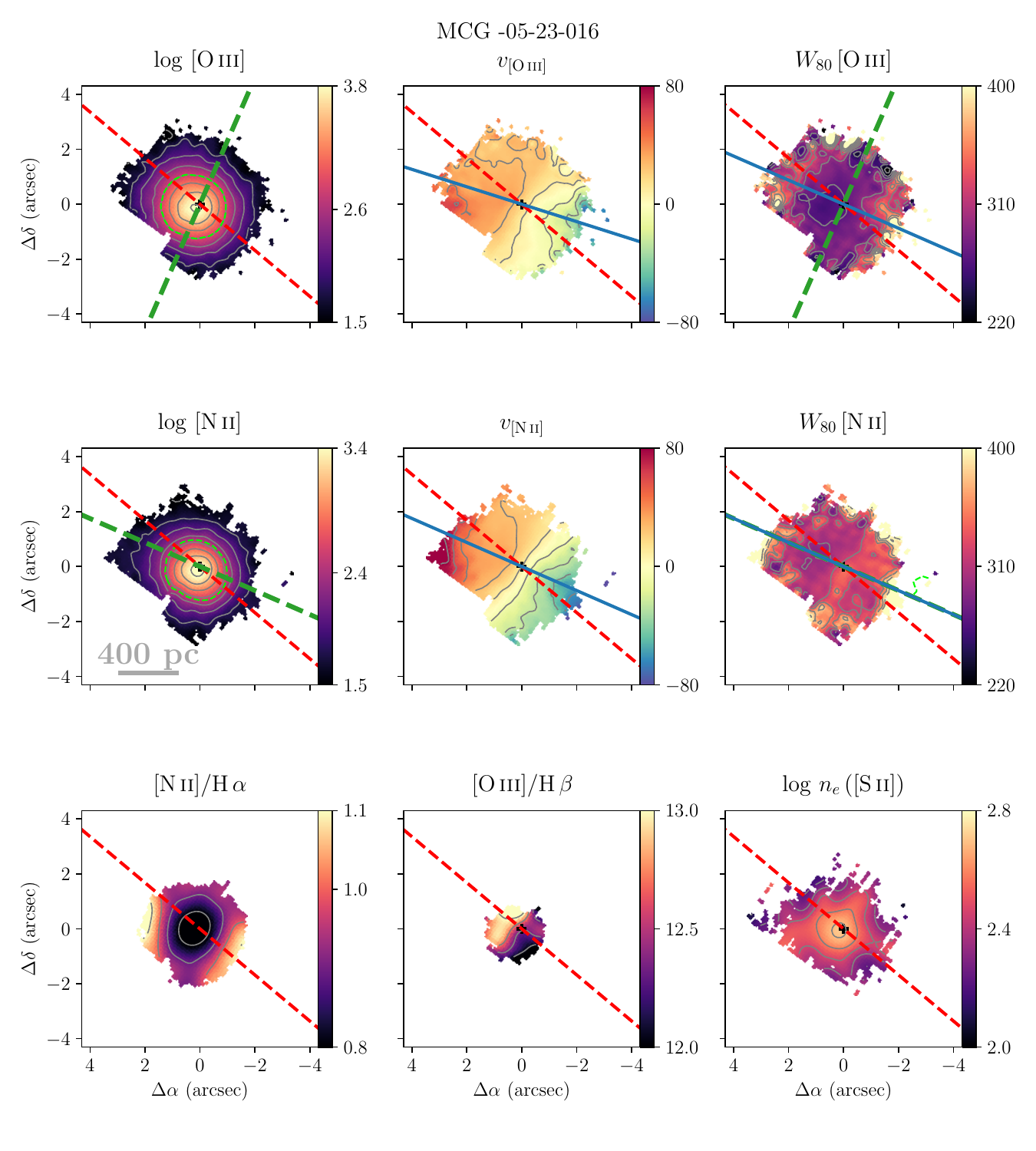}
    \caption{Continuation of figure \ref{fig:data_mrk348}.}
    \label{fig:data_mcg0523}
\end{figure*}

\begin{figure*}
    \centering
    \includegraphics[page=15,width=\textwidth]{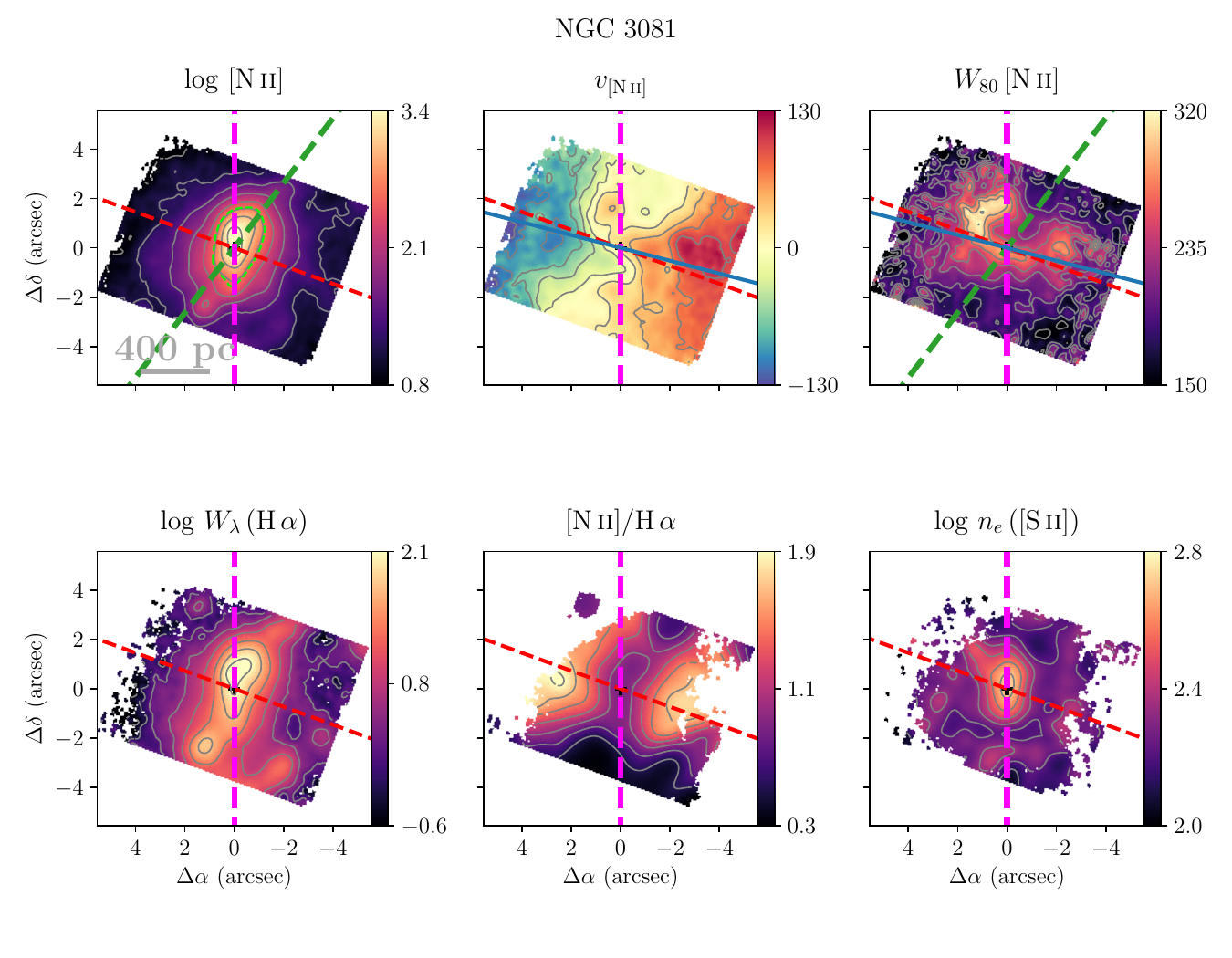}
    \caption{Continuation of figure \ref{fig:data_mrk348}.}
    \label{fig:data_ngc3081}
\end{figure*}

\begin{figure*}
    \centering
    \includegraphics[page=16,width=\textwidth]{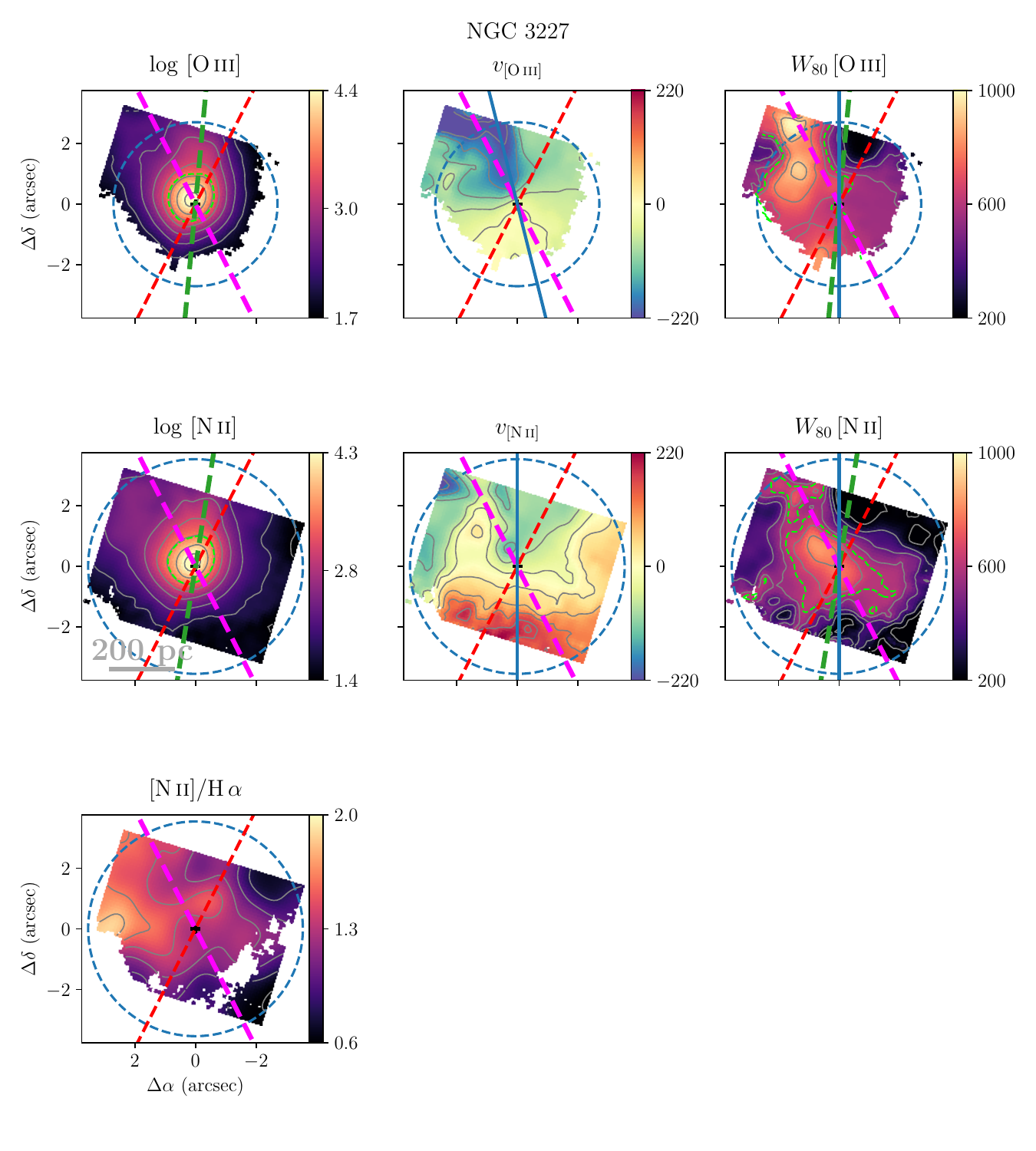}
    \caption{Continuation of figure \ref{fig:data_mrk348}.}
    \label{fig:data_ngc3227}
\end{figure*}

\begin{figure*}
    \centering
    \includegraphics[page=17,width=\textwidth]{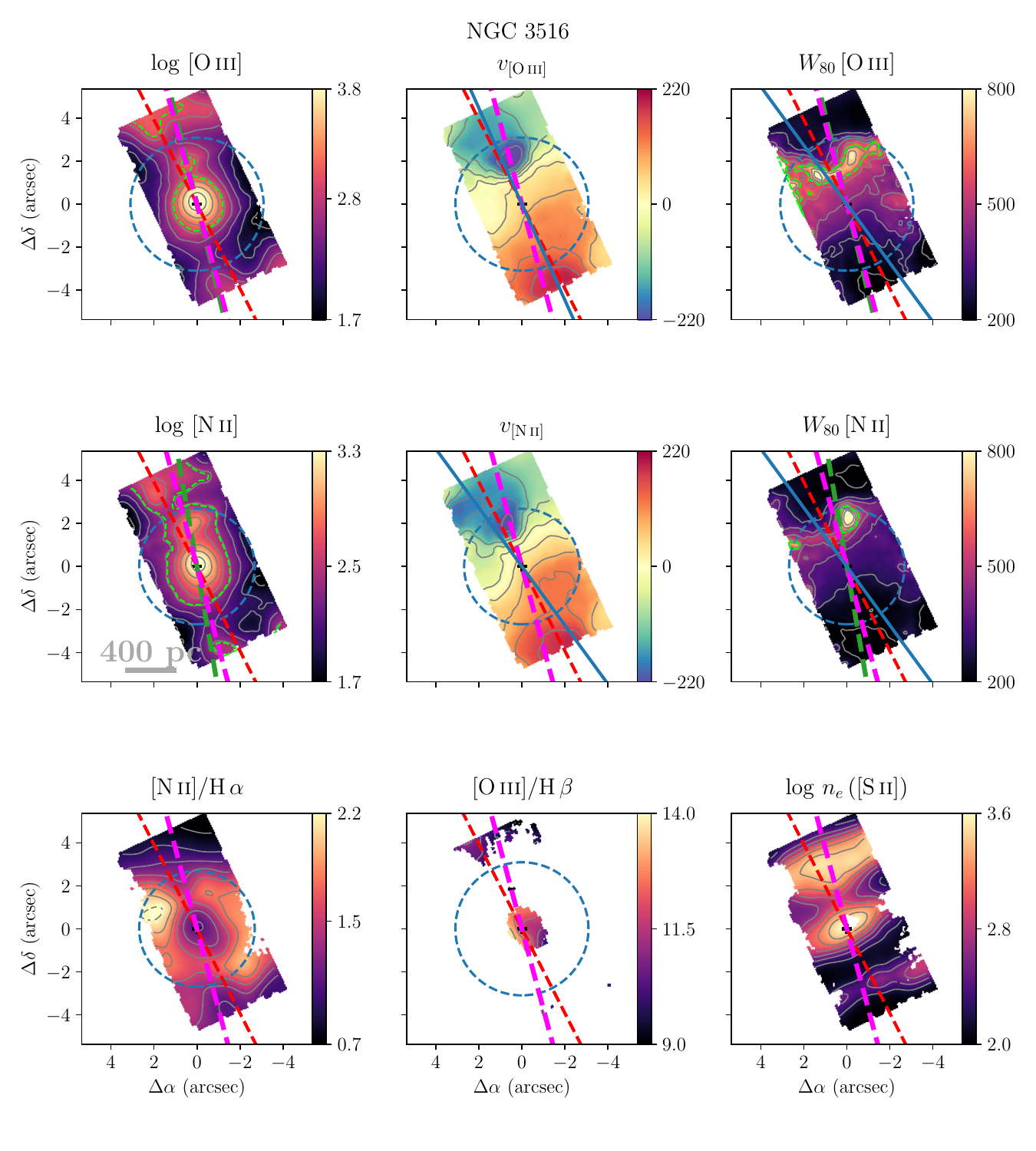}
    \caption{Continuation of figure \ref{fig:data_mrk348}.}
    \label{fig:data_ngc3516}
\end{figure*}

\begin{figure*}
    \centering
    \includegraphics[page=18,width=\textwidth]{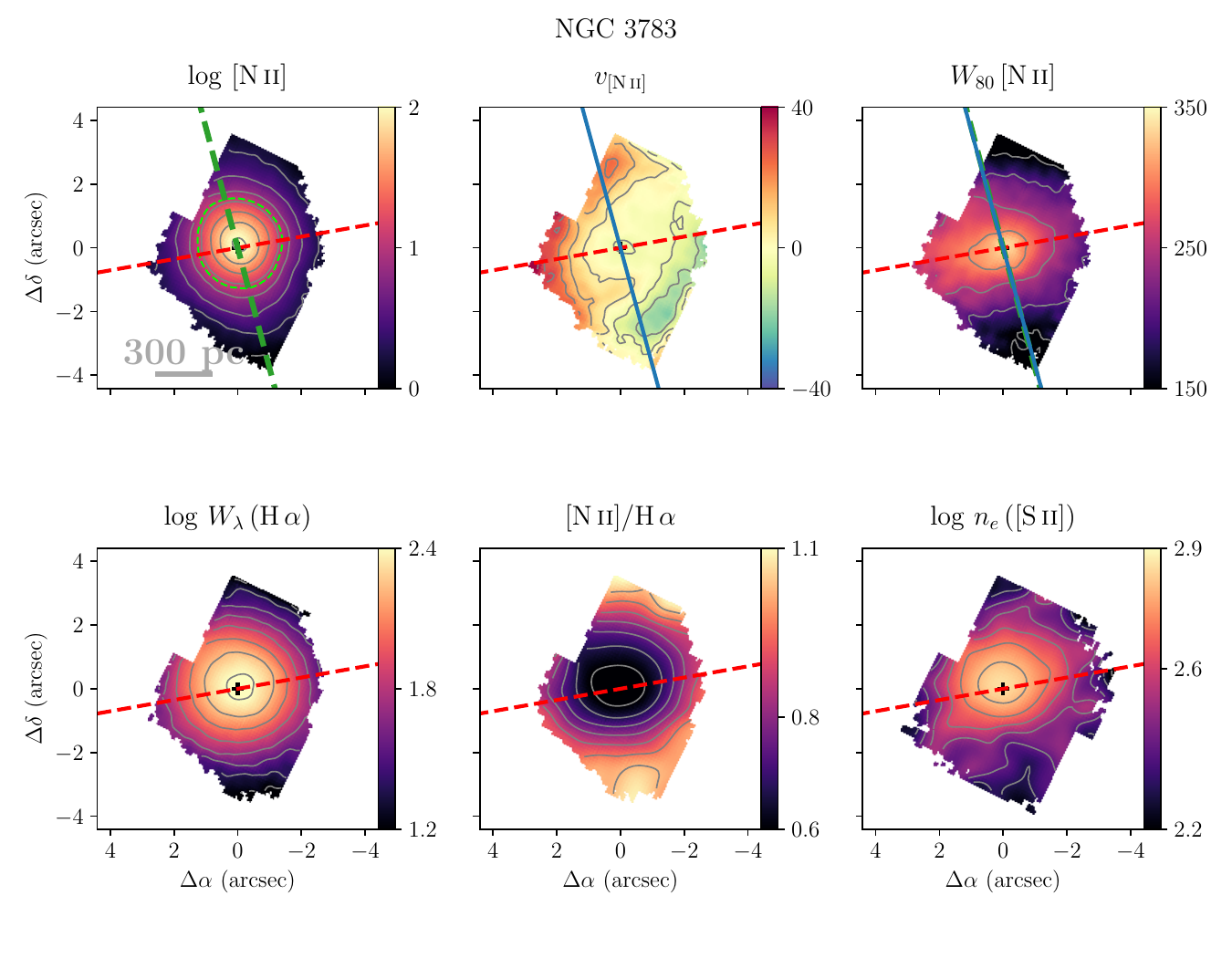}
    \caption{Continuation of figure \ref{fig:data_mrk348}.}
    \label{fig:data_ngc3783}
\end{figure*}

\begin{figure*}
    \centering
    \includegraphics[page=19,width=\textwidth]{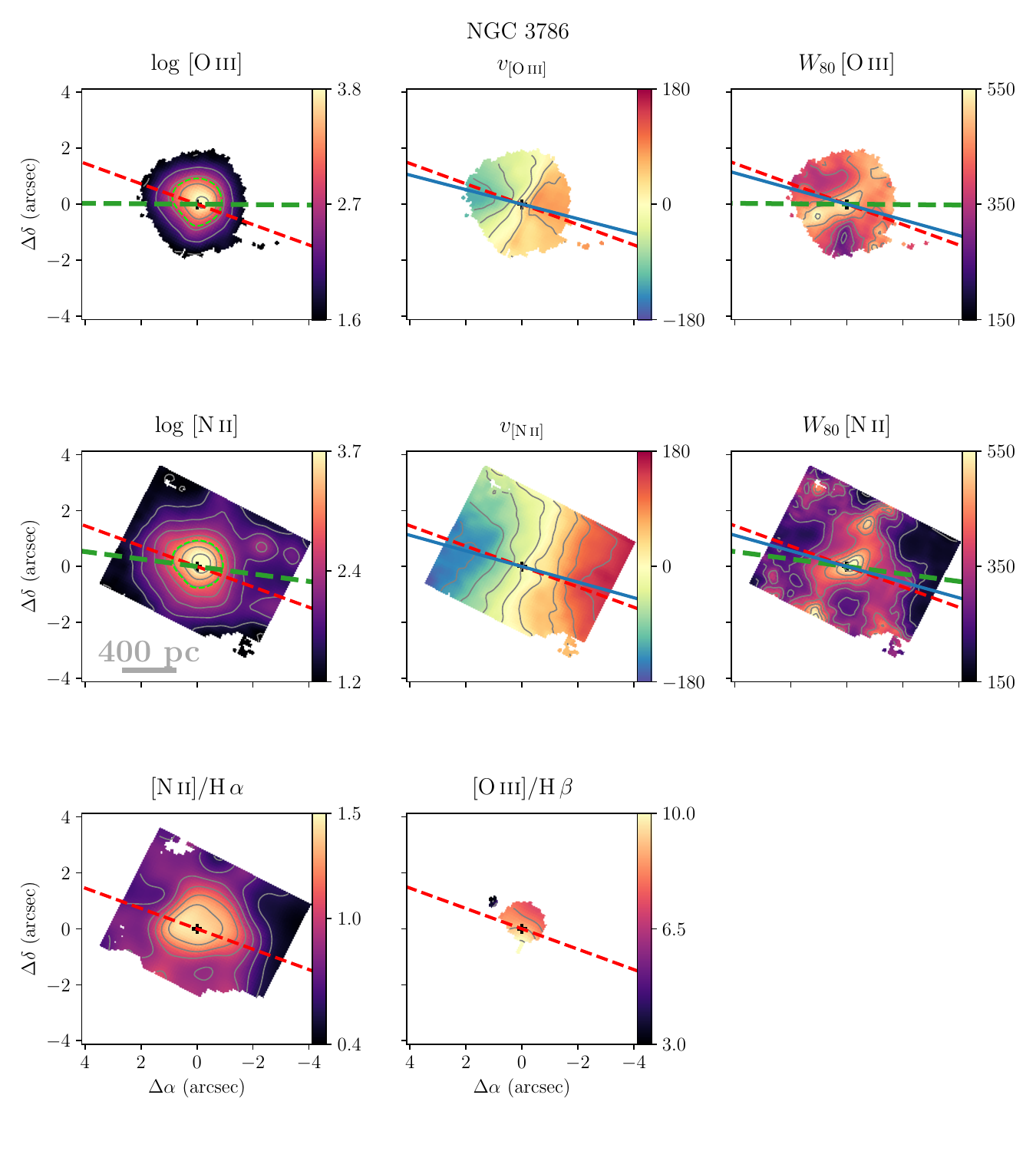}
    \caption{Continuation of figure \ref{fig:data_mrk348}.}
    \label{fig:data_ngc3786}
\end{figure*}

\begin{figure*}
    \centering
    \includegraphics[page=20,width=\textwidth]{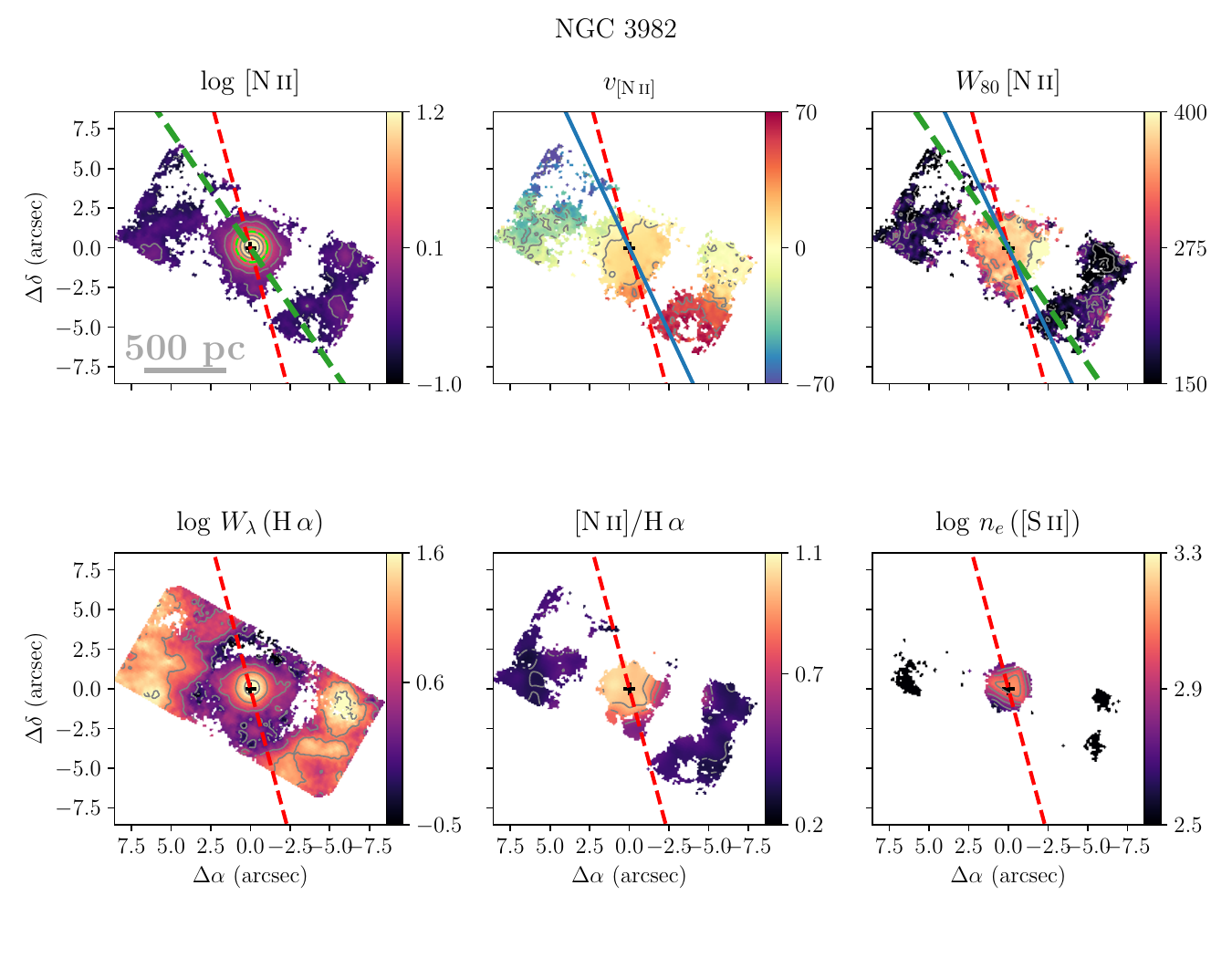}
    \caption{Continuation of figure \ref{fig:data_mrk348}.}
    \label{fig:data_ngc3982}
\end{figure*}

\begin{figure*}
    \centering
    \includegraphics[page=21,width=\textwidth]{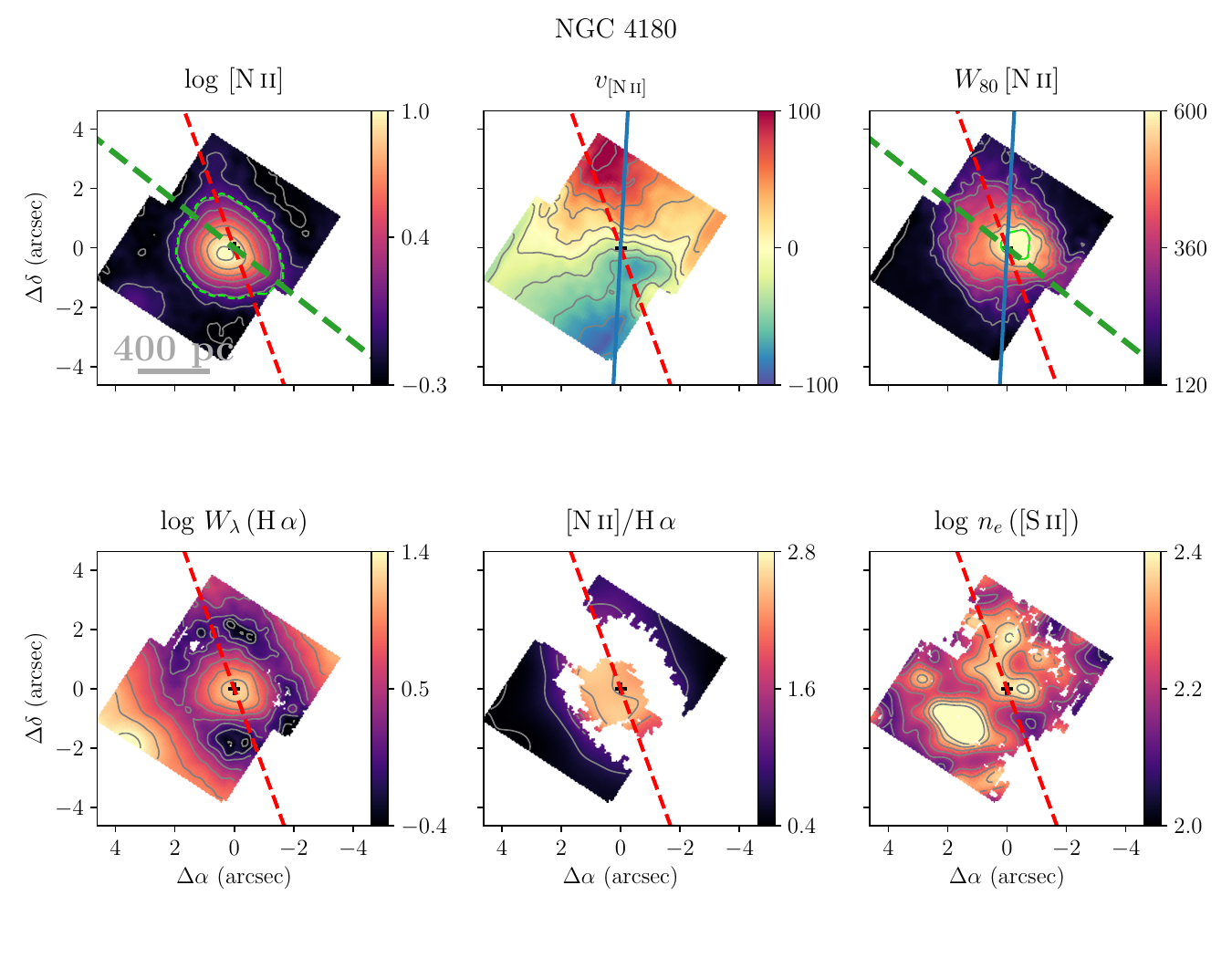}
    \caption{Continuation of figure \ref{fig:data_mrk348}.}
    \label{fig:data_ngc4180}
\end{figure*}

\begin{figure*}
    \centering
    \includegraphics[page=22,width=\textwidth]{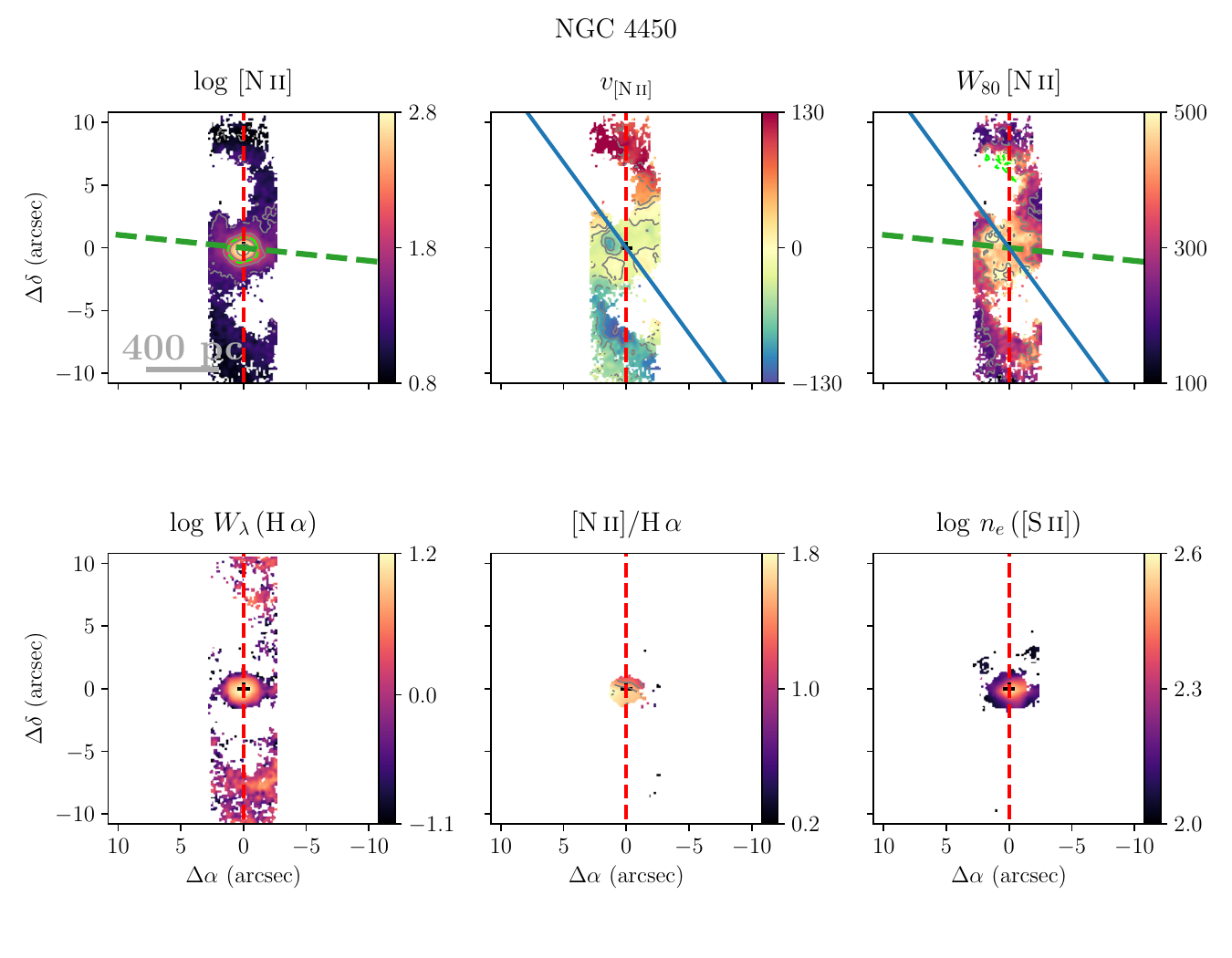}
    \caption{Continuation of figure \ref{fig:data_mrk348}.}
    \label{fig:data_ngc4450}
\end{figure*}

\begin{figure*}
    \centering
    \includegraphics[page=23,width=\textwidth]{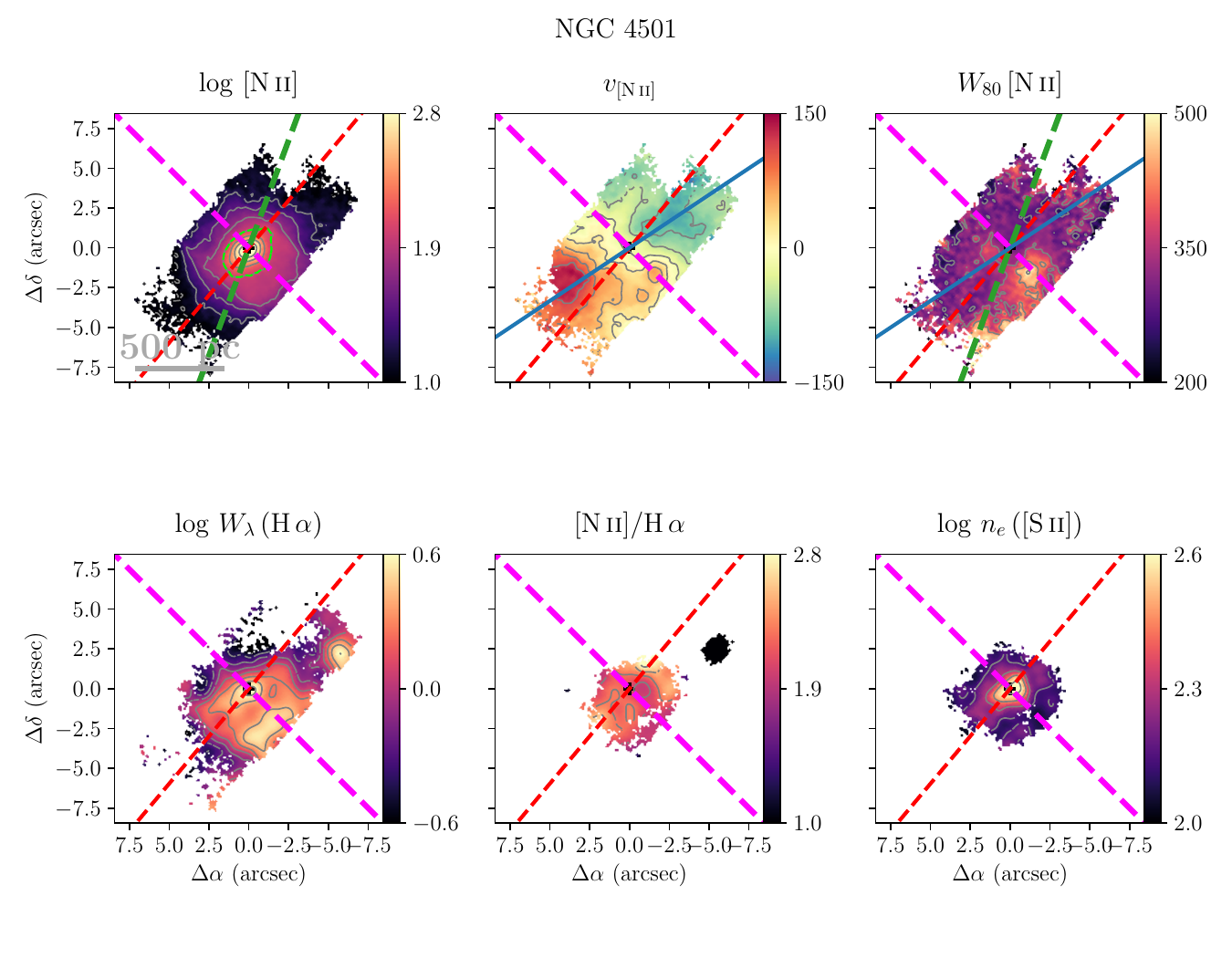}
    \caption{Continuation of figure \ref{fig:data_mrk348}.}
    \label{fig:data_ngc4501}
\end{figure*}

\begin{figure*}
    \centering
    \includegraphics[page=24,width=\textwidth]{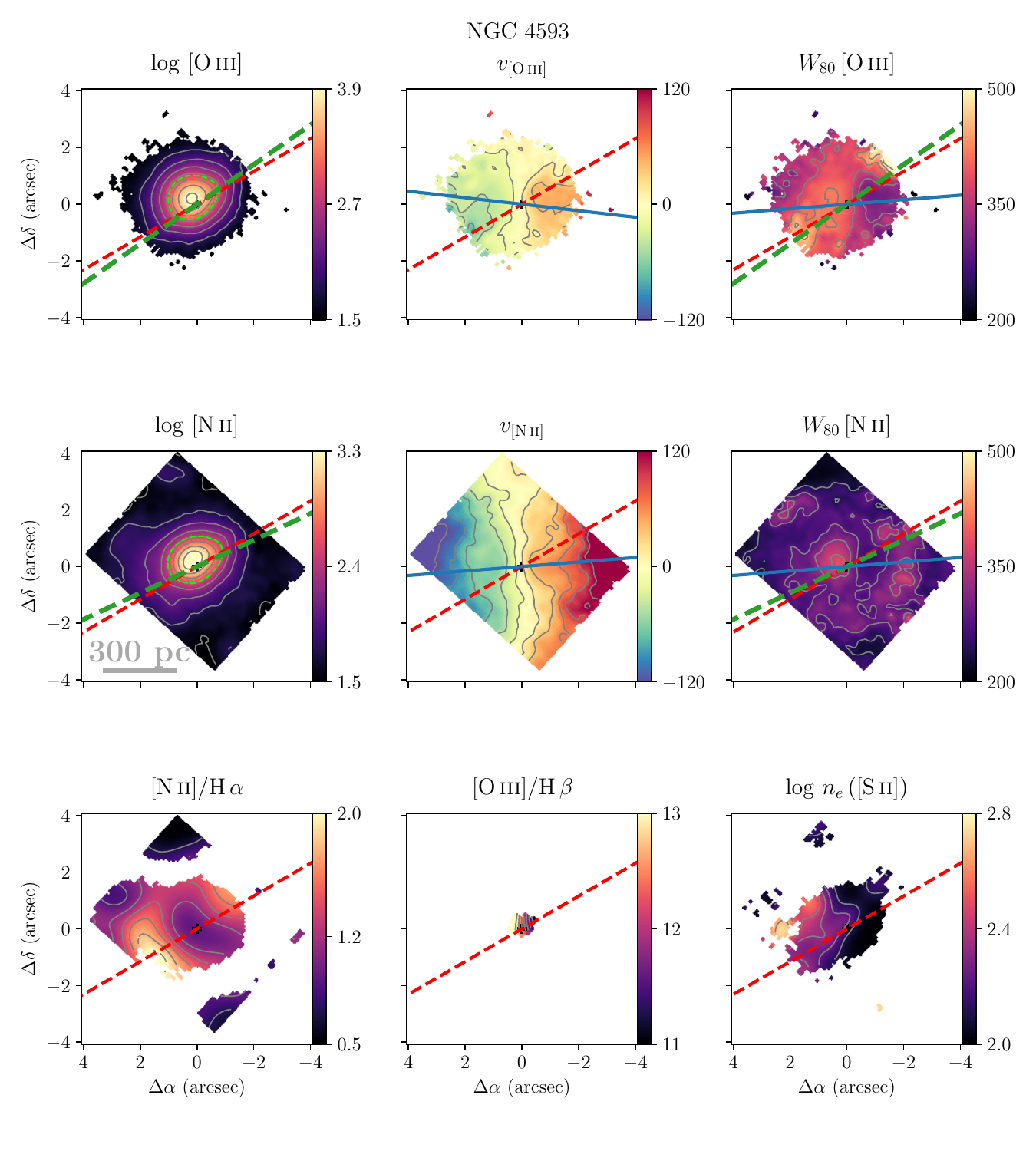}
    \caption{Continuation of figure \ref{fig:data_mrk348}.}
    \label{fig:data_ngc4593}
\end{figure*}

\begin{figure*}
    \centering
    \includegraphics[page=25,width=\textwidth]{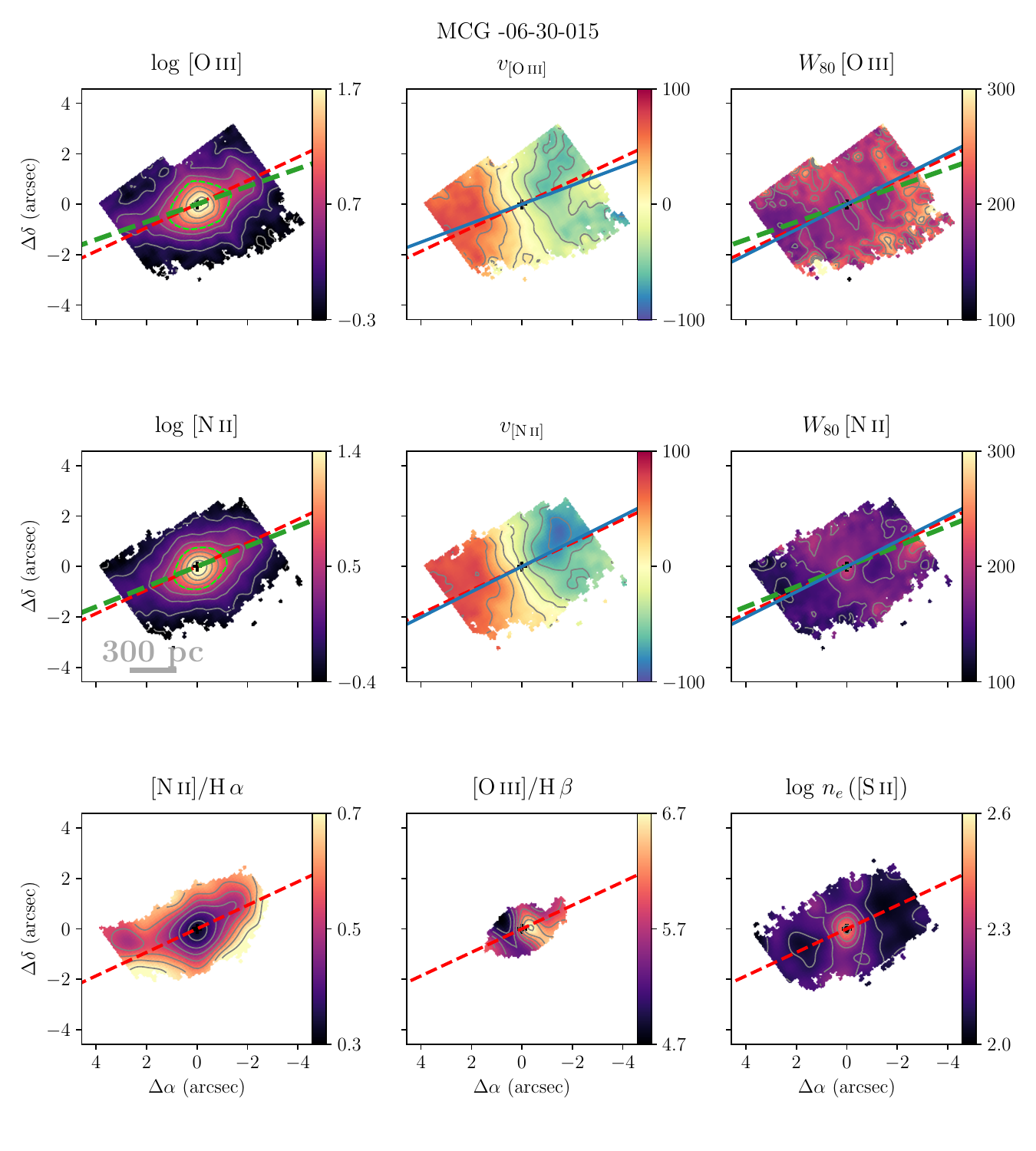}
    \caption{Continuation of figure \ref{fig:data_mrk348}.}
    \label{fig:data_mcg0630}
\end{figure*}

\begin{figure*}
    \centering
    \includegraphics[page=26,width=\textwidth]{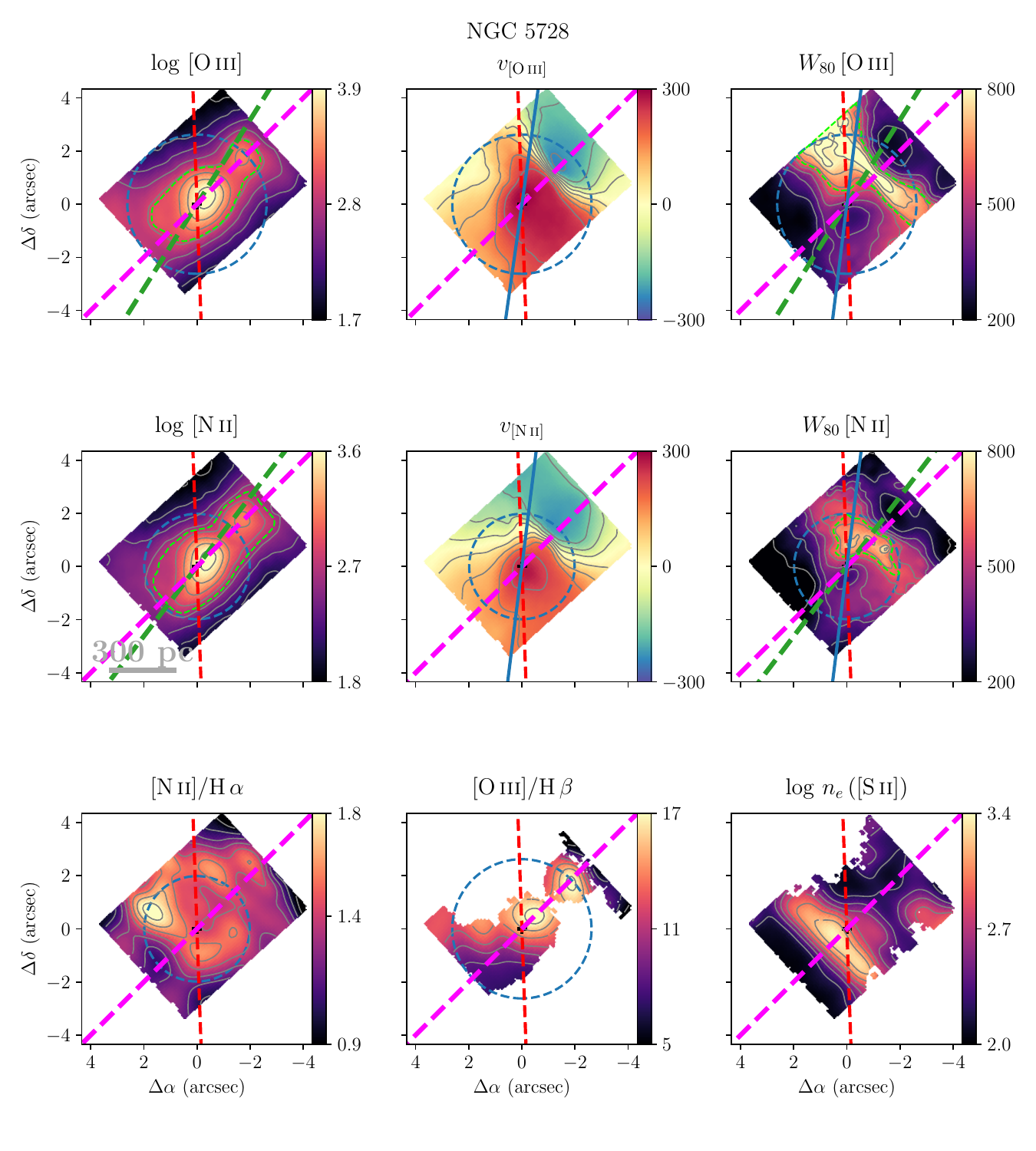}
    \caption{Continuation of figure \ref{fig:data_mrk348}.}
    \label{fig:data_ngc5728}
\end{figure*}

\begin{figure*}
    \centering
    \includegraphics[page=27,width=\textwidth]{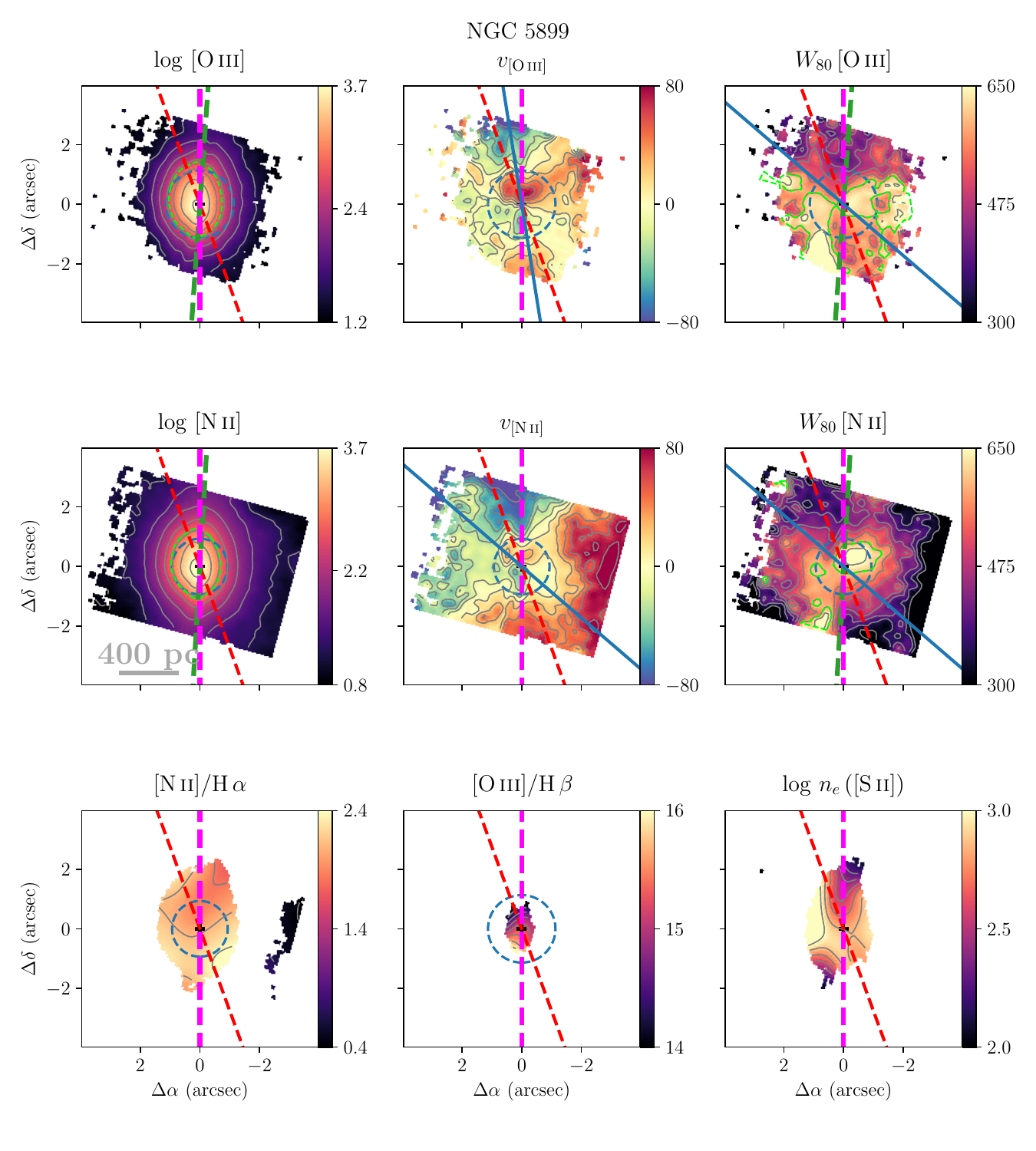}
    \caption{Continuation of figure \ref{fig:data_mrk348}.}
    \label{fig:data_ngc5899}
\end{figure*}

\begin{figure*}
    \centering
    \includegraphics[page=28,width=\textwidth]{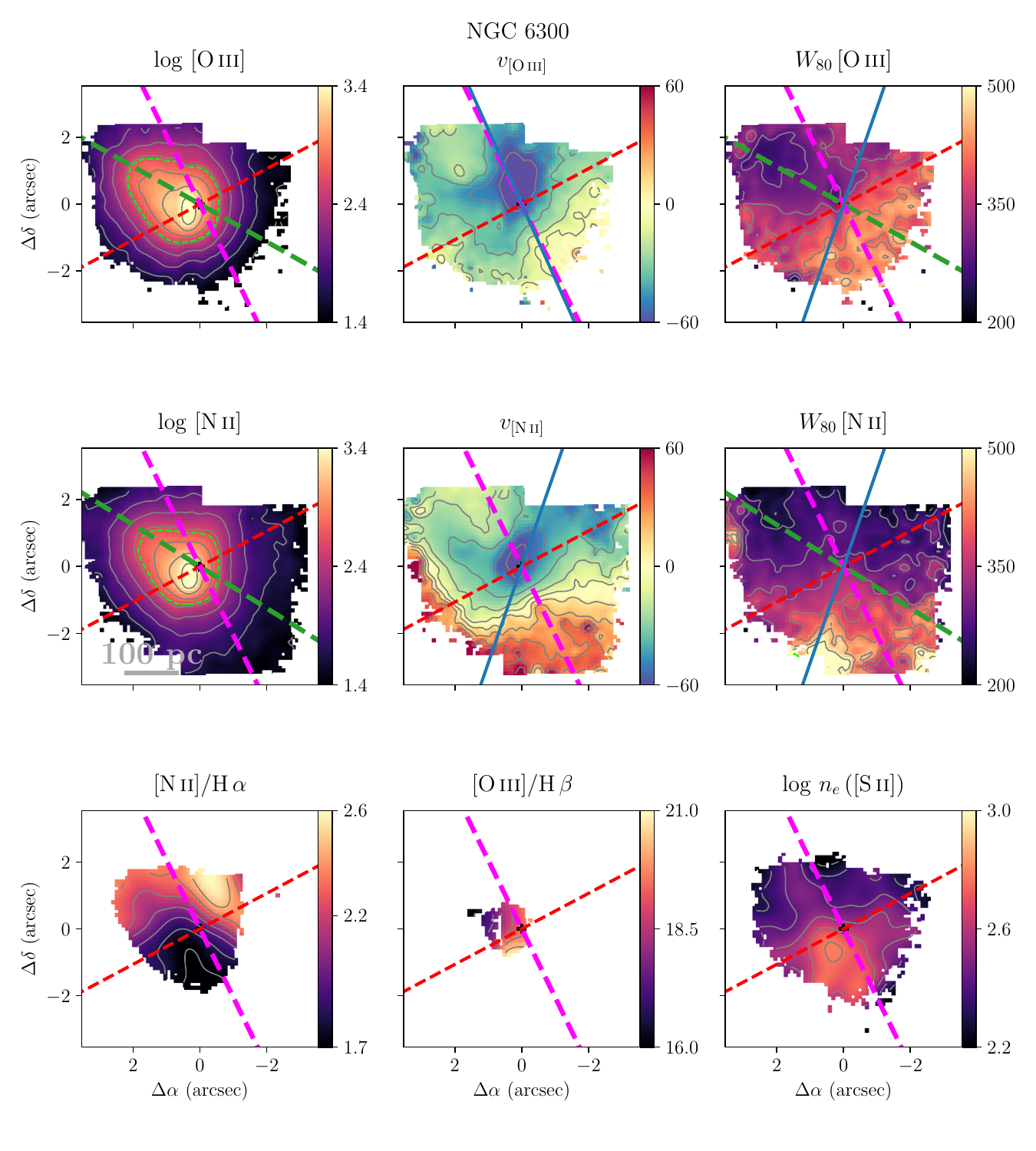}
    \caption{Continuation of figure \ref{fig:data_mrk348}.}
    \label{fig:data_ngc6300}
\end{figure*}

\begin{figure*}
    \centering
    \includegraphics[page=29,width=\textwidth]{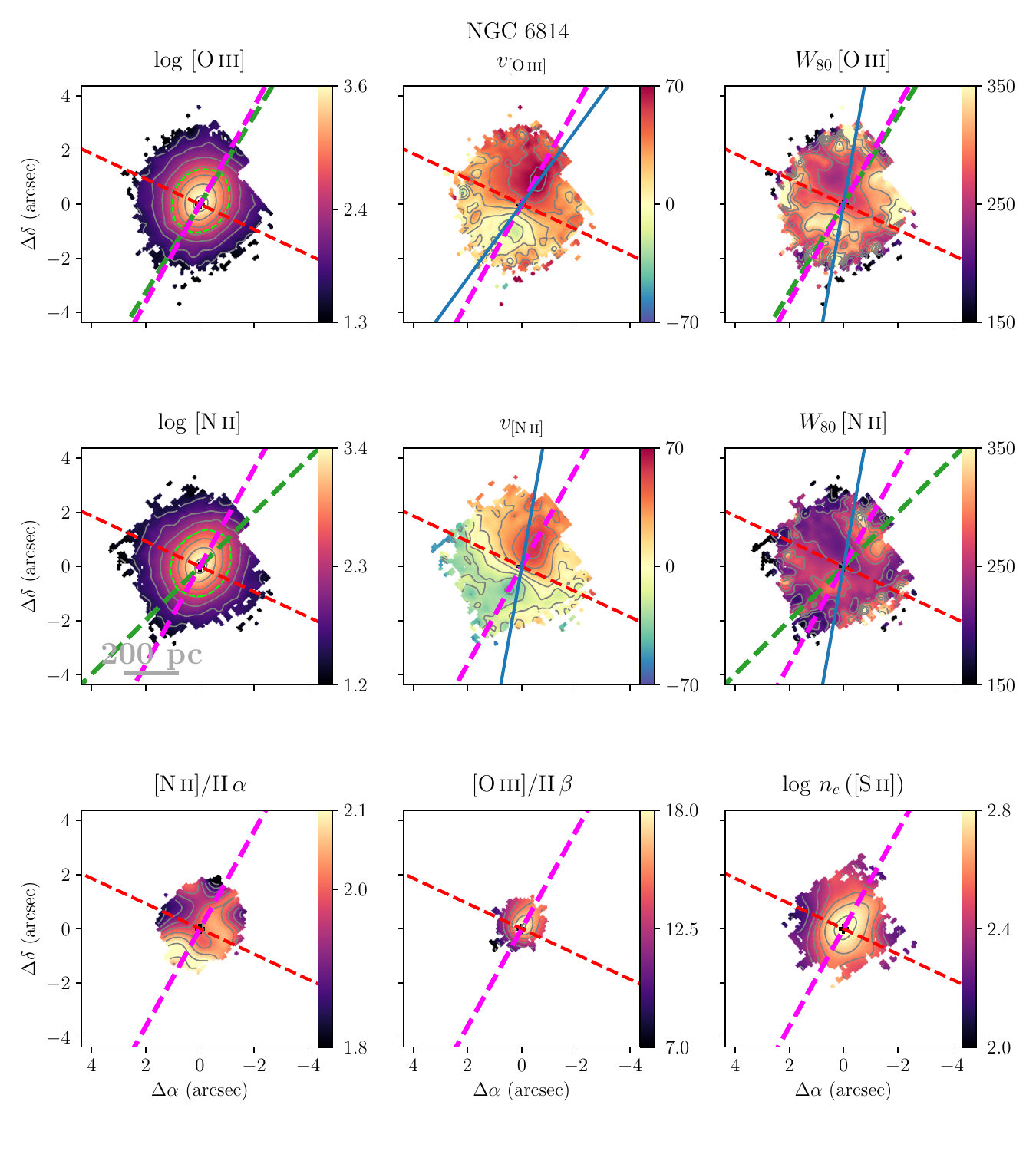}
    \caption{Continuation of figure \ref{fig:data_mrk348}.}
    \label{fig:data_ngc6814}
\end{figure*}

\begin{figure*}
    \centering
    \includegraphics[page=30,width=\textwidth]{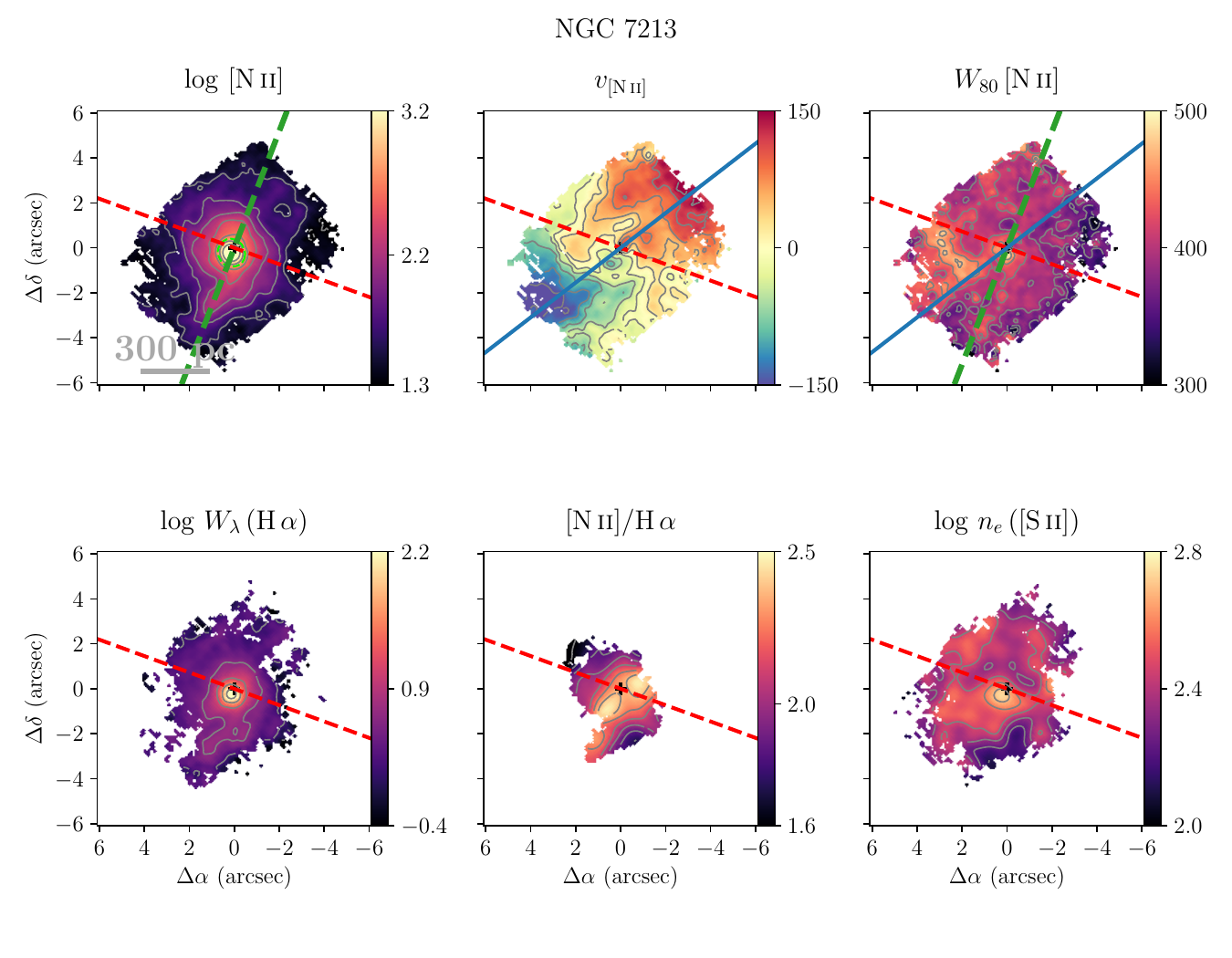}
    \caption{Continuation of figure \ref{fig:data_mrk348}.}
    \label{fig:data_ngc7213}
\end{figure*}

%% file: hermite_results.tex
\begin{figure*}
    \centering
    \includegraphics[page=1,width=\textwidth]{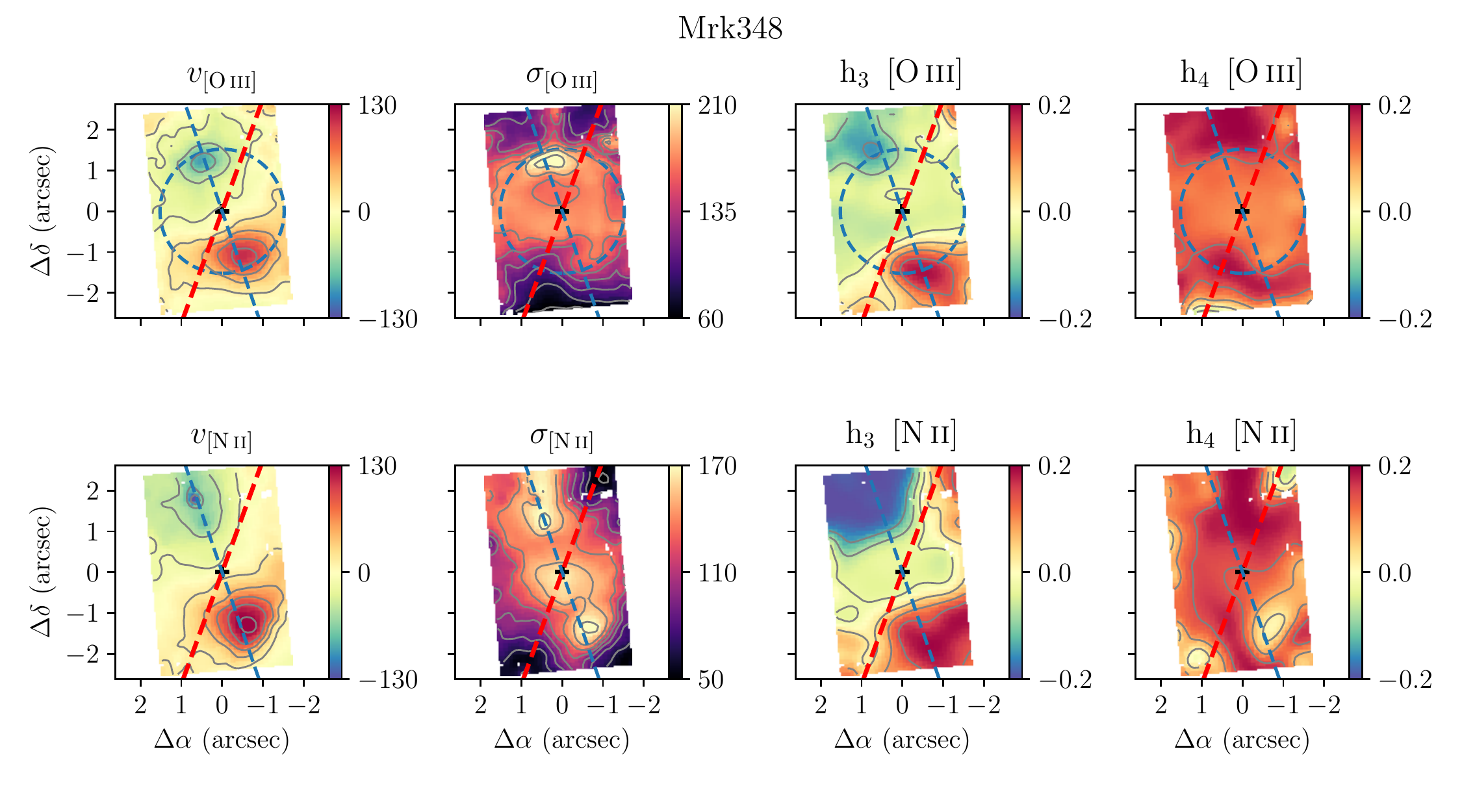}
    \caption{Summary of results for Mrk\,348. The bar in the lower left of
  each of the panels represents a projected proper distance of 100 pc. The
  colour coding is in units of ${\rm km\,s^{-1}}$ for the velocity and velocity
  dispersion. The ${\rm h_3}$ and ${\rm h_4}$ coefficients of the Gauss-Hermite
  polynomials are dimensionless quantities, and the limits of the colour scale
  reflect the actual bounds used in the fitting process. The systemic velocity
  of the galaxy, inferred from the stellar velocity field, has been subtracted
  from the radial velocity of the emission lines.}
    \label{fig:hermite_mrk348}
\end{figure*}

\begin{figure*}
    \centering
    \includegraphics[page=2,width=\textwidth]{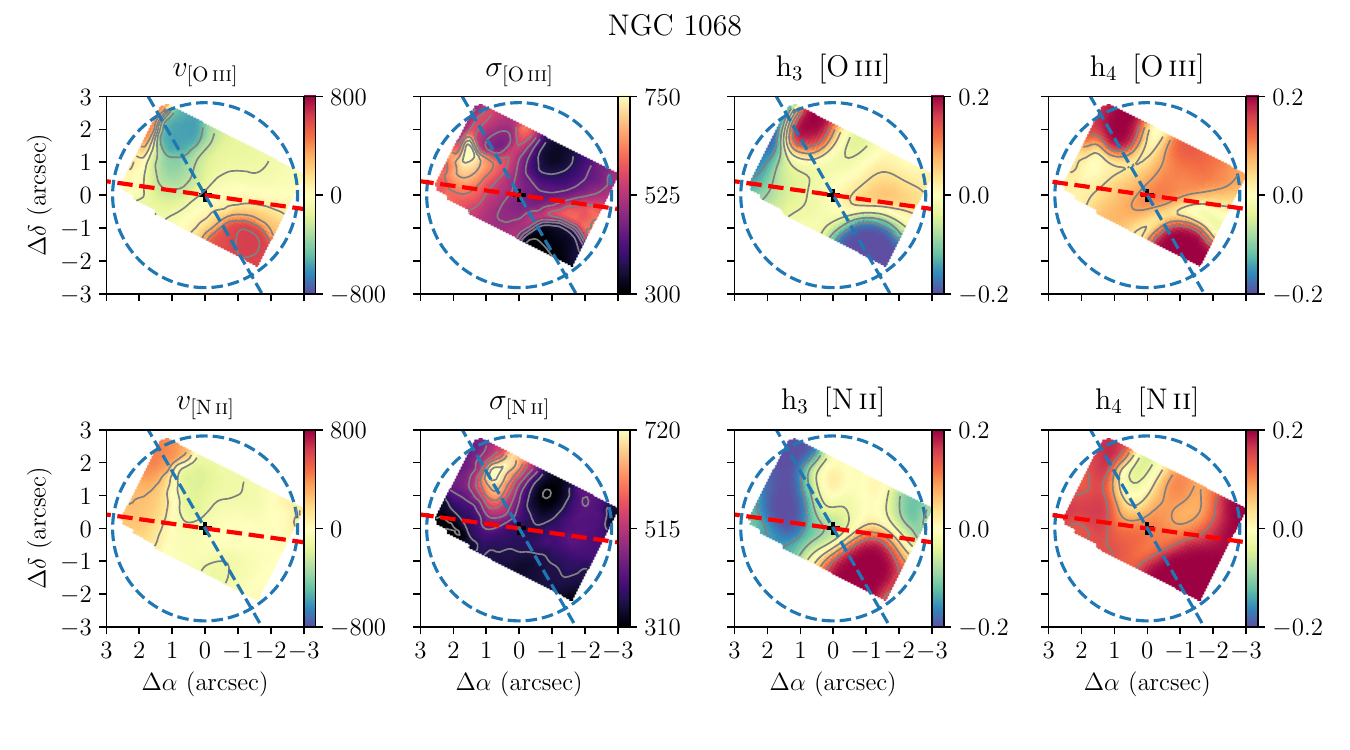}
    \caption{Same as \autoref{fig:hermite_mrk348}, but for NGC\,1068}
    \label{fig:hermite_ngc1068}
\end{figure*}

\clearpage

\begin{figure*}
    \centering
    \includegraphics[page=3,width=\textwidth]{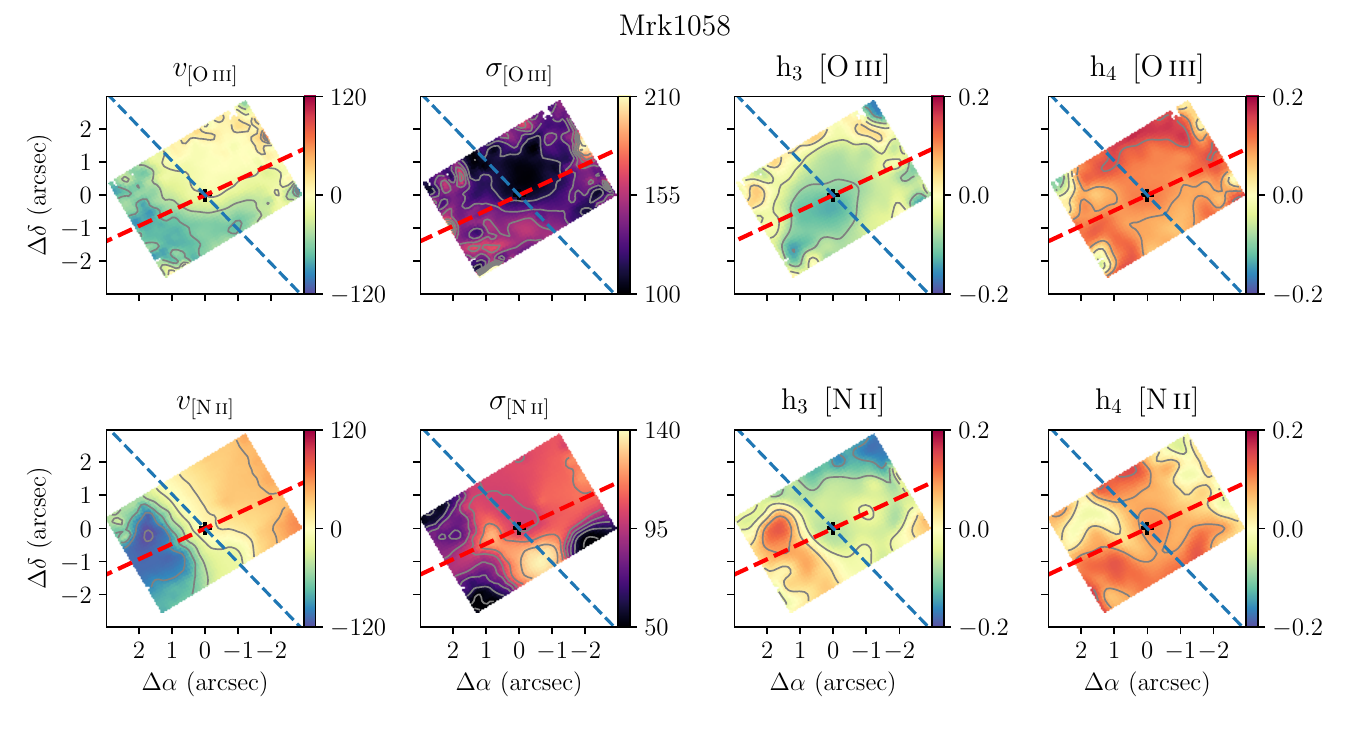}
    \caption{Same as \autoref{fig:hermite_ngc1068}, but for Mrk\,1058}
    \label{fig:hermite_mrk1058}
\end{figure*}

\begin{figure*}
    \centering
    \includegraphics[page=4,width=\textwidth]{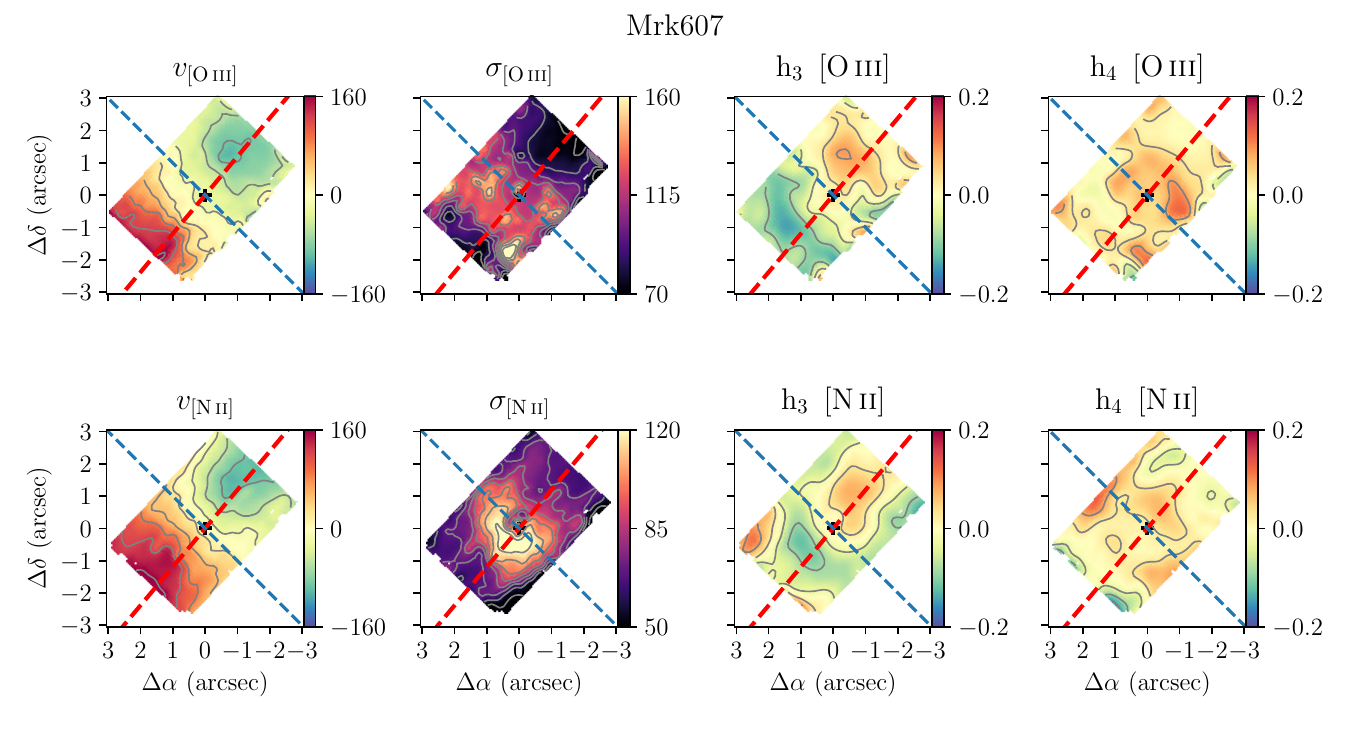}
    \caption{Same as \autoref{fig:hermite_ngc1068}, but for Mrk\,607}
    \label{fig:hermite_mrk607}
\end{figure*}


\begin{figure*}
    \centering
    \includegraphics[page=6,width=\textwidth]{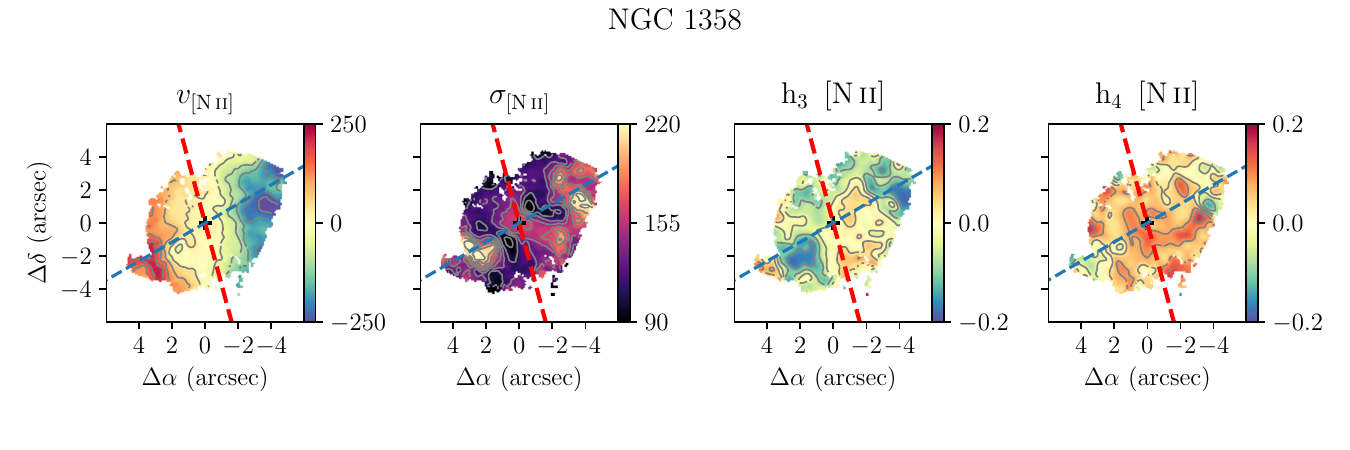}
    \caption{Same as \autoref{fig:hermite_ngc1068}, but for Mrk\,607}
    \label{fig:hermite_ngc1358}
\end{figure*}

\begin{figure*}
    \centering
    \includegraphics[page=7,width=\textwidth]{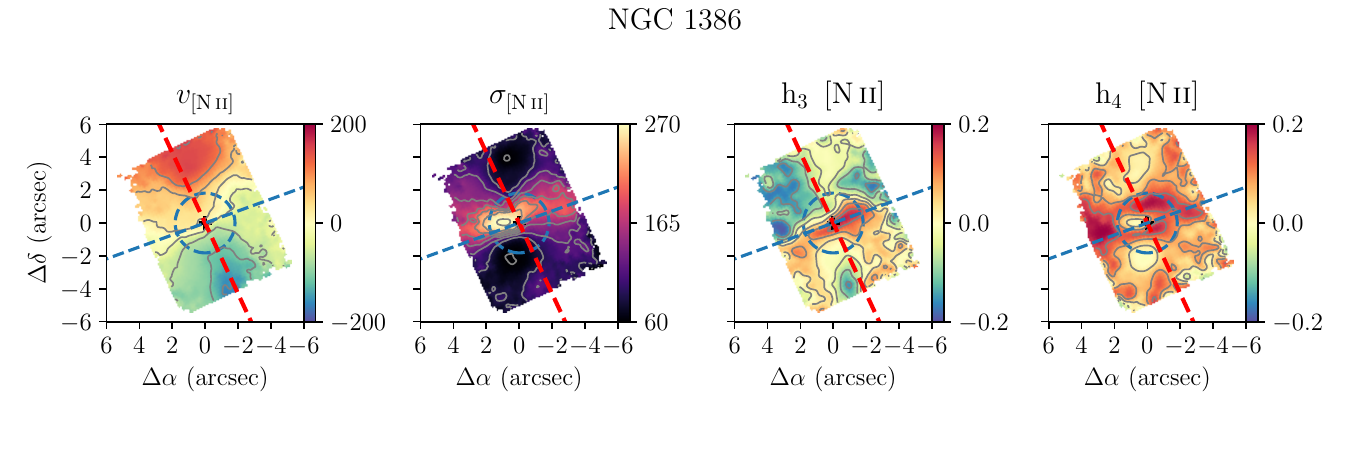}
    \caption{Same as \autoref{fig:hermite_ngc1068}, but for Mrk\,607}
    \label{fig:hermite_ngc1386}
\end{figure*}

\begin{figure*}
    \centering
    \includegraphics[page=8,width=\textwidth]{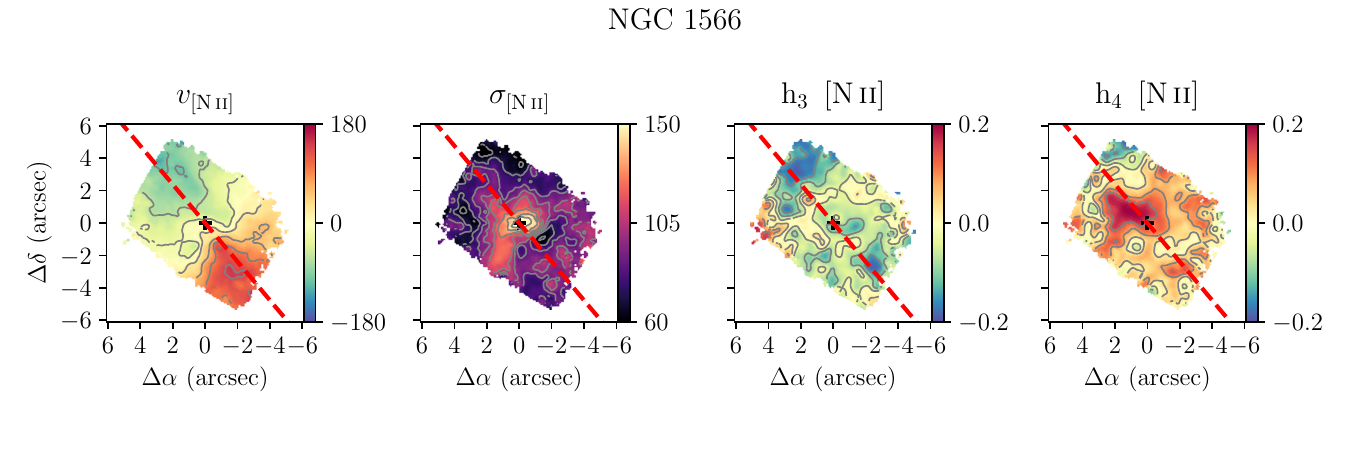}
    \caption{Same as \autoref{fig:hermite_ngc1068}, but for NGC\,1566}
    \label{fig:hermite_ngc1566}
\end{figure*}

\begin{figure*}
    \centering
    \includegraphics[page=9,width=\textwidth]{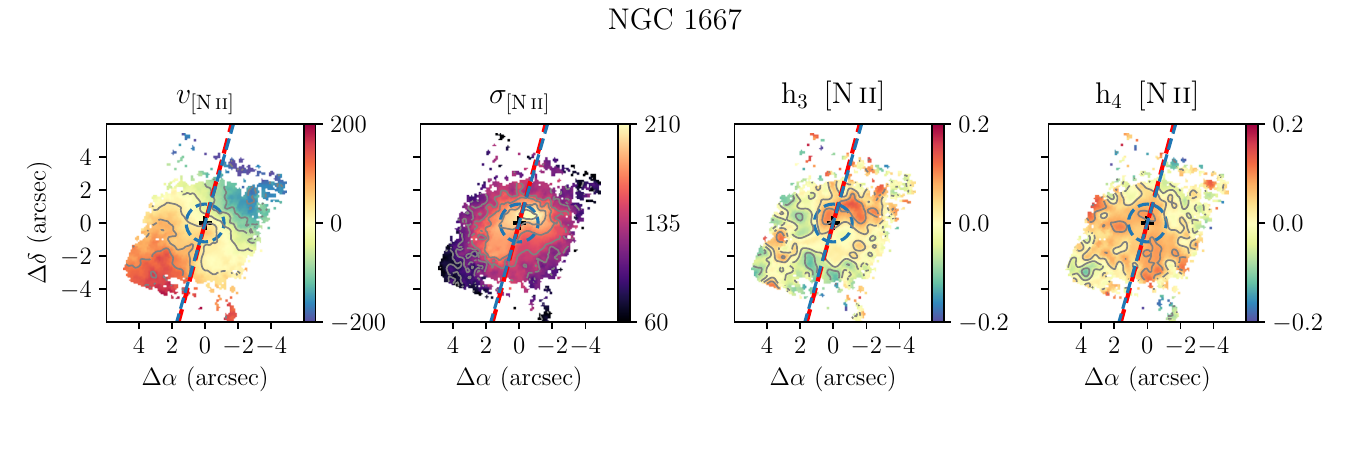}
    \caption{Same as \autoref{fig:hermite_ngc1068}, but for NGC\,1667}
    \label{fig:hermite_ngc1667}
\end{figure*}

\begin{figure*}
    \centering
    \includegraphics[page=10,width=\textwidth]{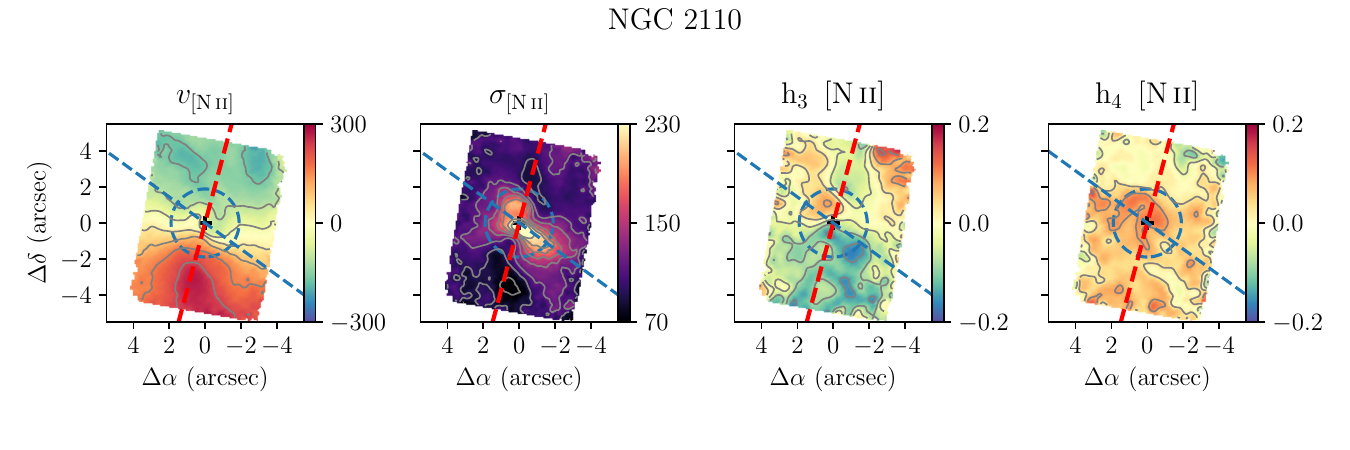}
    \caption{Same as \autoref{fig:hermite_ngc1068}, but for NGC\,2110}
    \label{fig:hermite_ngc2110}
\end{figure*}

\begin{figure*}
    \centering
    \includegraphics[page=11,width=\textwidth]{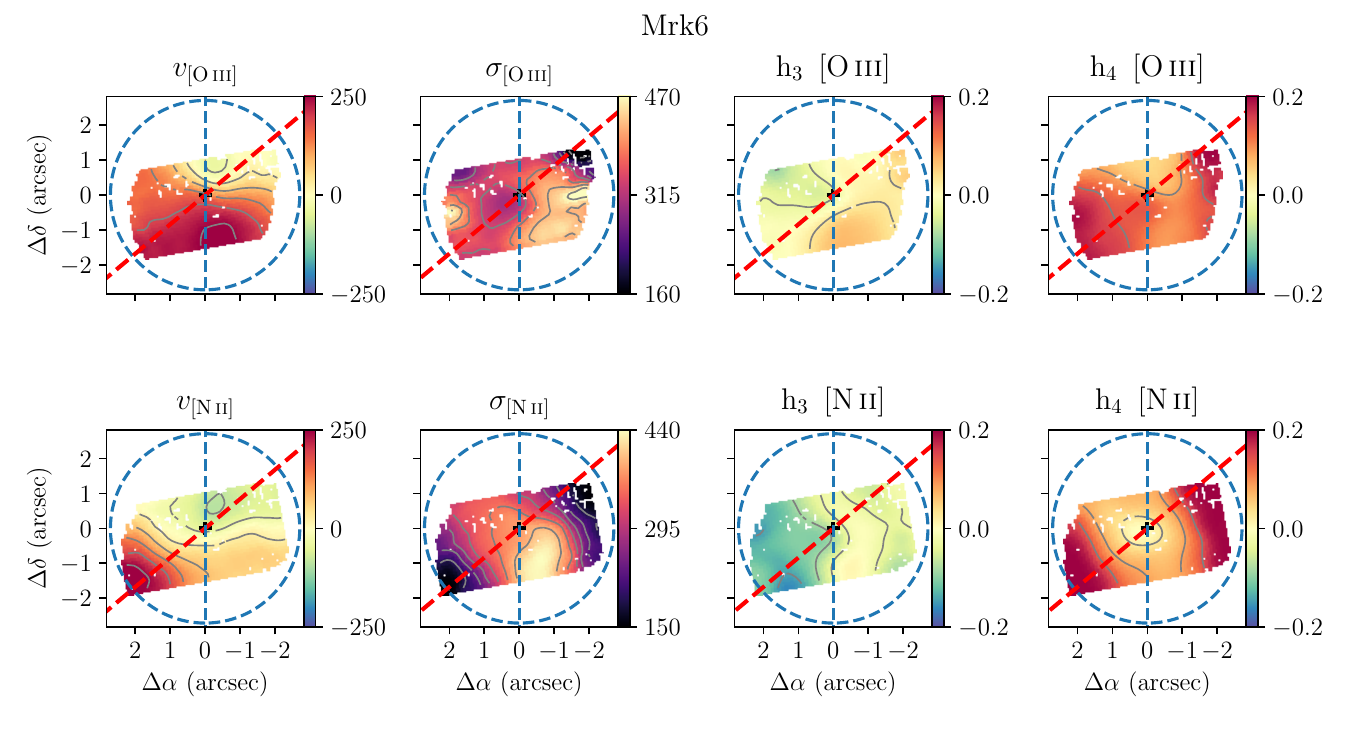}
    \caption{Same as \autoref{fig:hermite_ngc1068}, but for Mrk\,6}
    \label{fig:hermite_mrk6}
\end{figure*}

\begin{figure*}
    \centering
    \includegraphics[page=12,width=\textwidth]{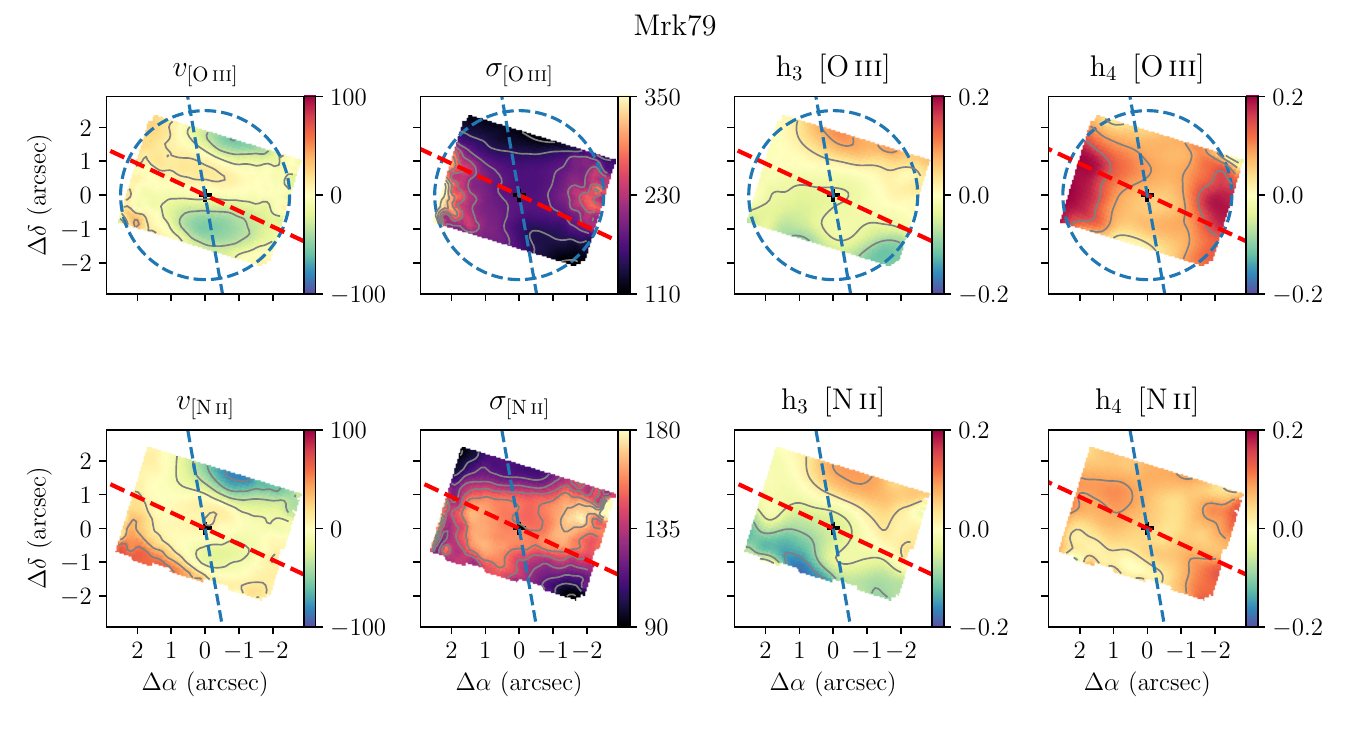}
    \caption{Same as \autoref{fig:hermite_ngc1068}, but for Mrk\,79}
    \label{fig:hermite_mrk79}
\end{figure*}

\clearpage

\begin{figure*}
    \centering
    \includegraphics[page=13,width=\textwidth]{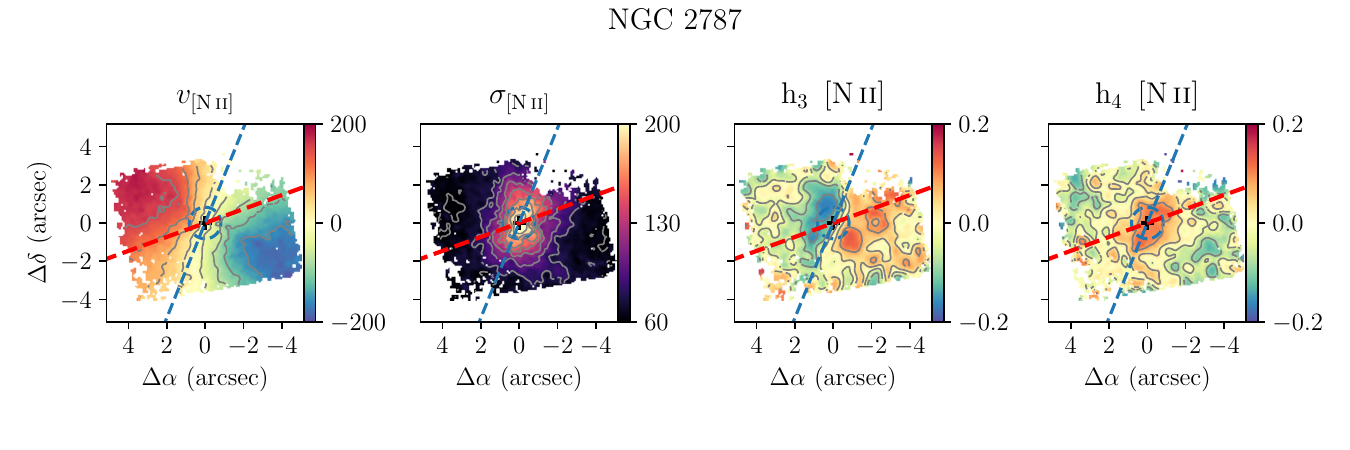}
    \caption{Same as \autoref{fig:hermite_ngc1068}, but for NGC\,2787}
    \label{fig:hermite_ngc2787}
\end{figure*}

\begin{figure*}
    \centering
    \includegraphics[page=14,width=\textwidth]{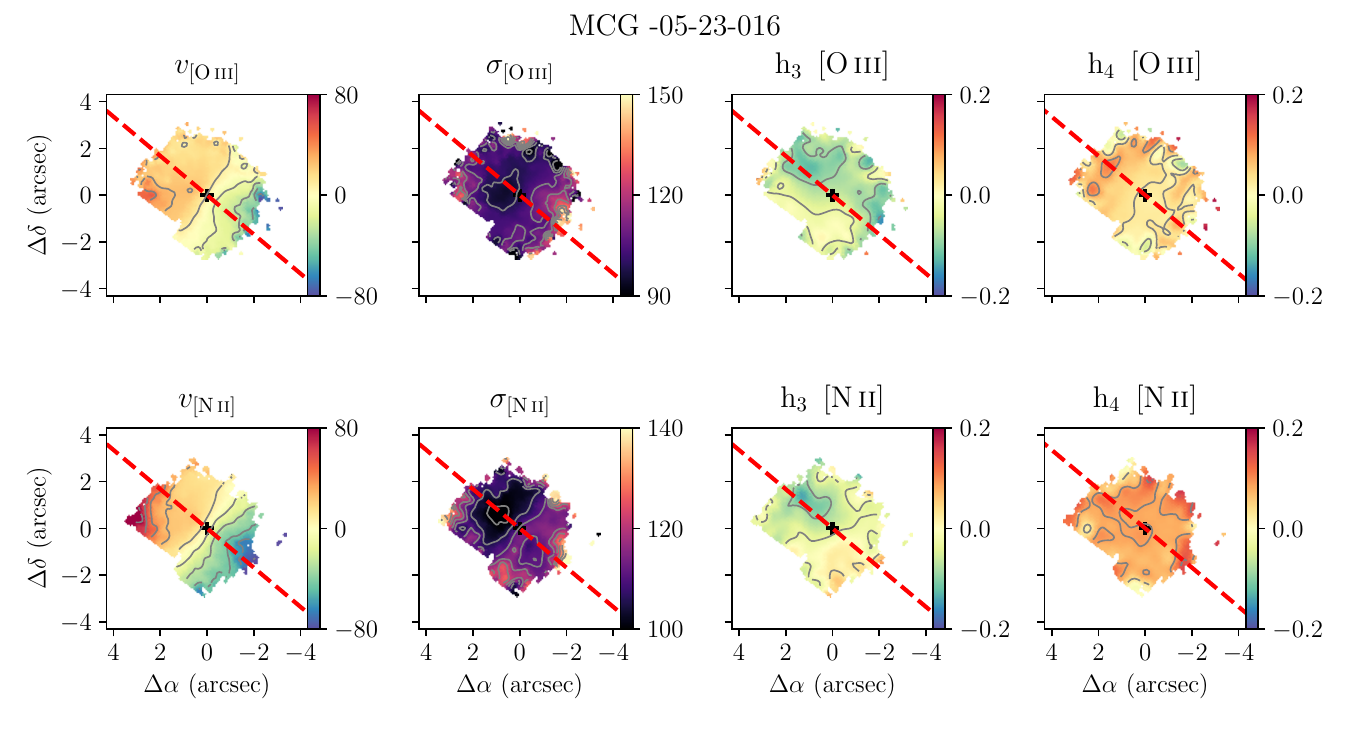}
    \caption{Same as \autoref{fig:hermite_ngc1068}, but for MCG\,-05-23-016}
    \label{fig:hermite_mcg0523}
\end{figure*}

\clearpage

\begin{figure*}
    \centering
    \includegraphics[page=15,width=\textwidth]{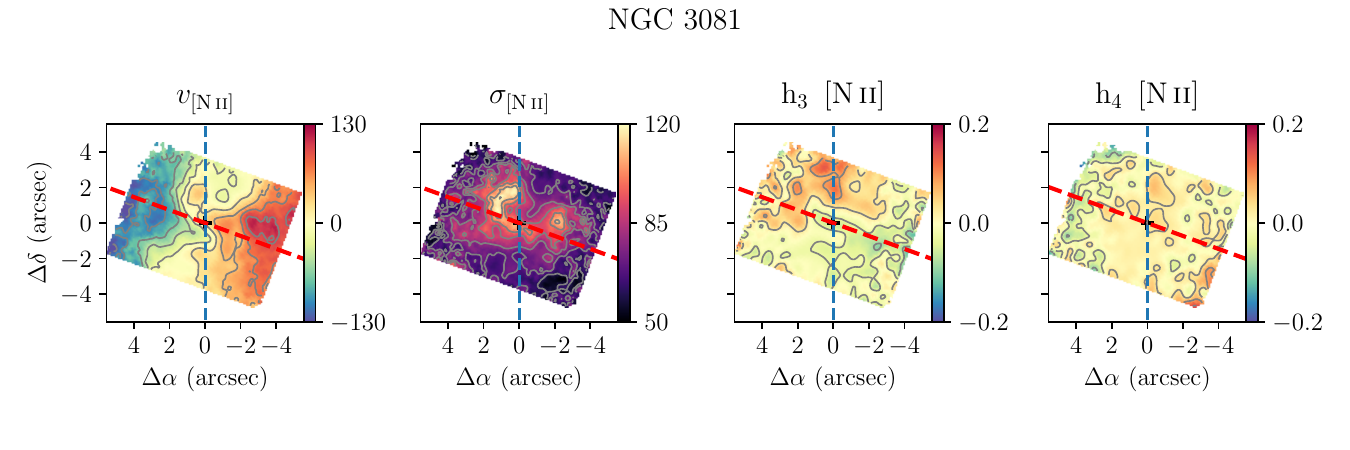}
    \caption{Same as \autoref{fig:hermite_ngc1068}, but for Mrk\,3081}
    \label{fig:hermite_ngc3081}
\end{figure*}


\begin{figure*}
    \centering
    \includegraphics[page=17,width=\textwidth]{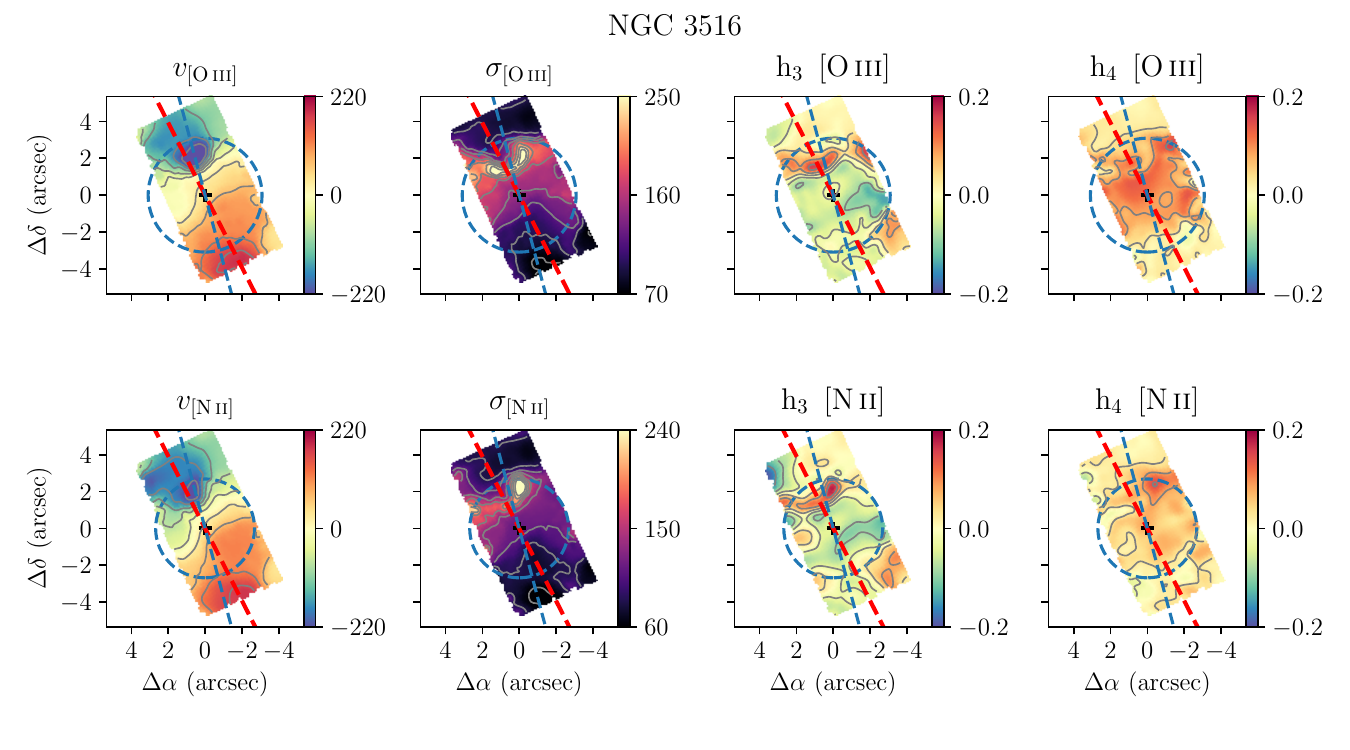}
    \caption{Same as \autoref{fig:hermite_ngc1068}, but for NGC\,3516}
    \label{fig:hermite_ngc3516}
\end{figure*}
%
%
%
%
\begin{figure*}
    \centering
    \includegraphics[page=19,width=\textwidth]{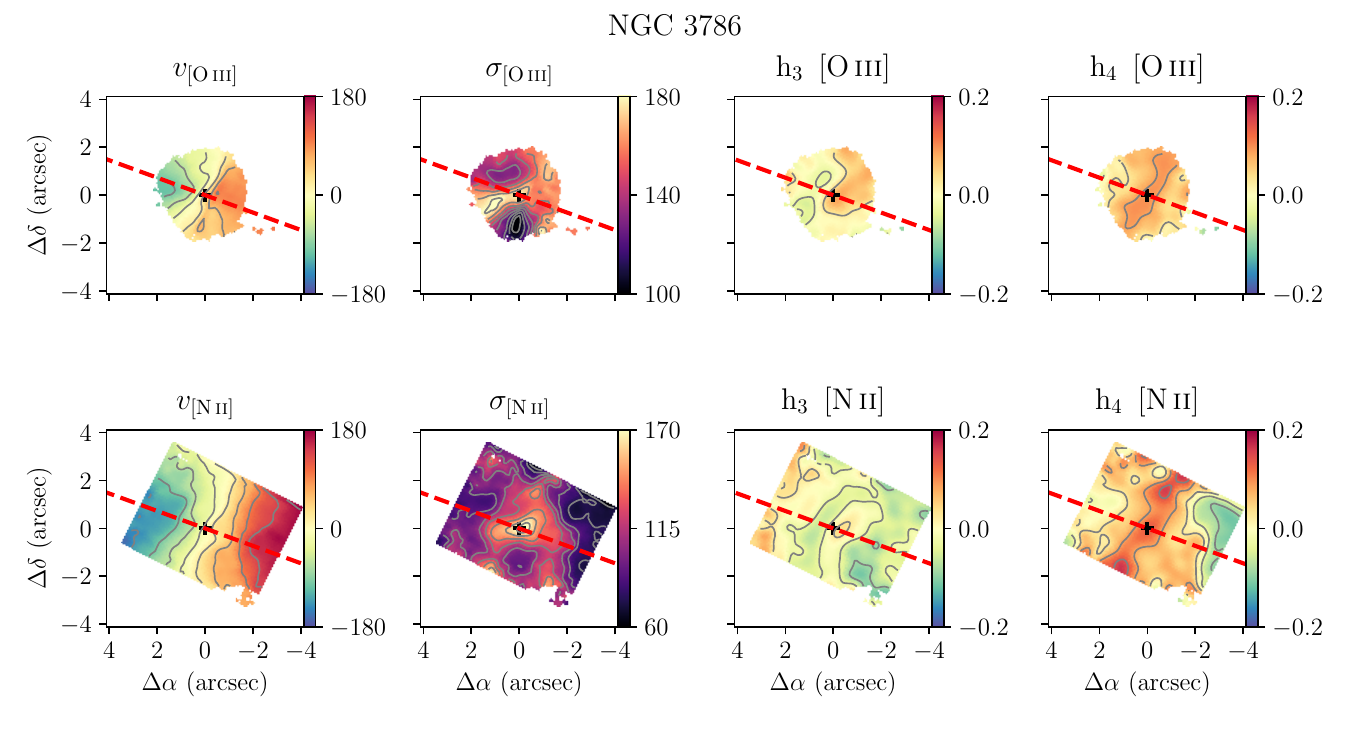}
    \caption{Same as \autoref{fig:hermite_ngc1068}, but for NGC\,3786}
    \label{fig:hermite_ngc3786}
\end{figure*}

\begin{figure*}
    \centering
    \includegraphics[page=20,width=\textwidth]{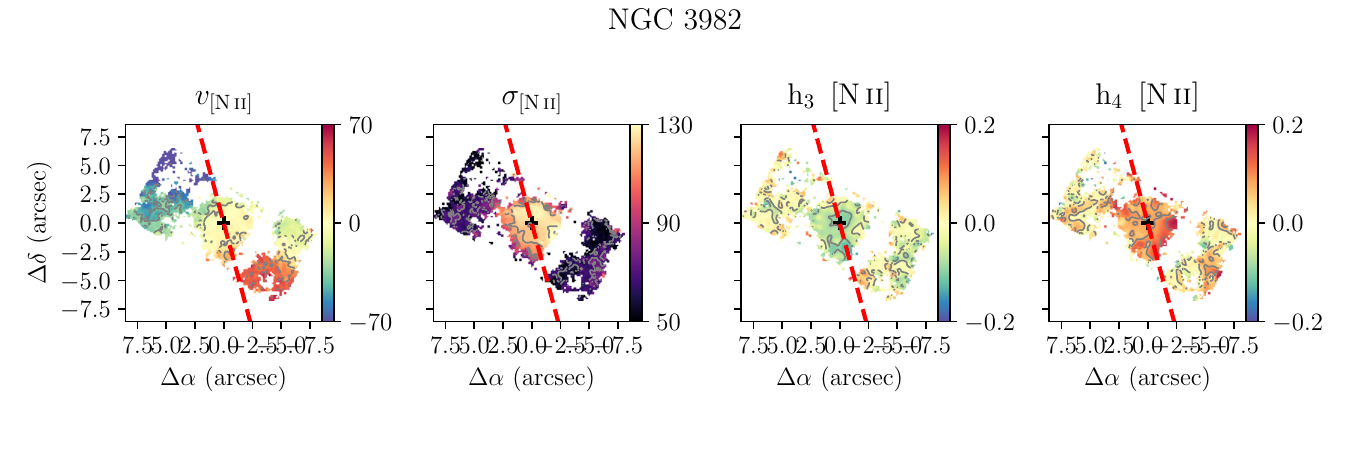}
    \caption{Same as \autoref{fig:hermite_ngc1068}, but for NGC\,3982}
    \label{fig:hermite_ngc3982}
\end{figure*}

\begin{figure*}
    \centering
    \includegraphics[page=21,width=\textwidth]{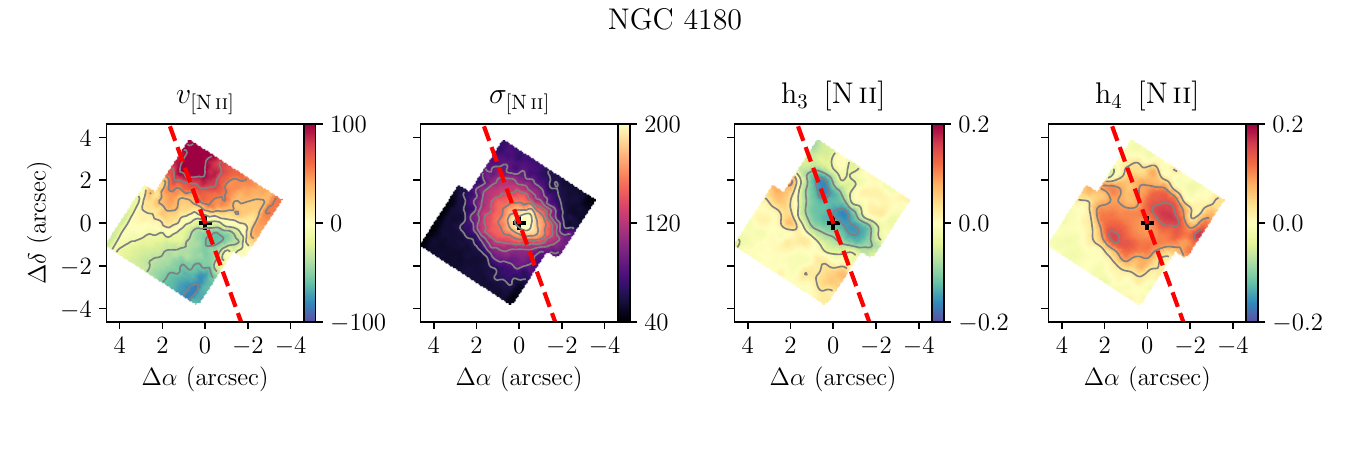}
    \caption{Same as \autoref{fig:hermite_ngc1068}, but for NGC\,4180}
    \label{fig:hermite_ngc4180}
\end{figure*}
\begin{figure*}
    \centering
    \includegraphics[page=22,width=\textwidth]{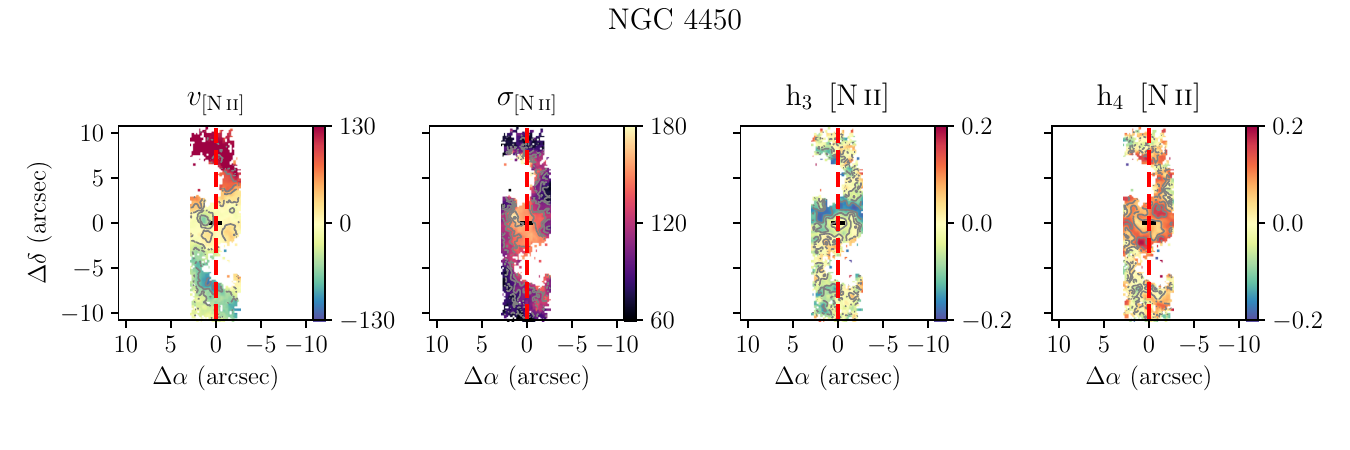}
    \caption{Same as \autoref{fig:hermite_ngc1068}, but for NGC\,4450}
    \label{fig:hermite_ngc4450}
\end{figure*}
\begin{figure*}
    \centering
    \includegraphics[page=23,width=\textwidth]{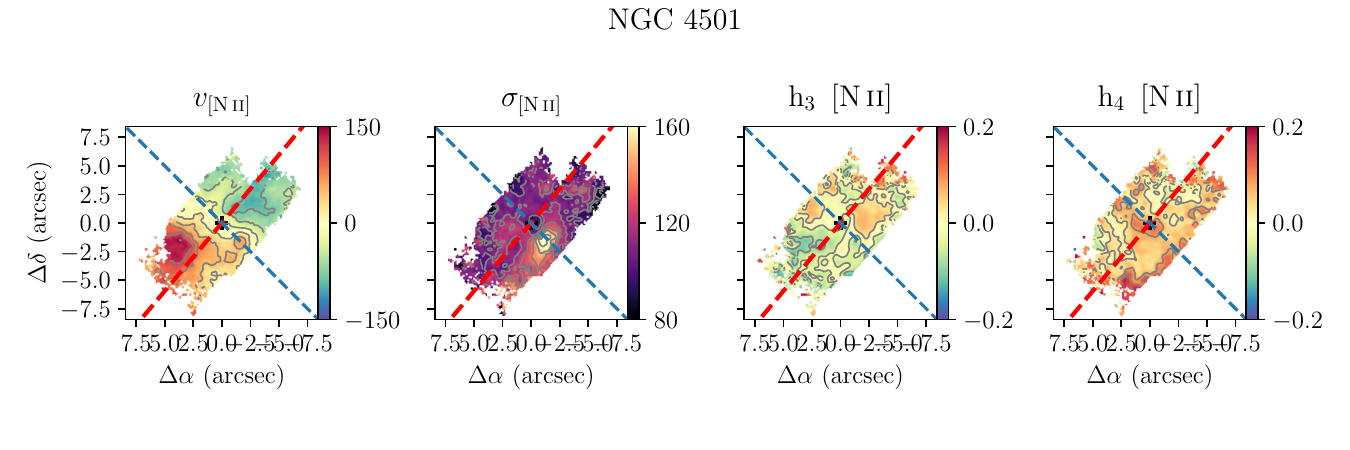}
    \caption{Same as \autoref{fig:hermite_ngc1068}, but for NGC\,4501}
    \label{fig:hermite_ngc4501}
\end{figure*}
%
%
%
%
%
%
\begin{figure*}
    \centering
    \includegraphics[page=26,width=\textwidth]{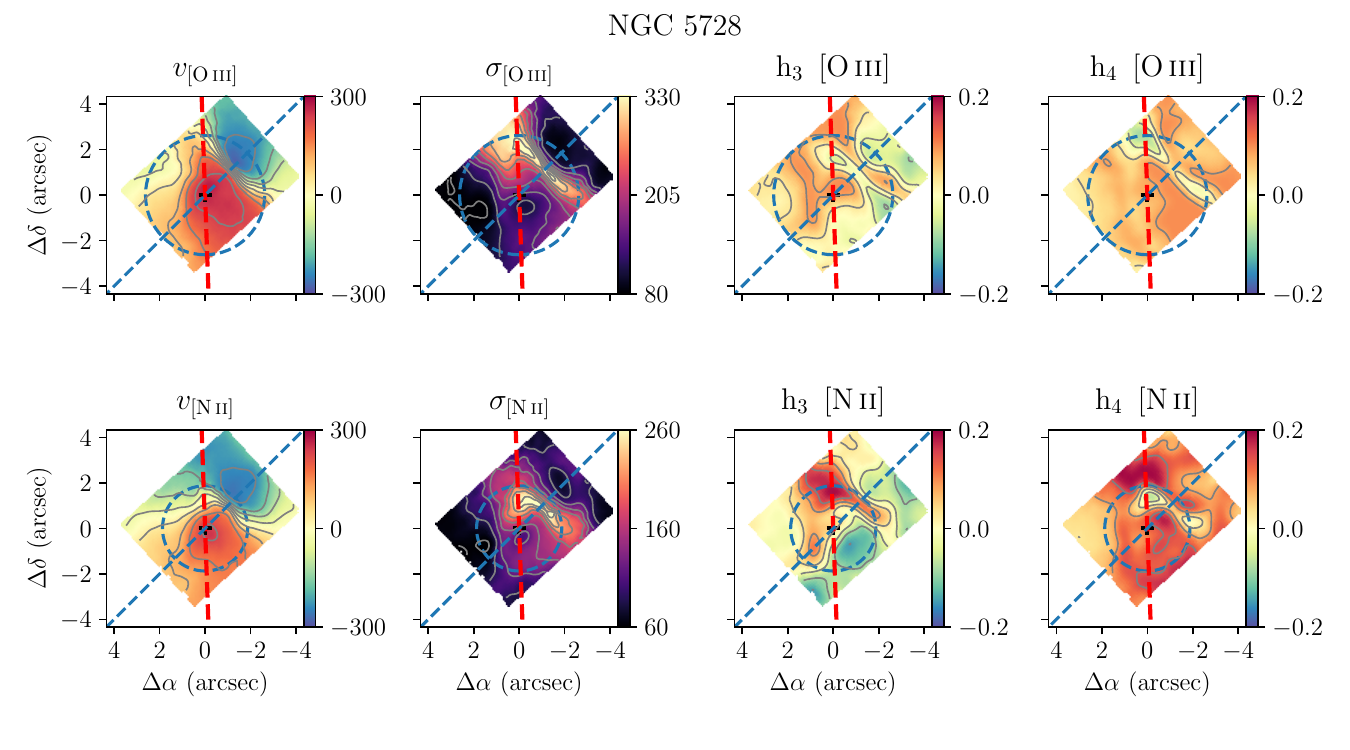}
    \caption{Same as \autoref{fig:hermite_ngc1068}, but for NGC\,5728}
    \label{fig:hermite_ngc5728}
\end{figure*}

\begin{figure*}
    \centering
    \includegraphics[page=27,width=\textwidth]{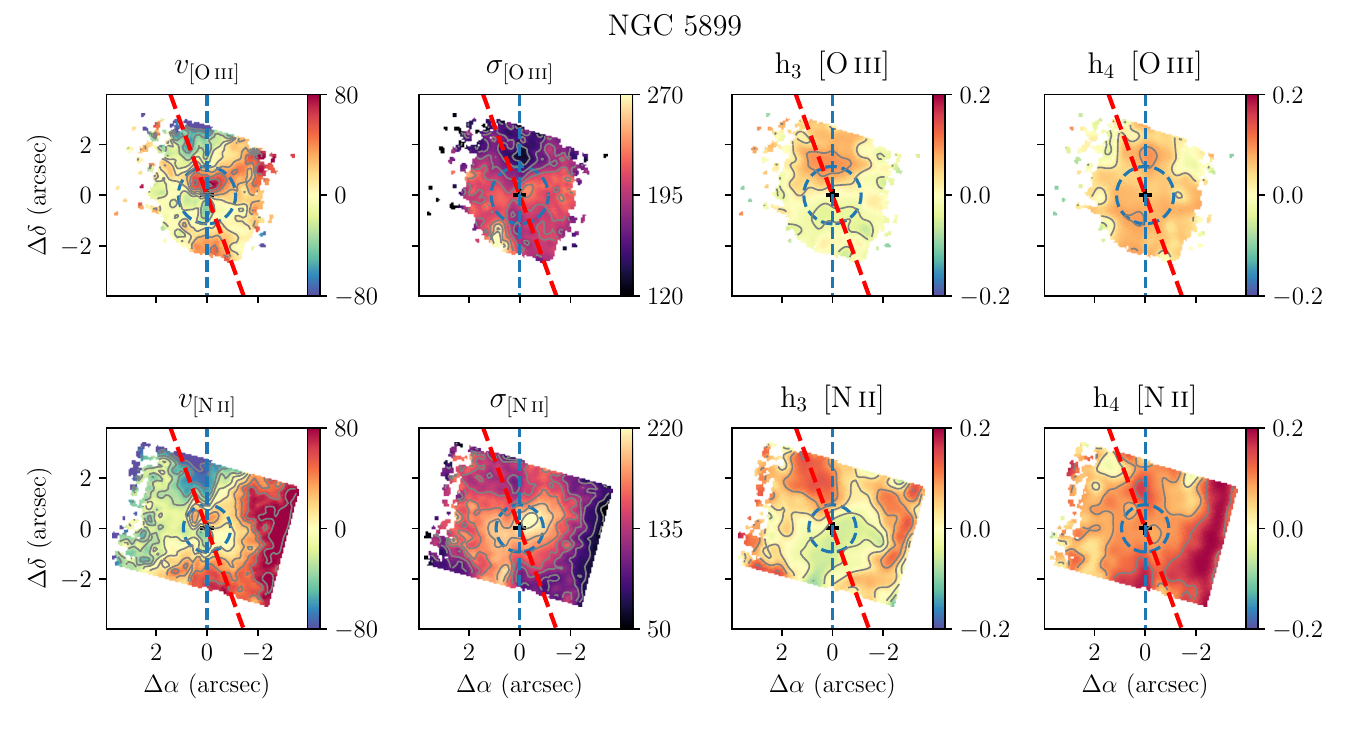}
    \caption{Same as \autoref{fig:hermite_ngc1068}, but for NGC\,5899}
    \label{fig:hermite_ngc5899}
\end{figure*}
\begin{figure*}
    \centering
    \includegraphics[page=28,width=\textwidth]{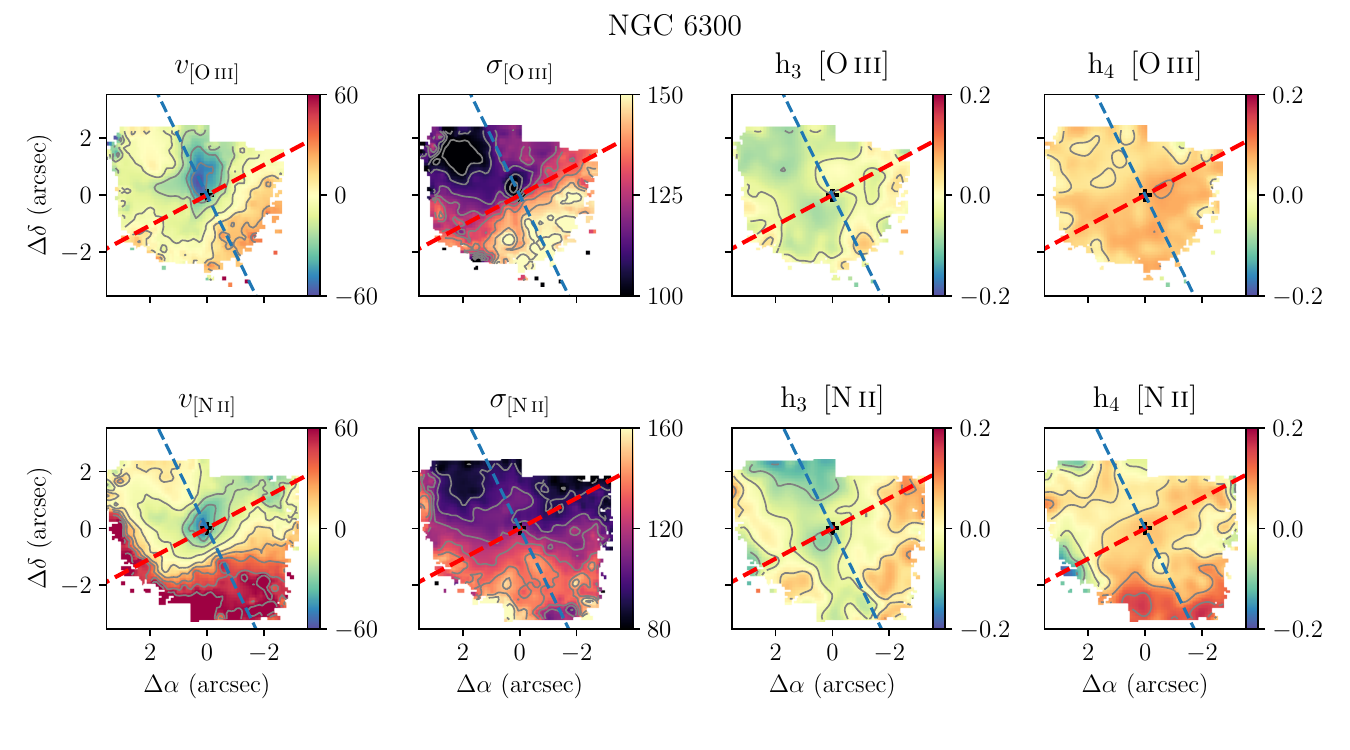}
    \caption{Same as \autoref{fig:hermite_ngc1068}, but for NGC\,6300}
    \label{fig:hermite_ngc6300}
\end{figure*}
%
%
%
%
%
%

%% file: whan.tex

\begin{figure*}
  \centering
  \includegraphics[page=1, width=\textwidth]{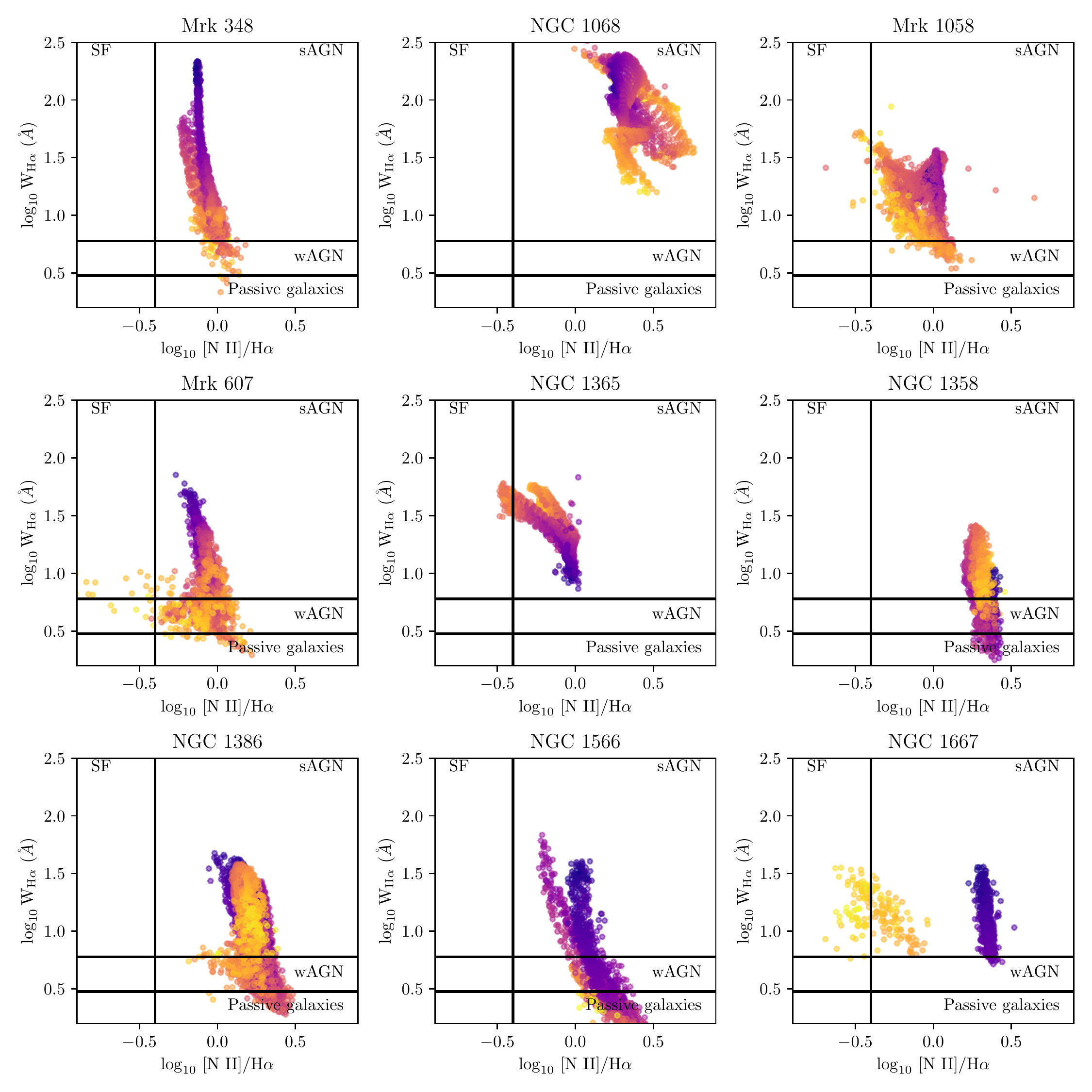}
  \caption{
      WHAN diagrams for all the galaxies in the sample.
      The defining lines for each category are based on \citet{cidfernandes2010b}.
      Colors represent the distance from the nucleus, increasing from blue to orange.
  }
  \label{fig:whan1}
\end{figure*}

\begin{figure*}
  \centering
  \includegraphics[page=2, width=\textwidth]{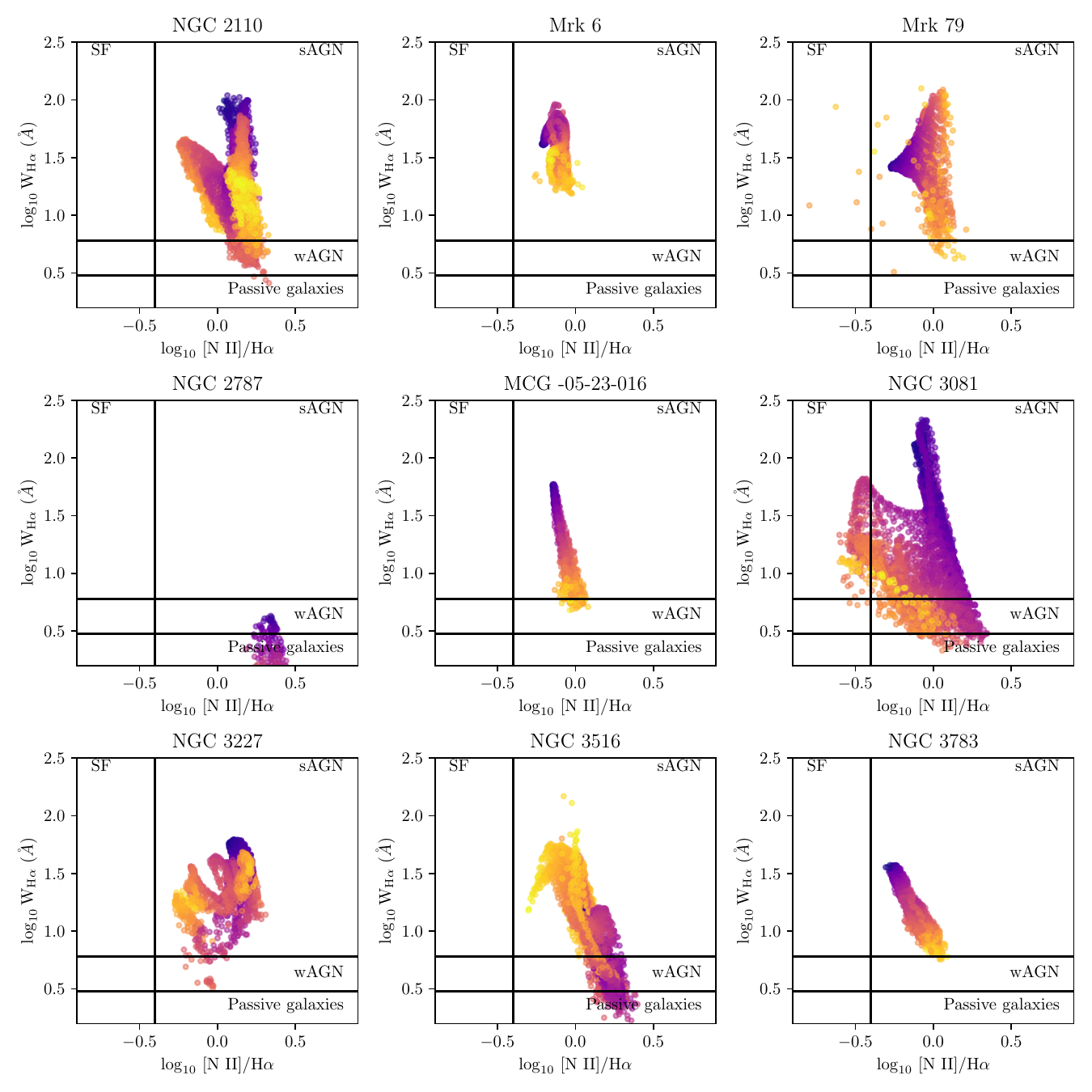}
  \caption{Continuation of \autoref{fig:whan1}.}
  \label{fig:whan2}
\end{figure*}

\begin{figure*}
  \centering
  \includegraphics[page=3, width=\textwidth]{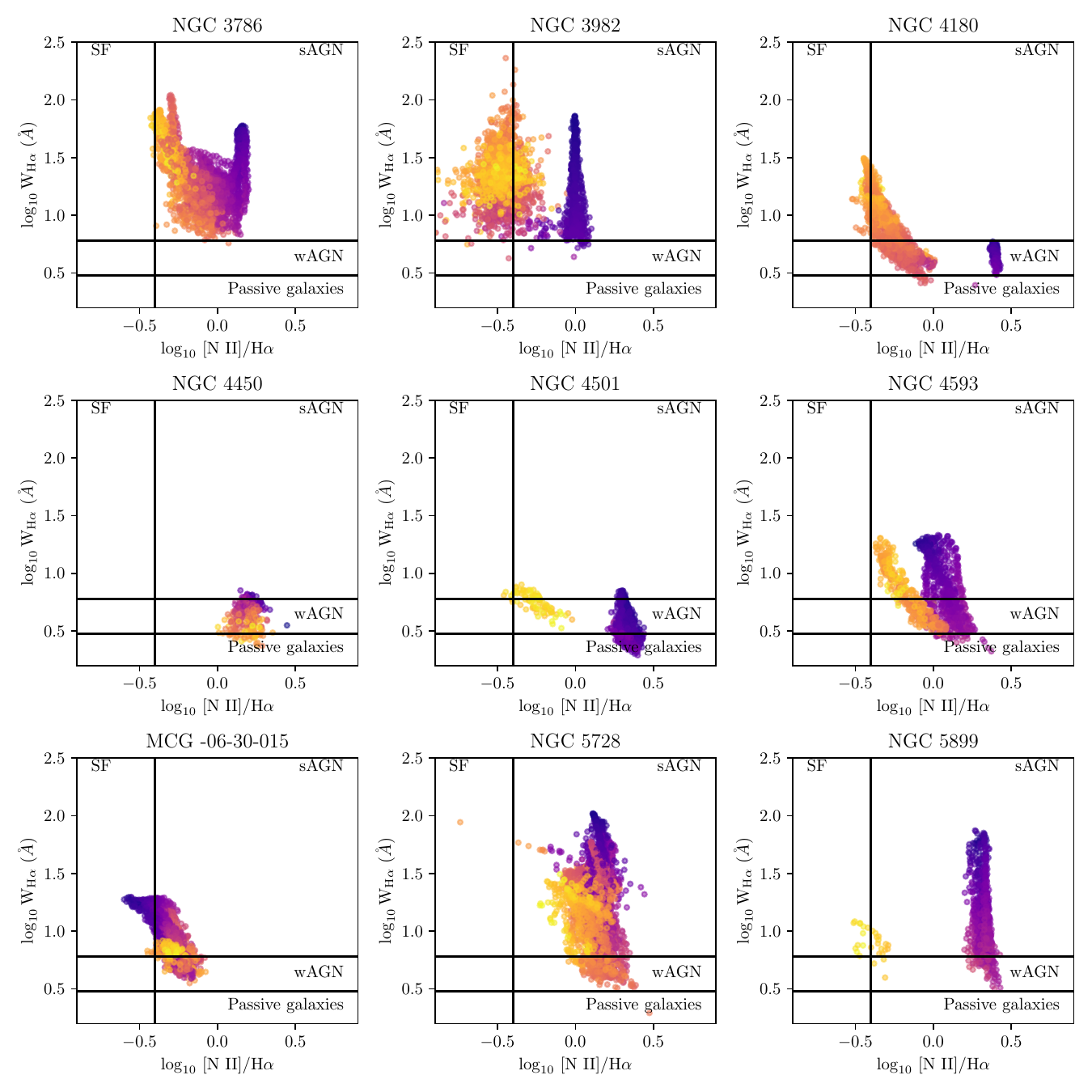}
  \caption{Continuation of \autoref{fig:whan1}.}
  \label{fig:whan3}
\end{figure*}

\begin{figure*}
  \centering
  \includegraphics[page=4, width=\textwidth]{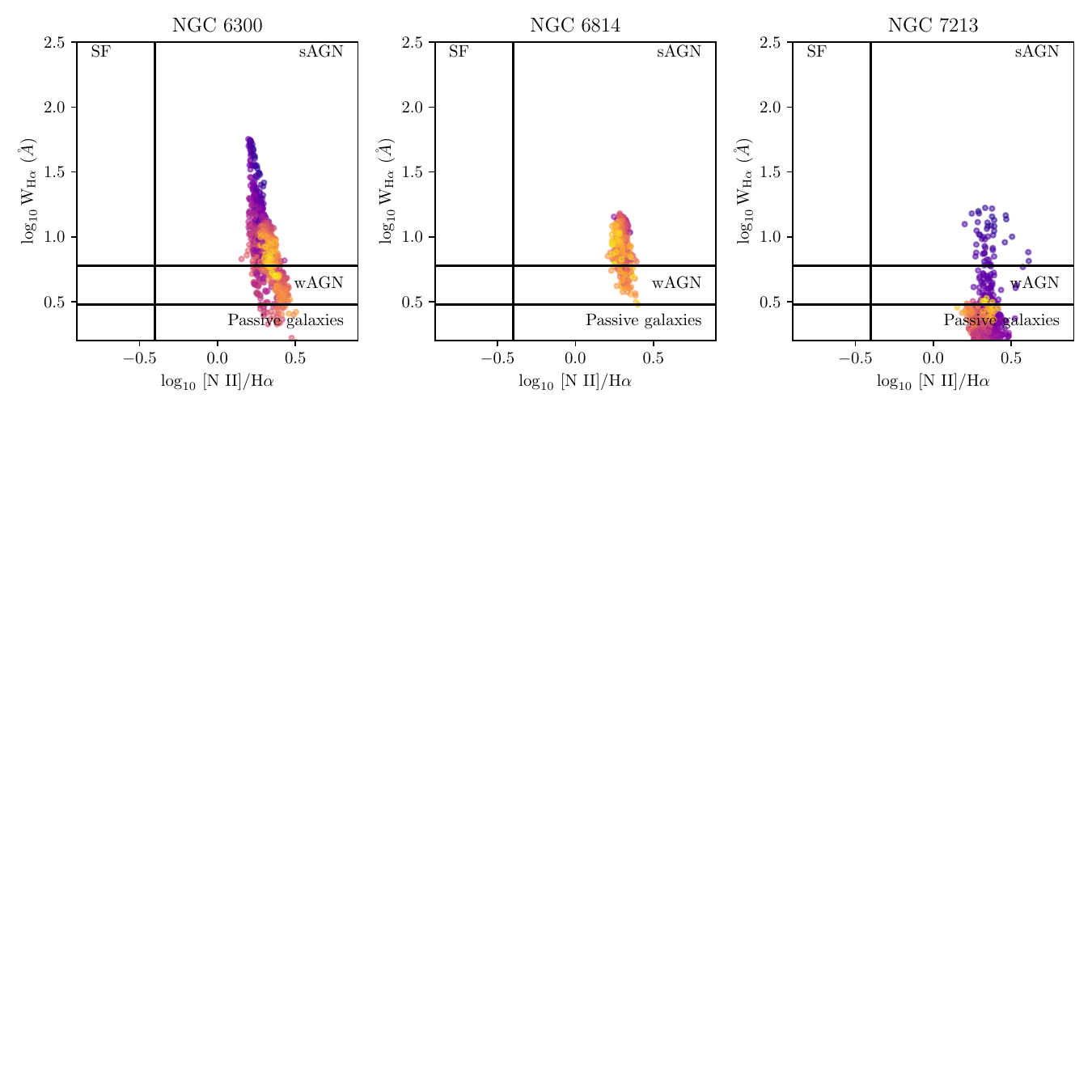}
  \caption{Continuation of \autoref{fig:whan1}.}
  \label{fig:whan4}
\end{figure*}